\documentclass[aps,prb,preprint,onecolumn,citeautoscript,superscriptaddress,footinbib,
eqsecnum]{revtex4-1}  
\usepackage{amsmath,amssymb,bm} 
\usepackage{graphicx}  
\usepackage{color} 
\usepackage[papersize={8.5in,11in}]{geometry}
\usepackage[colorlinks=true]{hyperref}  
\hypersetup{
    bookmarks=true,         % show bookmarks bar?
    unicode=false,          % non-Latin characters 
    pdftoolbar=true,        % show Acrobat
    pdfmenubar=true,        % show Acrobat 
    pdffitwindow=false,     % window fit to page when opened
    pdfstartview={FitH},    % fits the width of the page to the window
    pdftitle={Thermoelectric transport in SYK models and holography},    % title
    pdfauthor={Subir Sachdev},     % author
    pdfsubject={},   % subject of the document
    pdfcreator={},   % creator of the document
    pdfproducer={}, % producer of the document
    pdfkeywords={} {} {}, % list of keywords
    pdfnewwindow=true,      % links in new window
    colorlinks=true,       % false: boxed links; true: colored links
    linkcolor=magenta, %red,          % color of internal links (change box color with linkbordercolor)
    citecolor=blue,        % color of links to bibliography
    filecolor=magenta,      % color of file links
    urlcolor=blue           % color of external links
} 

\usepackage{amsfonts}
\usepackage{upgreek}
\usepackage{slashed}
\usepackage{latexsym}
\usepackage{todonotes}

\newcommand{\beq}{\begin{equation}}
\newcommand{\eeq}{\end{equation}}
\def\bea{\begin{eqnarray}}
\def\eea{\end{eqnarray}}

\newcommand{\nn}{\nonumber \\}
\newcommand{\eff}{\operatorname{eff.}}

\newcommand{\UU}{\operatorname{U}}

\newcommand{\PSL}{\operatorname{PSL}}

\renewcommand{\leq}{\leqslant}
\renewcommand{\geq}{\geqslant}
\renewcommand{\bar}{\overline}
\renewcommand{\tilde}{\widetilde}

\makeatletter
\newsavebox{\@brx}
\newcommand{\llangle}[1][]{\savebox{\@brx}{\(\m@th{#1\langle}\)}%
  \mathopen{\copy\@brx\kern-0.5\wd\@brx\usebox{\@brx}}}
\newcommand{\rrangle}[1][]{\savebox{\@brx}{\(\m@th{#1\rangle}\)}%
  \mathclose{\copy\@brx\kern-0.5\wd\@brx\usebox{\@brx}}}
\makeatother
\begin{document}
\preprint{\href{http://arxiv.org/abs/1612.00849}{arXiv:1612.00849}}

\title{Thermoelectric transport in disordered metals without quasiparticles:\\
the Sachdev-Ye-Kitaev models and holography}

\author{Richard A. Davison}
\affiliation{Department of Physics, Harvard University, Cambridge MA 02138, USA}

\author{Wenbo Fu}
\affiliation{Department of Physics, Harvard University, Cambridge MA 02138, USA}

\author{Antoine Georges}
\affiliation{Centre de Physique Th\'eorique, 
\'Ecole Polytechnique, CNRS, 91128 Palaiseau Cedex, France}
\affiliation{Coll\`ege de France, 11 place Marcelin Berthelot, 75005 Paris, France}
\affiliation{Department of Quantum Matter Physics, University of Geneva, 24 Quai Ernest-Ansermet, 1211 Geneva 4, Switzerland}

\author{Yingfei Gu}
\affiliation{Department of Physics, Stanford University, Stanford, CA 94305, USA}

\author{Kristan Jensen}
\affiliation{Department of Physics and Astronomy, San Francisco State University, San Francisco, CA 94132, USA}

\author{Subir Sachdev}
\affiliation{Department of Physics, Harvard University, Cambridge MA 02138, USA}
\affiliation{Perimeter Institute for Theoretical Physics, Waterloo, Ontario, Canada N2L 2Y5}

\date{\today
\\
\vspace{0.4in}}

\begin{abstract}
We compute the thermodynamic properties of the Sachdev-Ye-Kitaev (SYK) models of fermions with a conserved fermion number, $\mathcal{Q}$. We extend a previously proposed Schwarzian effective action to include a phase field, and this describes the low temperature energy and $\mathcal{Q}$ fluctuations. We obtain higher-dimensional generalizations of the SYK models which display disordered metallic states without quasiparticle excitations, and we deduce their thermoelectric transport coefficients. We also examine the corresponding properties of Einstein-Maxwell-axion theories on black brane geometries which interpolate from either AdS$_4$ or AdS$_5$ to an AdS$_2\times \mathbb{R}^2$ or AdS$_2\times \mathbb{R}^3$ near-horizon geometry. These provide holographic descriptions of non-quasiparticle metallic states without momentum conservation. We find a precise match between low temperature transport and thermodynamics of the SYK and holographic models. In both models the Seebeck transport coefficient is exactly equal to the $\mathcal{Q}$-derivative of the entropy. For the SYK models, quantum chaos, as characterized by the butterfly velocity and the Lyapunov rate, universally determines the thermal diffusivity, but not the charge diffusivity.

\end{abstract}
\maketitle

\tableofcontents

%********************************************************
\section{Introduction}
\label{sec:intro}
%********************************************************

Strange metal states are ubiquitous in modern quantum materials \cite{Bruin804}. Field theories of strange metals \cite{ssbook} have largely focused on disorder-free models of Fermi surfaces coupled to various gapless bosonic excitations which lead to breakdown of the quasiparticle excitations near the Fermi surface, but leave the Fermi surface intact. On the experimental side~\cite{Bruin804,Crossno1058,Lucas:2015sya,Zhang:2016ofh}, there are numerous indications that disorder effects are important, even though many of the measurements have been performed in nominally clean materials. Theories of strange metals with disorder have only examined the consequences of dilute impurities perturbatively~\cite{MaslovChubukov11,Hartnoll:2014gba,Patel:2014jfa}.

Disordered metallic states have been extensively studied~\cite{PALTVR85,1985electron} 
under conditions in which the quasiparticle excitations survive. The quasiparticles are no longer plane wave states as they undergo frequent elastic scattering from impurities, and spatially random and extended quasiparticle states have been shown to be stable under electron-electron interactions \cite{PhysRevB.24.6783}. In contrast, the literature on quantum electronic transport is largely silent on the possibility of disordered conducting metallic states at low temperatures without quasiparticle excitations, when the electron-electron scattering length is of order or shorter than the electron-impurity scattering length. (Previous studies include a disordered doped antiferromagnet in which quasiparticles eventually reappear at low temperature \cite{PG98}, and a partial treatment of weak 
disorder in a model of a Fermi surface coupled to a gauge 
field \cite{HLR}.)

On the other hand, holographic methods do yield many examples of conducting quantum states in the presence of disorder and with no quasiparticle excitations~\cite{Hartnoll:2016apf,Hartnoll:2007ih,Hartnoll:2008hs,Davison:2013txa,lucas1401,Hartnoll:2014cua,Donos:2014yya,SAH15,Rangamani:2015hka,Hartnoll:2015faa,2015PhRvL.115v1601G}. If we assume that the main role of disorder is to dissipate momentum, and  we average over disorder to obtain a spatially homogeneous theory, then we may consider homogeneous holographic models which do not conserve momentum. Many such models have been studied~\cite{Vegh:2013sk,Davison:2013jba,Blake:2013bqa,Andrade:2013gsa,Donos:2013eha,Gouteraux:2014hca,Donos:2014uba,Bardoux:2012aw,Davison:2014lua,Davison:2015bea,Blake:2015epa}, and their transport properties have been worked out in 
%some 
detail. However, there is not a clear quantum matter interpretation of these 
disordered, non-quasiparticle, metallic states. 

The Sachdev-Ye-Kitaev (SYK) models~\cite{SY92,PG98,GPS99,GPS01,kitaev2015talk} are theories of fermions with a label $i=1 \ldots N$ and random all-to-all interactions. They have many interesting features, including 
the absence of quasiparticles in a non-trivial,
soluble limit in the presence of disorder at low temperature ($T$). 
In the limit where $N$ is first taken to infinity and the temperature is subsequently taken to zero, the entropy$/N$ remains non-zero. Note, however, that such an entropy does not imply an exponentially large 
ground state degeneracy: it can be achieved
by a many-body level spacing that is of the same
order near the ground state as at a typical excited state 
energy \cite{WFSS16}.
The SYK models
were connected holographically to black holes with AdS$_2$ horizons, and the $T \rightarrow 0$ limit of the entropy was identified as the Bekenstein-Hawking entropy in Refs.~\onlinecite{SS10,SS10b,SS15}. Many recent studies have taken a number of perspectives, including the connections to  two-dimensional quantum gravity~\cite{kitaev2015talk,AAJP15,SS15,Hosur15,JPRV16,YLX16,Anninos:2016szt,WFSS16,Jevicki16,JMDS16,KJ16,JMDS16b,HV16,DHT16,AABK16,BAK16,Garcia-Alvarez:2016wem,CP16,Jevicki16b,GQS16,Gross:2016kjj,Berkooz:2016cvq,Garcia-Garcia:2016mno,Fu:2016vas,Cotler:2016fpe}. The gravity duals of the SYK models are not in the category of the familiar AdS/CFT correspondence \cite{StanfordStrings}, and their low-energy physics is controlled by a symmetry-breaking pattern~\cite{JMDS16} which also arises in a generic two-derivative theory of dilaton gravity on a nearly AdS$_2$ spacetime~\cite{KJ16,JMDS16b,HV16}. 

With disordered metallic states in mind, in this paper we will study a class of SYK models~\cite{SY92,PG98,GPS99,GPS01} which have a conserved fermion number\footnote{A word about global symmetries is in order. The Majorana SYK model~\cite{kitaev2015talk} with $2N$ Majorana fermions has a SO($2N$) symmetry only after averaging over disorder. However, this symmetry is not generated by a conserved charge.
The model in Eq.~(\ref{h}) has a global U(1) symmetry for each realization of the disorder, and so this symmetry corresponds to a conserved charge. It is this symmetry which is of interest in this work. For completeness, we note that this model acquires an additional SU($N$) symmetry after averaging over disorder.}. The SYK models have recently been extended to lattice models in one or more spatial dimensions~\cite{GQS16} (see also~\cite{Berkooz:2016cvq}), which has opened an exploration into their transport properties. In this work we shall further extend these higher-dimensional models to include a conserved fermion number. This will allow us to describe the thermoelectric response functions of a solvable metallic state without quasiparticles and in the presence of disorder. We will then study thermoelectric transport in holographic theories which have a conserved charge and which break translational symmetry homogeneously via ``axion'' fields~\cite{Andrade:2013gsa,Donos:2013eha,Gouteraux:2014hca,Donos:2014uba}. These momentum-dissipating holographic theories have black brane solutions with AdS$_2\times \mathbb{R}^{d-1}$ near-horizon geometries. Many of their transport properties have been computed earlier, but some crucial features have gone unnoticed; we will highlight these features and show that they imply a precise match between the low-temperature thermodynamic and transport properties of the higher-dimensional SYK models and the Einstein-Maxwell-axion holographic theories.

The zero-dimensional SYK model of interest to us has canonical complex fermions $f_i$ labeled by $i=1\ldots N$. We refer to them as the complex SYK models.
%.
The Hamiltonian is
\beq
H_0 = \sum_{\substack{1\leq i_1 < i_2 \ldots i_{q/2} \leq N,\\ 1\leq i_{q/2+1} < i_{q/2+2} \ldots < i_{q} \leq N}} J_{i_1, i_2 \ldots i_{q}} \, 
f^\dagger_{i_1}  f^\dagger_{i_2} \ldots f^\dagger_{i_{q/2}}   f_{i_{q/2 + 1}}  \ldots f_{i_{q-1}} f_{i_{q}}
%\YG{-\mu \sum_{1\leq i \leq N} f^\dagger_i f_i}
\,. \label{h}
\eeq %\YG{\bf[The chemical potential term was lost somehow, and do we need the ``0'' in subscript of $H_0$? ]}
Here $q $ is an even integer, and the couplings $J_{i_1, i_2 \ldots i_{q}}$ are random complex numbers with zero mean obeying
\begin{align}
\begin{split}
J_{i_1i_2\ldots i_{q/2},i_{q/2+1}\ldots i_{q-1}i_q} &= J^*_{i_{q/2+1}\ldots i_{q-1}i_q, i_1i_2\ldots i_{q/2}} \,,\\
\overline{\left|J_{i_1, i_2 \ldots i_{q}} \right|^2} &=  \frac{ J^2 (q/2)!^2}{N^{q-1}}\,.
\end{split}
\label{eqn: hermitian}
\end{align}
Note that the case $q=2$ is special and does have quasiparticles: this describes
free fermions and the eigenstates of the random matrix $J_{i_1, i_2}$ obey
Wigner-Dyson statistics \cite{1997RvMP...69..731B}. Our attention will be focused on $q \geq 4$, and $N\gg q$, when 
the model flows to a phase without quasiparticles 
with an emergent conformal symmetry at low energies. We define the fermion number $-1/2 < \mathcal{Q} < 1/2$ by 
\beq
\mathcal{Q} = \frac{1}{N} \sum_i  \langle f_i^\dagger f_i \rangle  - 1/2\,.
\eeq
We also define
\beq
\Delta = \frac{1}{q}\,, \label{Deltaq}
\eeq
which will be the low-energy scaling dimension of the fermion $f$.

%********************************************************
\subsection{Thermodynamics}
\label{sec:thermo}
%********************************************************

We will show in Section~\ref{sec:thermosyk} that the 
%low temperature ($T$) limit of the 
canonical free energy, $NF$, of the Hamiltonian $H_0$ in Eq.~\eqref{h} 
%obeys
has a low temperature ($T)$ expansion 
\beq
F(\mathcal{Q}, T) = E_0 (\mathcal{Q}) - T \mathcal{S} (\mathcal{Q}) + \ldots. \label{funiv}
\eeq
where the ground state energy, $E_0 (\mathcal{Q})$, is not universal, but the zero-temperature entropy $\mathcal{S}(\mathcal{Q})$ is {\em universal\/}, %{\em i.e.\/} 
meaning that it depends only on the scaling dimension $\Delta$ and is independent of high energy  (``UV'') details, such as higher order fermion interactions that could be added to Eq.~\eqref{h}. The value of $E_0 (\mathcal{Q})$ is not known analytically, but can only be computed numerically or in a large $q$ expansion as we perform in Appendix~\ref{app:largeq}. However, remarkably, we can obtain exact results for the universal function $\mathcal{S}(\mathcal{Q})$ for all $0 < \Delta < 1/2$, and for all $-1/2 < \mathcal{Q} < 1/2$ (see Fig.~\ref{fig:entropy_Q}). These results agree with those obtained earlier for the special cases $\Delta=1/4$ and all $\mathcal{Q}$ in Ref.~\onlinecite{GPS01}, and for $\mathcal{Q}=0$ and all $\Delta$ in Ref.~\onlinecite{kitaev2015talk}. The higher-dimensional complex SYK models also have a free energy of the form Eq.~\eqref{funiv}, where $NF$ is now understood to be the free energy per site of the higher-dimensional lattice.
%The free energy in Eq.~(\ref{funiv}) also applies to the higher-dimensional complex SYKmodels, where $NF$ is now defined as the free energy per site of the higher dimensional lattice.

Because of the non-universality of $E_0(\mathcal{Q})$, the universal properties of the thermodynamics
%$H$
are more subtle in the grand canonical ensemble. The chemical potential, $\mu = (\partial F/\partial \mathcal{Q})_T$, has both universal and non-universal contributions. Consequently it requires a delicate computation to extract the universal portions of the grand potential$/N$, 
\beq
\Omega (\mu, T) = F- \mu \mathcal{Q}\,.
\eeq 
It is interesting to note that this universal dependence on $\mathcal{Q}$, and not $\mu$, is similar to that in the Luttinger theorem for a Fermi liquid: there the Fermi volume is a universal function of $\mathcal{Q}$, but the connection with $\mu$ depends upon many UV details. And indeed, the computation of $\mathcal{S} (\mathcal{Q})$ in Ref.~\onlinecite{GPS01} for $\Delta=1/4$ employs an analysis which parallels that used to prove the Luttinger theorem in Fermi liquids; see also Appendix~\ref{app:georges}.

Section~\ref{sec:AdSthermodyn} will examine in detail the thermodynamics of the simplest 
%charged 
holographic axion theory. This theory has a planar black brane solution whose geometry interpolates between AdS$_4$ near the boundary and AdS$_2\times R^2$ near the horizon.
%, and AdS$_4$ near the boundary. 
The holographic dictionary would suggest that the IR
%low energy (``IR'') 
properties of the dual field theory state should be controlled by the near-horizon AdS$_2$ geometry. We will find that the free energy of the holographic theory also has a low-temperature expansion of the form~\eqref{funiv}, where the universal part of the free energy is determined by the AdS$_2$ part of the geometry, while the non-universal part depends upon the details of its embedding into the UV AdS$_4$ geometry. The universal `equation of state', $\mathcal{S}(\mathcal{Q})$ will, however, be different between the SYK and holographic models. The holographic theory we are studying was chosen as it is the simplest theory with momentum dissipation and an AdS$_2$ horizon -- it is not the precise holographic dual of the SYK model.

A quantity that will play a central role in our analyses of 
%both 
the SYK and holographic models is
\beq
2 \pi \mathcal{E} = -  \lim_{T\to 0}\frac{\partial^2 F}{\partial \mathcal{Q} \partial T} =    \frac{d \mathcal{S}}{d \mathcal{Q}} = - \lim_{T\to 0}\left(\frac{\partial \mu}{\partial T} \right)_{\mathcal{Q}} \,.\label{defE0}
\eeq
Note that $\mathcal{E}$ is also universal. The factor of $2 \pi$ has been inserted because then, for theories dual to gravity with an AdS$_2$ near-horizon geometry, $\mathcal{E}$ 
%also has the interpretation as 
is the electric field in the AdS$_2$ region
% of the holographic theories 
\cite{Sen05,Sen08,SS15}.

%*****************************
\subsection{Effective action}
\label{sec:effective}
%*****************************

We will also examine aspects of $1/N$ fluctuations about the saddle point which led to the thermodynamic results in Section~\ref{sec:thermo}. Here we will follow Ref.~\onlinecite{JMDS16}, who argued that the dominant fluctuations of the Majorana SYK model at low $T$ are controlled by a Schwarzian effective action with $\PSL (2, \mathbb{R})$ time reparameterization symmetry. This effective action can be used to compute energy fluctuations, and hence the specific heat, in the canonical ensemble. In our analysis of the complex SYK model, we find that an additional U(1) phase field, $\phi$, is needed; $\phi$ is conjugate to 
$\mathcal{Q}$ fluctuations in the grand canonical ensemble. A similar phase field also appeared in a recent analysis of SYK models with $\mathcal{N}=2$ supersymmetry \cite{Fu:2016vas} with the 
mean $\mathcal{Q}$ close to zero. 
We propose a combined action for energy and 
$\mathcal{Q}$ fluctuations at a generic mean $\mathcal{Q}$, with both $\PSL (2, \mathbb{R})$ and U(1) symmetry; for the zero-dimensional complex SYK models, the action is 
by
\beq
\frac{S_{\phi,\epsilon}}{N} = \frac{K}{2} \int_0^{1/T} d \tau \left[ \partial_\tau \phi + i (2 \pi \mathcal{E} T) \partial_\tau \epsilon \right]^2 -\frac{\gamma}{4 \pi^2} \int_0^{1/T} d \tau \, \{\tan(\pi T (\tau + \epsilon(\tau)), \tau\}.
\label{pf1i}
\eeq
Here $\tau$ is imaginary time, $\tau \rightarrow \tau + \epsilon (\tau)$ is the time reparameterization, 
$\{f, \tau\}$ is the Schwarzian derivative (given explicitly in Eq.~(\ref{Schwarzian})), and $K$ and $\gamma$ 
are non-universal thermodyamic parameters determining the compressibility and the specific heat respectively.
The off-diagonal coupling between energy and
$\mathcal{Q}$ fluctuations is controlled by the value of $\mathcal{E}$. Our effective action will play a central role in the structure of thermoelectric transport, as described in the next subsection. 

%********************************************************
\subsection{Transport}
\label{sec:transport}
%********************************************************

We will characterize transport by two-point correlators of the conserved number density, which we continue to refer to as $\mathcal{Q}$, and the conserved energy density $E = H_0/N $. For the zero-dimensional SYK model in (\ref{h}), both of these quantities are constants of the motion, and so have no interesting dynamics. So we consider here the higher dimensional SYK models, for which 
%the densities
 $\mathcal{Q}$ and $E$
are defined per site of the higher-dimensional lattice. Then their correlators do have an interesting dependence of wavevector, $k$, and frequency, $\omega$. We define the dynamic susceptibility
%as a 
matrix, $\chi (k, \omega)$, where
\beq
\chi (k, \omega) = \left( \begin{array}{cc}
\left\langle  \mathcal{Q};  \mathcal{Q} \right\rangle_{k, \omega} &
\left\langle  E - \mu  \mathcal{Q};  \mathcal{Q} \right\rangle_{k, \omega} /T \\
\left\langle  E - \mu  \mathcal{Q};  \mathcal{Q} \right\rangle_{k, \omega}  & 
\left\langle  E - \mu  \mathcal{Q};  E - \mu  \mathcal{Q}  \right\rangle_{k, \omega}/T
\end{array} \right),
 \label{ti1}
\eeq
and we use the notation
\beq
\left\langle A ; B \right\rangle_{k, \omega} \equiv -i \int_0^{\infty} dt \int d^d x \left\langle [ A(x, t), B(0,0)] \right\rangle e^{-i k x + i \omega t}\,.
\eeq
As in the standard analysis of Kadanoff and Martin \cite{KM63}, we expect the low energy and long distance form of these correlators to be fully dictated by the hydrodynamic equations of motion for a diffusive metal \cite{SAH15}. From such an analysis, we obtain, at low frequency and wavenumber,
\beq
\chi (k, \omega)  = \left[ i \omega (-i \omega + D k^2)^{-1}  + 1 \right] \chi_s,
 \label{ti2}
\eeq
where $D$ and $\chi_s$ are $2 \times 2$ matrices. The diffusivities are specified by $D$, and the static susceptibilities are, as usual, $\chi_s = \lim_{k \rightarrow 0} \lim_{\omega \rightarrow 0} \chi (k, \omega)$. The values of $\chi_s$ are related by standard thermodynamic identities to second derivatives of the grand potential $\Omega$, as shown in Eq.~(\ref{defchis}).

One of our main results is that the low $T$ limit of the diffusivity matrix, $D$, takes a specific form
\beq
D = \left( \begin{array}{cc}
D_1 & 0 \\ 2 \pi \mathcal{E} T (D_1 - D_2) & D_2
\end{array} \right). \label{pf5}
\eeq
where $D_1$ and $D_2$ are temperature-independent constants. We will show that the result in Eq.~(\ref{pf5}) is obeyed both in the higher-dimensional SYK models, and in the holographic theories. It is a consequence of the interplay between the global U(1) fermion number symmetry and the emergent $\PSL (2, \mathbb{R})$ symmetry of the scaling limit of the SYK model. In holography, $\PSL (2, \mathbb{R})$ is the isometry group of AdS$_2$; while transport properties of the holographic theories have been computed earlier, the specific form the diffusivity matrix in Eq.~(\ref{pf5}) was not noticed. This form will be crucial for the mapping between the holographic and SYK models.

We can use the Einstein relation to define a matrix of conductivities
\beq
\left( \begin{array}{cc}
 \sigma   & \alpha  \\
 \alpha T &  \overline{\kappa} \end{array} \right) = D \chi_s , \label{ti3}
 \eeq
where $\sigma$ is the electrical conductivity, $\alpha$ is the thermoelectric conductivity, and $\kappa = \overline{\kappa} - T \alpha^2/\sigma$ is the thermal conductivity. The matrix in Eq.~(\ref{ti3}) is constrained by Onsager reciprocity. From Eqs.~(\ref{pf5}) and (\ref{ti3}) we find the following result for the low $T$ limit of the thermopower; the Seebeck coefficient $S$ is given in both the SYK and holographic models by
\beq
\lim_{T \rightarrow 0} S \equiv \frac{\alpha}{\sigma} =  \frac{ d \mathcal{S}}{d \mathcal{Q}} \,. \label{seebeck}
\eeq
Since $d\mathcal{S}/d\mathcal{Q}=2\pi\mathcal{E}$, we see that the Seebeck coefficient is entirely determined by the particle-hole asymmetry of the fermion spectral function. 

Eq.~(\ref{seebeck}) has been proposed earlier as the `Kelvin formula' by Peterson and Shastry \cite{Shastry} using very different physical arguments. Earlier holographic computations of transport did not notice the result in Eq.~(\ref{seebeck}). The remarkable aspect of this expression is that it relates a transport quantity to a thermodynamic one, the derivative 
of the entropy with respect to particle number. Such a relation is in general only approximate, see Ref.~\onlinecite{Shastry} and the discussion 
and applications in Ref~\onlinecite{mravlje}. Remarkably, this relation holds exactly here: it is an exact consequence of the $\PSL (2, \mathbb{R})$ symmetry of both SYK and holographic models. We note that the form in Eq.~(\ref{pf5}) is implied by Eq.~(\ref{seebeck}) and Onsager reciprocity.

We also obtain an interesting result for the Wiedemann-Franz ratio, $L$, of the SYK model. For the particular higher-dimensional generalization in Eq.~(\ref{tt3}), we find the exact result
\begin{equation}
L \equiv \lim_{T \rightarrow 0} \frac{\kappa}{T \sigma} = \frac{4 \pi^2}{3 q^2}.    
\label{eq:WFforSYK}
\end{equation}
We comments on aspects of this result:\\
$\bullet$ For the free fermion case, $q = 2$, this reduces
to the universal Fermi liquid Lorenz number $L_0 = \pi^2 k_B^2/(3 e^2)$ (re-inserting fundamental constants). 
Although expected, this agreement with $L_0$ at $q=2$ is 
non-trivial and remarkable: rather than the usual arguments based upon integrals over the Fermi function, Eq.~(\ref{eq:WFforSYK}) arises from the structure of {\em bosonic\/} normal modes of the $1/N$ fluctuations, as discussed in Appendix~\ref{app:normalmode}.\\
$\bullet$ The decrease of $L$ for large $q$ can be understood as follows. As we will see in the large $q$ solution in Appendix~\ref{app:largeq}, the energy bandwidth for fermion states vanishes as $q \rightarrow \infty$. Consequently, fermion hopping transfers little energy, and the thermal conductivity $\kappa$ is suppressed. In contrast, fermion hopping continues to transfer unit charge, and hence the conductivity does not have a corresponding suppression.\\
$\bullet $ Although the result in Eq.~(\ref{eq:WFforSYK}) appears 
universal, it is not so: there are other higher-dimensional generalizations of the inter-site coupling term in Eq.~(\ref{tt3}) which will lead to corrections to the value of $L$; see Section~\ref{sec:SYKd}. 
Needless to say, these corrections
will not modify the universal value $L_0$ at $q=2$. The non-universality of $L$ for higher $q$ is connected to the non-renormalization of inter-site disorder in the present large $N$ limit.\\ 
$\bullet$ Results for the values of $L$ for the holographic models appear in the body of the paper.

\emph{Note:} While we were completing this work we were made aware of Ref.~\onlinecite{Blake:2016jnn}, which has some overlap with our holographic analysis.

%********************************************************
\section{Complex SYK model}
\label{sec:SYK}
%********************************************************

%********************************************************
\subsection{Large $N$ saddle point}
%********************************************************
In this section, we employ Green's functions in the grand canonical ensemble at a chemical potential $\mu$.
Starting from a perturbative expansion of $H_0$ in Eq.~(\ref{h}), and averaging term-by-term, we obtain the following equations
for the Green's function and self energy in the large $N$ limit:
\bea
\Sigma (\tau) &=& -(-1)^{q/2} q J^2 \left[G (\tau)\right]^{q/2} \left[ G (-\tau) \right]^{q/2-1} \label{e2} \\
G (i\omega_n)  &=& \frac{1}{i \omega_n + \mu - \Sigma (i \omega_n)} ; \label{e3}
\eea
where $\omega_n$ is a Matsubara frequency $\omega_n = 2\pi (n + \frac{1}{2})$ and $\tau$ is imaginary time. As in Refs.~\onlinecite{SY92,GPS01}, we make the following IR ansatz at a complex frequency $z$
\beq
G (z) = C \frac{e^{-i (\pi \Delta + \theta)}}{z^{1-2 \Delta}} \quad, \quad \mbox{Im}(z) >0\,, \label{ansatz}
\eeq
which is expressed in terms of three real parameters, $C$, $\Delta$ and $\theta$. Here the complex frequency is small compared to the disorder, $|z|\ll J$. As we describe in Appendix~\ref{app:conformal}, this is the appropriate form for the two-point function of a charged operator at nonzero chemical potential in a limit where there is an approximate conformal invariance.
%The positivity constraint on the spectral weight ({\em i.e.\/} unitarity) constrains
Unitarity implies that the spectral weight is positive, which in turn implies that
\beq
-\pi\Delta < \theta < \pi \Delta\,.
\eeq
The particle-hole symmetric value is $\theta = 0$. Inserting Eq.~(\ref{ansatz}) into (\ref{e3}), a straightforward analysis described in Appendix~\ref{app:saddle} shows that $\Delta$ is given by (\ref{Deltaq}), while
\beq
C = \left[\frac{\Gamma (2(1-1/q))}{\pi q J^2}\right]^{1/q} \left[ \frac{\pi}{\Gamma(2/q)} \right]^{1 - 1/q}
\left[\sin(\pi/q + \theta) \sin(\pi/q - \theta) \right]^{1/q-1/2}\,. \label{Cval}
\eeq
The value of $\theta$ remains undetermined in this IR analysis. Below, in Eq.~(\ref{Qtheta}), we find an exact relationship between  $\theta$ and the density $\mathcal{Q}$, as was first found
in Ref.~\onlinecite{GPS01} for the $\Delta = 1/4$ theory.

%********************************************************
\subsection{Non-zero temperature}
%********************************************************

%For the $T>0$ discussion, 
%At nonzero temperature $T>0$ it is useful to introduce the ``electric field,'' $\mathcal{E}$,
%\beq
%e^{2 \pi \mathcal{E}} = \frac{\sin (\pi \Delta + \theta)}{\sin(\pi \Delta - \theta)}\,. \label{theta}
%\eeq
%This is defined here by the asymmetry in the fermion Green's function at $T=0$, 
%\beq
%G (\tau) \sim \left\{
%\begin{array}{ccc} - \, \tau^{-2 \Delta} & , & \tau > 0 \\
%e^{-2 \pi \mathcal{E}} \, |\tau|^{-2 \Delta} &,& \tau < 0 . \label{twist}
%\end{array}
%\right. ,
%\eeq
Fourier transforming the fermion Green's function Eq.~\eqref{ansatz} gives
\beq
G(\tau) \sim \begin{cases} - |\tau|^{-2\Delta} \,, & \tau >0\,, \\ e^{-2\pi \mathcal{E}} |\tau|^{-2\Delta}\,, & \tau <0\,,\end{cases}\label{twist}
\eeq
where the ``spectral asymmetry'' $\mathcal{E}$ is related to $\theta$ as
\beq
e^{2\pi\mathcal{E}} = \frac{\sin(\pi\Delta + \theta)}{\sin(\pi \Delta - \theta)}\,. \label{theta}
\eeq
The asymmetry in (\ref{twist}) was argued in \cite{PGKS97} (and reviewed in \cite{SS15}) to fix
the $T$ derivative of $\mu$
\beq
\lim_{T\to 0}\left( \frac{\partial \mu}{\partial T} \right)_\mathcal{Q} = - 2 \pi \mathcal{E} \,.
%\quad \mbox{as $T \rightarrow 0$.}
\label{dmdt}
\eeq
So this $\mathcal{E}$ is the same as that introduced in Eq.~(\ref{defE0}). See Appendix~\ref{app:conformal} for an independent argument for this relation from low-energy conformal symmetry. 

The $T>0$ generalization of Eq.~(\ref{twist}) is a saddle point of the action in Eq.~(\ref{action})
\beq
G_s (\tau) =  - C \left( \frac{\Gamma(2 \Delta) \sin (\pi \Delta + \theta)}{\pi} \right) \, e^{-2 \pi \mathcal{E} T \tau} \, \left( \frac{\pi T}{\sin (\pi T \tau)} \right)^{2 \Delta} 
\quad,\quad 0 < \tau < \frac{1}{T}, \label{Gsigma2}
\eeq
with $\mathcal{E}$ remaining independent of $T$ provided $\mathcal{Q}$ is held fixed. This result was found both in \cite{PGKS97,SS15}, and in the AdS$_2$ computation in \cite{Faulkner09}. After using the KMS condition and (\ref{theta}), the $T \rightarrow 0$ limit of (\ref{Gsigma2}) agrees with (\ref{Gftau}).

%********************************************************
\subsection{Thermodynamics}
\label{sec:thermosyk}
%********************************************************

Ref.~\onlinecite{JMDS16} has given an expression for the free energy of the Majorana SYK models as a functional of the Green's function and the self energy; related expressions were given earlier for the complex SYK models \cite{GPS01,SS15}. It is straightforward to obtain a similar result for the grand potential of the complex SYK models, which we give in Eq.~(\ref{action}). Here, we will only compute the $\Delta$ derivative of the grand potential in the $T \rightarrow 0$ limit, 
and then integrate with respect to $\Delta$ to obtain the low-temperature grand potential.

The only term 
%with the explicit $q$ 
%$\Delta $ 
%dependence 
in the grand potential which explicitly depends on $q$ is (see Eq.~(\ref{action}))
\beq
\Omega = \ldots - J^2 \int_0^{1/T} d \tau \left[ G(\tau) \right]^{q/2}  \left[ G(1/T - \tau) \right]^{q/2}\,.
\eeq
Substituting $\Delta = 1/q$ and using (\ref{Cval}), (\ref{theta}) and (\ref{Gsigma2}) the leading low-temperature derivative of $\Omega$ with respect to $\Delta$ comes from this term,
\begin{align}
\nonumber
\frac{d \Omega}{d \Delta} &= -\frac{2(2 \Delta-1)}{\pi} \frac{\sin(\pi\Delta + \theta) \sin(\pi \Delta - \theta)}{\sin (2 \pi \Delta)}
\int_0^{1/T} d \tau \left( \frac{\pi T}{\sin (\pi T \tau)} \right)^{2} \ln \left( \frac{\Lambda}{\sin (\pi T \tau)} \right)
\\
&= -2\pi T (2 \Delta - 1) \frac{\sin(\pi\Delta + \theta) \sin(\pi \Delta - \theta)}{\sin (2 \pi \Delta)}  + \mbox{a term of order $J$}\,,
\end{align}
where $\Lambda$ is some $\tau$-independent constant. We subtract the term in $\Omega$ of order $J$, which we call $\Omega_0$. Using Eq.~(\ref{theta}) to express the result in terms of $\mathcal{E}$ we obtain
\begin{align}
\begin{split}
\frac{d(\Omega- \Omega_0)}{d \Delta} &= -\pi T (2 \Delta -1) \frac{\sin(2 \pi \Delta)}{\cos (2 \pi \Delta) + \cosh (2 \mathcal{E} \pi)}
\\
&= -\frac{\pi T (2 \Delta - 1)}{2} \left[ \tan(\pi (\Delta - i \mathcal{E})) + \tan(\pi (\Delta + i \mathcal{E})) \right]\,. \label{dFdD}
\end{split}
\end{align}
The relation~\eqref{defE0} between $\mathcal{E}$, $\mu$, and $T$, implies that $\mathcal{E}$ is fixed when $\mu$ and $T$ are also fixed. Thus in writing $d\Omega/d\Delta$ we treat $\mathcal{E}$ and $\Delta$ as independent variables, in particular, we keep $\mathcal{E}$ fixed when integrating over $\Delta$.
Note that the relationship between $\mathcal{E}$ and $\mathcal{Q}$ in (\ref{QE}) depends upon $q$, and so varying $q$ at fixed $\mathcal{E}$ implies that $\mathcal{Q}$ will vary. We can now integrate (\ref{dFdD}) to obtain
\bea
&&\Omega  - \Omega_0 =  \frac{(2 \Delta - 1)T}{2} \ln \left[2 ( \cos (2 \pi \Delta) + \cosh (2 \pi \mathcal{E}))\right]  \nn
&&~~~ +\frac{i T}{4 \pi}\left[{\rm Li}_2 \left(- e^{2 \pi (\mathcal{E}- i \Delta)} \right) + {\rm Li}_2 \left(- e^{2 \pi (-\mathcal{E} - i \Delta)} \right) 
-{\rm Li}_2 \left(- e^{2 \pi (\mathcal{E}+ i \Delta)} \right) - {\rm Li}_2 \left(- e^{2 \pi (-\mathcal{E} + i \Delta)} \right) 
\right] \nn
&& ~~~\equiv - T \mathcal{G}(\mathcal{E}).
\label{Fres}
\eea
The integration constant is fixed by the boundary conditions that the singular part of the grand potential, $\Omega-\Omega_0$ vanishes at the free fermion point $\Delta = 1/2$. The last line in Eq.~(\ref{Fres}) defines the function $\mathcal{G}(\mathcal{E})$.

%********************************************************
\subsubsection{Separating the universal and non-universal parts}
\label{sec:univ}
%********************************************************

The grand potential $\Omega$ computed in (\ref{Fres}) depends upon $\mathcal{E}$ and $T$. But, in the grand canonical ensemble at fixed $\mu$ and $T$, $\mathcal{E}$ has an unknown dependence upon $\mu$ and $T$. It is therefore better to convert to the canonical ensemble at fixed $\mathcal{Q}$ and $T$, where we know the $\mathcal{Q}$ and $T$ dependence of $\mu$ from (\ref{dmdt}), 
\beq
\mu (\mathcal{Q}, T) = \mu_0 - 2 \pi \mathcal{E} (\mathcal{Q}) T + \ldots \label{mu1}
\eeq
as $T \rightarrow 0$. Here $\mathcal{E}$ depends only on $\mathcal{Q}$, 
and $\mu_0$ is the contribution from the ground state energy, $E_0$, with
\beq
\mu_0 = \frac{d E_0}{d \mathcal{Q}}\,. \label{t1}
\eeq
The complete grand potential, including the contribution of the ground state energy, is 
\beq
\Omega = E_0 - \mu_0 \mathcal{Q} - T \mathcal{G} (\mathcal{E})+\hdots\,, \label{t2}
\eeq
where the functional form of the singular term $ \mathcal{G} (\mathcal{E})$ was given in  (\ref{Fres}).
The free energy in the canonical ensemble, $F$, is
\beq
F(\mathcal{Q}, T) = \Omega + \mu \mathcal{Q}\,. \label{t3}
\eeq

Now we use the thermodynamic identity
\beq
\mu = \left( \frac{\partial F}{\partial \mathcal{Q}} \right)_{T}, \label{t4}
\eeq
to obtain an expression for the density, $\mathcal{Q}$. Using (\ref{mu1}-\ref{t3}), (\ref{t4}) becomes
\beq
\mu_0 - 2 \pi \mathcal{E} T = \frac{d E_0}{d \mathcal{Q}} - T \frac{d \mathcal{G}}{d \mathcal{E}} \frac{d \mathcal{E}}{d \mathcal{Q}}  - 2 \pi \mathcal{E} T  - 2 \pi T\frac{d \mathcal{E}}{d \mathcal{Q}} \mathcal{Q}, \label{t5}
\eeq
which gives us 
\beq
\mathcal{Q} = - \frac{1}{2 \pi} \frac{d \mathcal{G}}{d \mathcal{E}}. \label{t6}
\eeq

Similarly, the entropy is 
\begin{align}
\begin{split}
\label{t7}
\mathcal{S} &= -\left( \frac{\partial F}{\partial T} \right)_{\mathcal{Q}}
\\
&= \mathcal{G} + 2 \pi \mathcal{E} \mathcal{Q} 
\end{split}
\end{align}
Eqs.~(\ref{t6}) and (\ref{t7}) show that $\mathcal{S} (\mathcal{Q})$ and $\mathcal{G} (\mathcal{E})$ are a Legendre pair,
and so
\beq
\frac{d \mathcal{S}}{d \mathcal{Q}} = 2 \pi \mathcal{E}. \label{edsdq}
\eeq
This equality is equivalent to Eq.~(\ref{dmdt}) by the Maxwell relation in Eq.~(\ref{defE0}), and this supports the validity of our analysis.

Appendix~\ref{app:largeq} presents a computation of the thermodynamics at large $q$.
%large $q$ solution: 
In this limit, we explicitly verify the above decompositions into universal and non-universal components.

%********************************************************
\subsubsection{Charge}
%********************************************************

\begin{figure}
\begin{center}
\includegraphics[height=7.5cm]{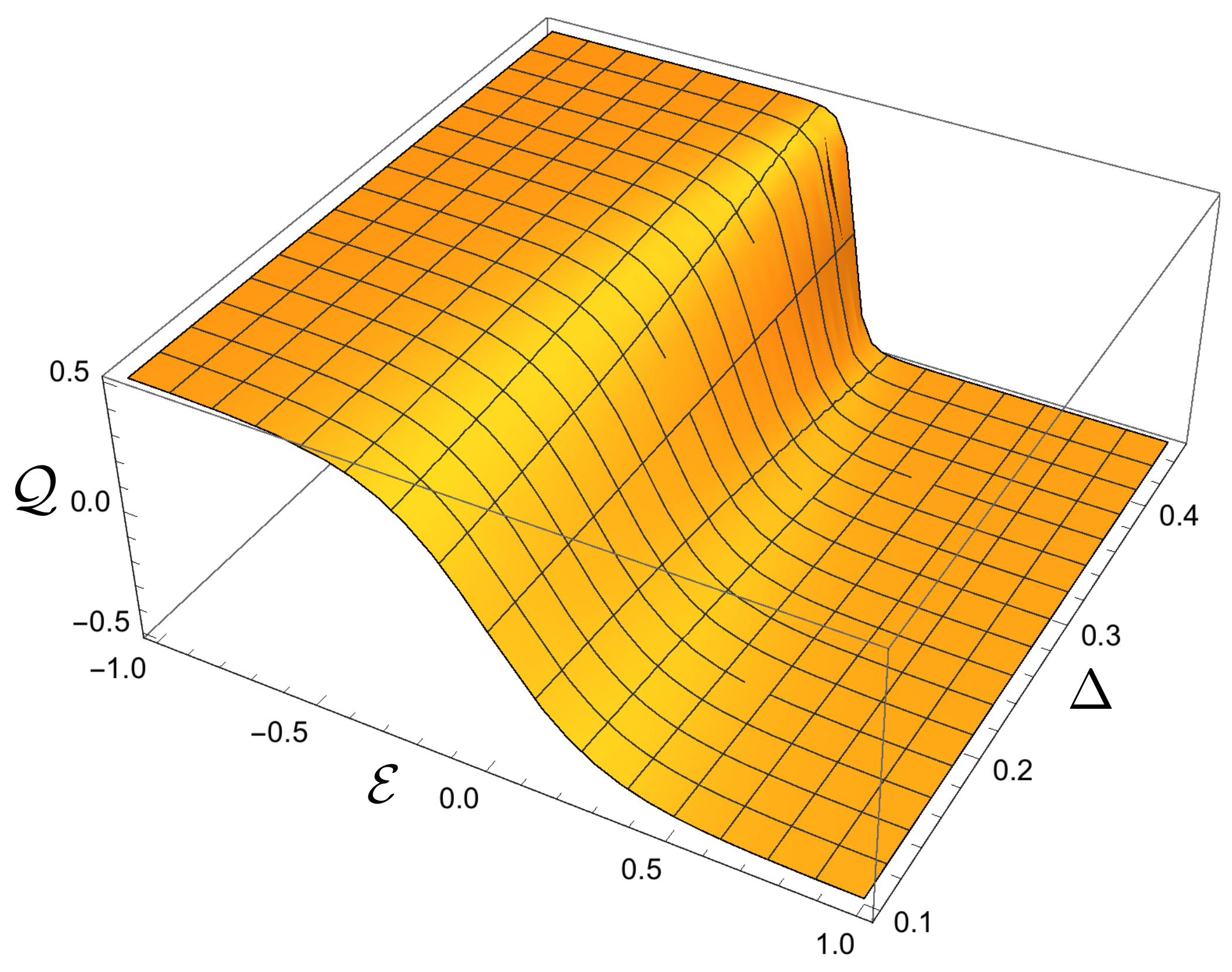}
\end{center}
\caption{The density $\mathcal{Q}$ as a function of $\mathcal{E}$ and $\Delta$}
\label{fig:density}
\end{figure}

We compute the density from Eqs.~(\ref{Fres}) and (\ref{t6}) to obtain
\beq
\mathcal{Q}= \frac{(2 \Delta -1) \sinh(2 \pi \mathcal{E})}{2 (\cos (2 \pi \Delta) + \cosh(2 \pi \mathcal{E}))}
- \frac{i}{4 \pi} \ln \left[ \frac{(1 + e^{2 \pi (\mathcal{E}- i \Delta)})(1 + e^{2 \pi (-\mathcal{E}+ i \Delta)})}{(1 + e^{2 \pi (\mathcal{E}+ i \Delta)})(1 + e^{2 \pi (-\mathcal{E}- i \Delta)})}\right]\,. \label{QE}
\eeq
This simplifies considerably when expressed in terms of $\theta$ via (\ref{theta})
\beq
\mathcal{Q} = - \frac{\theta}{\pi} + \left(\Delta - \frac{1}{2} \right) \frac{\sin (2 \theta)}{\sin (2 \pi \Delta)}\,. \label{Qtheta}
\eeq
This agrees with Appendix A of Ref.~\onlinecite{GPS01} at $q=4$. In Appendix~\ref{app:georges}, we generalize the Luttinger-Ward
argument of Ref.~\onlinecite{GPS01} to arbitrary $q$, and provide further evidence for Eq.~(\ref{Qtheta}).
Note that $\mathcal{Q}=\pm 1/2$
at the limiting values $\theta = \mp \pi \Delta$. A plot of the density appears in Fig.~\ref{fig:density}.

%********************************************************
\subsubsection{Entropy}
\label{sec:entropy}
%********************************************************

\begin{figure}
\begin{center}
\includegraphics[height=7.5cm]{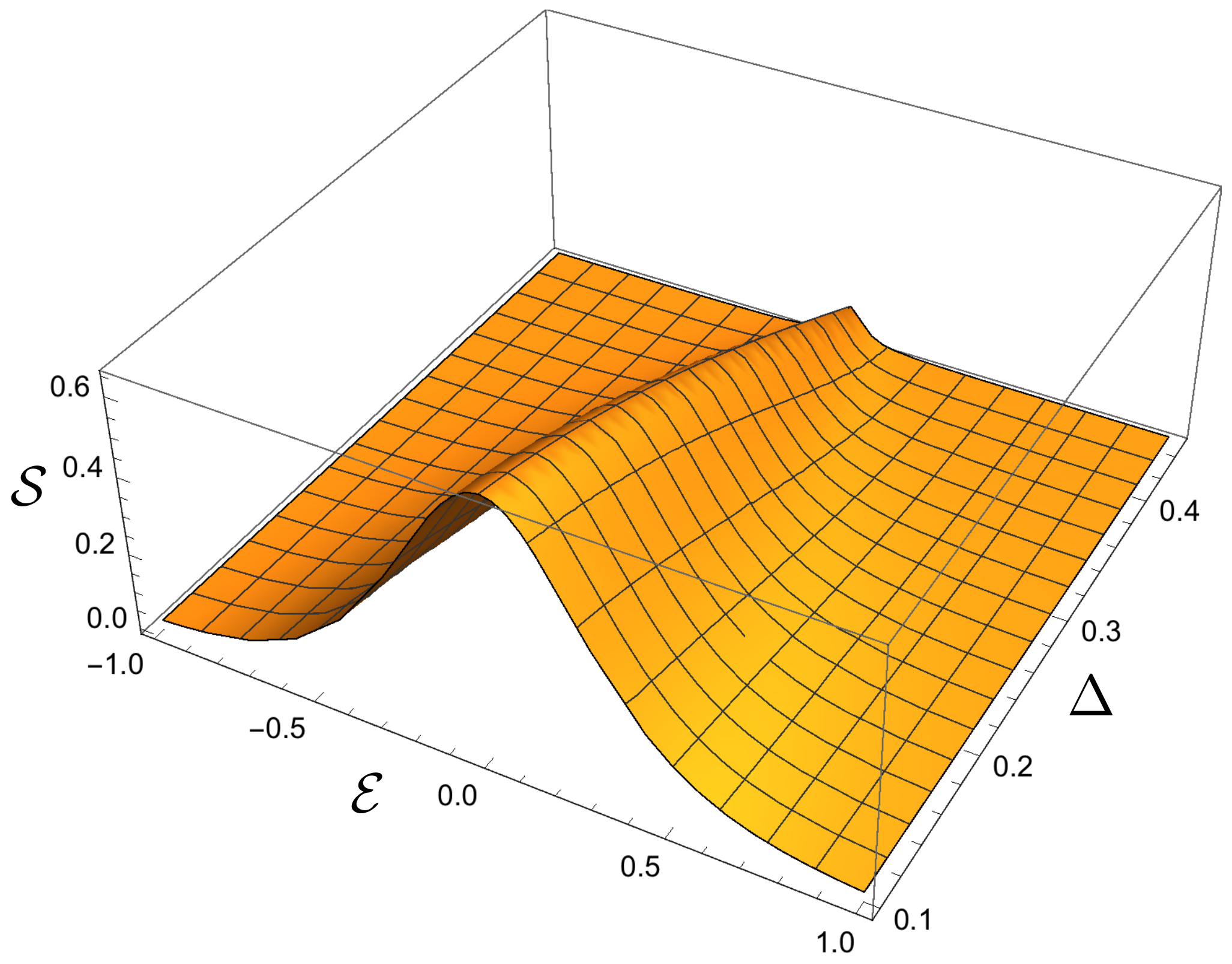}
\end{center}
\caption{The entropy $\mathcal{S}$ as a function of $\mathcal{E}$ and $\Delta$}
\label{fig:entropy}
\end{figure}

\begin{figure}
\begin{center}
\includegraphics[height=7.5cm]{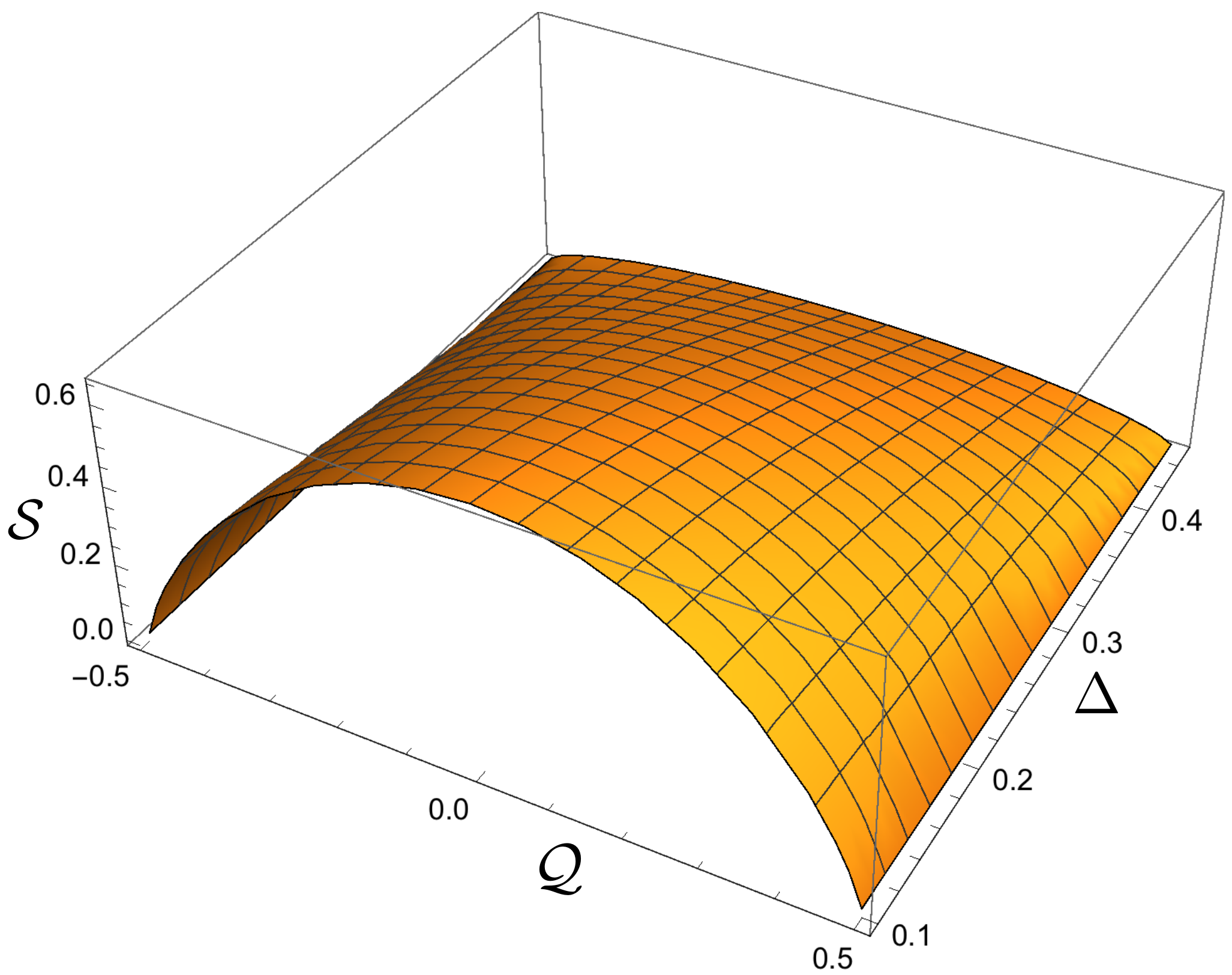}
\end{center}
\caption{The entropy $\mathcal{S}$ as a function of $\mathcal{Q}$ and $\Delta$}
\label{fig:entropy_Q}
\end{figure}

We can compute the entropy $\mathcal{S}$ from (\ref{Fres}), (\ref{t7}) and (\ref{QE}).
It can be verified that $\mathcal{S} \rightarrow 0$ as $\mathcal{E} \rightarrow \pm \infty$.
A plot of the entropy as a function of $\mathcal{E}$ appears in Fig.~\ref{fig:entropy}.
We can combine Figs.~\ref{fig:density} and \ref{fig:entropy} to obtain the entropy as a function of density, and this is shown in Fig.~\ref{fig:entropy_Q}.
In Appendix~\ref{app:numerics}, we present the results of the numerical solution of the saddle point equations in Eqs.~(\ref{e2}) and (\ref{e3}) for $q=4$, and find good agreement with the analytic results above.

At the particle-hole symmetric point, $\mathcal{E}=\mathcal{Q}=0$, this yields from (\ref{Fres})
\beq
\mathcal{S}(0) = 
\frac{(1-2\Delta)}{2} \ln \left[ 4 \cos^2 (\pi \Delta) \right] - \frac{i}{2 \pi} 
\left[{\rm Li}_2 \left(- e^{-2 \pi i \Delta} \right) - {\rm Li}_2 \left(- e^{2 \pi i \Delta} \right) \right],
\eeq
which agrees with Kitaev's result \cite{kitaev2015talk}.

%********************************************************
\subsection{Fluctuations}
\label{sec:SYKf1}
%********************************************************

This subsection presents an analysis of the ``zero mode'' fluctuations about the large $N$ saddle point found above. We will generalize the Schwarzian effective action, proposed in Ref.~\onlinecite{JMDS16}, to non-zero $\mu$, and relate its coupling constants to thermodynamic derivatives.

While solving the equations for the Green's function and the self energy, Eqs.~(\ref{e2}) and (\ref{e3}), we found that, at $\omega, T \ll J$, the $i \omega + \mu$ term in the inverse Green's function could be ignored in determining the IR solution Eq.~(\ref{ansatz}). After dropping the $i \omega + \mu$ term, it is not difficult to show that Eqs.~(\ref{e2}) and (\ref{e3}) have remarkable, emergent, time reparameterization and U(1) invariances. This is clearest if we write the Green's function in a two-time notation, i.e.
%$G(\tau_1 - \tau_2) = G(\tau_1, \tau_2)$
$G(\tau_1,\tau_2)$
; then Eqs.~(\ref{e2}) and (\ref{e3}) are invariant under \cite{kitaev2015talk,SS15} 
\begin{align}
\begin{split}
\label{repara}
%\tau &= f (\sigma)
%\\
G(\tau_1 , \tau_2) &= \left[ f' (\tau_1) f' (\tau_2) \right]^{\Delta} \frac{ g (\tau_2)}{g (\tau_1)} \, G(f(\tau_1), f(\tau_2))
\\
{\Sigma} (\tau_1, \tau_2) &= \left[ f' (\tau_1) f' (\tau_2) \right]^{1-\Delta} \frac{ g (\tau_1)}{g (\tau_2)} \, {\Sigma} (f(\tau_1), f(\tau_2)) 
\end{split}
\end{align}
where $f(\tau)$ and $g(\tau)$ are arbitrary functions representing the reparameterizations of time and U(1) transformations respectively. 

Next, we observe that these approximate symmetries are broken by the saddle-point solution, $G_s$ in Eq.~(\ref{Gsigma2}). So, following Ref.~\onlinecite{JMDS16}, we deduce an effective action for the associated Nambu-Goldstone modes by examining the action of the symmetries on the saddle-point solution,
\beq
G(\tau_1, \tau_2) = [f'(\tau_1) f'(\tau_2)]^{\Delta} G_s (f(\tau_1) , f(\tau_2)) e^{i \phi (\tau_1) - i \phi (\tau_2)},. \label{GGs}
\eeq
Here, we find it convenient to parameterize $g(\tau) = e^{-i \phi (\tau)}$ in terms of a phase field $\phi$. We will soon see that its derivative is conjugate to density fluctuations.
%which will turn out to be conjugate to density fluctuations. 
Our remaining task \cite{JMDS16} is to ({\em i\/}) find the set of $f(\tau)$ and $\phi (\tau)$ which leave Eq.~(\ref{GGs}) invariant {\em i.e.\/} Eq.~(\ref{GGs}) holds after we replace the l.h.s. by $G_s (\tau_1, \tau_2)$; and ({\em ii\/}) propose an effective action which has the property of remaining invariant under the set of $f(\tau)$ and $\phi (\tau)$ which leave Eq.~(\ref{GGs}) invariant.

For the first task, we find \cite{kitaev2015talk,JMDS16} that only reparameterizations, $f(\tau)$ belonging to 
$\PSL (2, \mathbb{R})$ leave Eq.~(\ref{GGs}) invariant. 
At $T>0$, we need $\PSL (2, \mathbb{R})$ transformations which map the thermal circle $0 < \tau < 1/T$
to itself. These are given by
\beq
\frac{1}{\pi T} \tan (\pi T f(\tau)) = \frac{a \tan(\pi T \tau) + b \pi T}{c \tan(\pi T \tau) + d \pi T} \quad, \quad ad - bc =1,
\eeq
where $a,b,c,d$ are real numbers.
This transformation is more conveniently written in terms of unimodular complex numbers 
\beq
z = e^{2 \pi i T \tau} \quad, \quad z_f = e^{2 \pi i T f(\tau)}
\eeq
as
\beq
z_f = \frac{ w_1 \, z + w_2 }{w_2^\ast \, z + w_1^\ast} \quad, \quad |w_1|^2 - |w_2|^2 = 1, \label{zsl2r}
\eeq
where $w_{1,2}$ are complex numbers.
Applying Eq.~(\ref{zsl2r}) to Eqs.~(\ref{Gsigma2}) and (\ref{GGs}), we find that Eq.~(\ref{GGs}) remains invariant only for the particle-hole symmetric case $\mathcal{E}=0$, which was considered previously \cite{JMDS16}. However, Eq.~\eqref{GGs} is invariant under $\PSL(2,\mathbb{R})$ transformations when the phase field $\phi(\tau)$ is related to the $\PSL(2,\mathbb{R})$ transformation $f(\tau)$ as
\beq
-i \phi (\tau) = 2 \pi \mathcal{E} T ( \tau - f(\tau)) \label{phif}
\eeq
The effective action is required to vanish when $\phi (\tau)$ satisfies Eq.~(\ref{phif}): this is the key result of this subsection, and is the origin of the constraints on thermoelectric properties described in this paper.

Now we can turn to the second task of obtaining an effective for $f(\tau)$ and $\phi (\tau)$ 
which is invariant Eqs.~(\ref{zsl2r}) and (\ref{phif}). It is more convenient to use the parameterization
\beq
f(\tau) \equiv \tau + \epsilon (\tau),
\eeq
and express the action in terms $\phi (\tau)$ and $\epsilon (\tau)$. 
Generalizing the reasoning in Ref.~\onlinecite{JMDS16}, we propose the action
\beq
\frac{S_{\phi,\epsilon}}{N} = \frac{K}{2} \int_0^{1/T} d \tau \left[ \partial_\tau \phi + i (2 \pi \mathcal{E} T) \partial_\tau \epsilon \right]^2 -\frac{\gamma}{4 \pi^2} \int_0^{1/T} d \tau \, \{\tan(\pi T (\tau + \epsilon(\tau)), \tau\},
\label{pf1}
\eeq
which appeared earlier in Eq.~(\ref{pf1i}).
Higher powers of the first term in square brackets can also be present, but we do not consider them here. 
The curly brackets in the second term represent a Schwarzian derivative
\beq
\{f, \tau\} \equiv \frac{f'''}{f'} - \frac{3}{2} \left( \frac{f''}{f'} \right)^2 , \label{Schwarzian}
\eeq
which has the important property of vanishing under $\PSL(2, \mathbb{R})$ transformations.

Our reasoning above falls short of a complete derivation of the structure of the effective action in Eq.~(\ref{pf1}).
The missing ingredient is our assumption that it is permissible to expand the action in gradients of $\phi$ and $\epsilon$,
when the saddle-point action contains long-range power-law interactions in time. If this assumption was not valid, then the 
phenomenological couplings
$K$ and $\gamma$ would diverge in the $T \rightarrow 0$ limit. We compute the values of $K$ and $\gamma$ in Appendix~\ref{app:largeq}
using a large $q$ expansion and find that they are finite as $T \rightarrow 0$. This a posteriori justifies our gradient expansion. 
We also present a normal-mode analysis of fluctuations of the underlying path integral for the SYK model in Appendix~\ref{app:normalmode};
this follows the analysis of Ref.~\onlinecite{JMDS16}, and uses their reasoning to provide an alternative motivation of Eq.~(\ref{pf1}).

We now relate the phenomenological couplings, $\gamma$ and $K$, to 
thermodynamic quantities by computing the fluctuations of energy and number
density implied by $S_{\phi ,\epsilon}$ in the large $N$ limit.
The energy and density operators are defined by
\beq
\delta E(\tau) - \mu \delta \mathcal{Q} (\tau) = \frac{1}{N} \frac{\delta S_{\phi,\epsilon}}{\delta\epsilon' (\tau)} \quad , \quad 
 \delta \mathcal{Q} (\tau) = \frac{i}{N}\frac{\delta S_{\phi,\epsilon}}{\delta\phi' (\tau)}. \label{dsde}
\eeq
Introducing,
\beq
\widetilde{\phi} (\tau) = \phi (\tau) + i 2 \pi \mathcal{E} T \epsilon (\tau) \label{deftildephi}
\eeq
and expanding (\ref{pf1}) to quadratic order in $\phi$ and $\epsilon$, we obtain the Gaussian action
\beq
\frac{S_{\phi,\epsilon}}{N} = \frac{K T}{2} \sum_{\omega_n \neq 0} \omega_n^2 \left| \widetilde{\phi} (\omega_n) \right|^2 
+ \frac{T \gamma}{8 \pi^2} \sum_{|\omega_n| \neq 0, 2 \pi T} \omega_n^2 (\omega_n^2 - 4 \pi^2 T^2) |\epsilon (\omega_n)|^2 + \ldots
\label{pf2}
\eeq
where $\omega_n$ is a Matsubara frequency. Note the restrictions on $n=0,\pm 1$ frequencies in (\ref{pf2}), which are needed to eliminate
the zero modes associated with $\PSL(2,\mathbb{R})$ and U(1) invariances. In terms of $\widetilde{\phi} (\tau) $ and $\epsilon (\tau)$, Eq.~(\ref{dsde}) is
\bea
\delta \mathcal{Q} (\tau) &=& i K  \widetilde{\phi}' (\tau) \nn
\delta E (\tau) - \mu_0 \delta \mathcal{Q} (\tau)   &=& - \frac{\gamma}{4 \pi^2} \left[ \epsilon''' (\tau) + 4 \pi^2 T^2 \epsilon' (\tau) \right]
+ i 2 \pi K \mathcal{E} T \widetilde{\phi}' (\tau). \label{s2b}
\eea
Now we compute the correlators of these observables in the Gaussian action in Eq.~(\ref{pf2}), following
the methods of Ref.~\onlinecite{JMDS16}.
We have for the two-point correlator of $\widetilde{\phi} (\tau)$
\begin{align}
\begin{split}
\label{z6}
\left\langle \widetilde{\phi} (\tau) \widetilde{\phi} (0) \right\rangle 
 &= \frac{T}{NK} \sum_{\omega_n \neq 0} \frac{e^{i \omega_n \tau}}{\omega_n^2}
 \\
&= \frac{1}{NKT} \left[ \frac{1}{2} \left( T \tau - \frac{1}{2} \right)^2 - \frac{1}{24} \right] \quad \mbox{for $0 < T \tau < 1$},
\end{split}
\end{align}
and extended periodically for all $\tau$ with period $1/T$. Similar for $\epsilon (\tau)$
\bea
\left\langle \epsilon (\tau) \epsilon (0) \right\rangle 
&=& \frac{4 \pi^2 T}{N\gamma} \sum_{|\omega_n| \neq 0, 2\pi T} \,
\frac{e^{i \omega_n \tau}}{\omega_n^2 (\omega_n^2 - 4 \pi^2 T^2)} \nn
&=& \frac{1}{N\gamma T^3} \left[ \frac{1}{24} + \frac{1}{4 \pi^2} - \frac{1}{2} \left( T \tau - \frac{1}{2} \right)^2 + 
\frac{5}{8 \pi^2} \cos (2 \pi T \tau) + \frac{1}{2 \pi} \left( T\tau - \frac{1}{2} \right) \sin (2 \pi T \tau) \right] \nn
&~&~~~~~~~~~~~~~~~~~~~~~~~\mbox{for $0 < T \tau  < 1$.} \label{s3}
\eea
Inserting Eqs.~(\ref{z6}) and (\ref{s3}) into Eq.~(\ref{s2b}), we confirm that the correlators of the conserved
densities are $\tau$-independent; their second moment correlators, which define the matrix of static
susceptibility correlators by (\ref{ti1}), are given by
\begin{align}
\begin{split}
\label{defchis}
\chi_s &= \frac{1}{N} \left( \begin{array}{cc}
-(\partial^2 \Omega/\partial \mu^2)_T &  -(\partial^2 \Omega/ \partial \mu \partial T)_\mu \\
-T (\partial^2 \Omega/ \partial \mu \partial T)_\mu & -T (\partial^2 \Omega/\partial T^2)_\mu
\end{array} \right)
\\
&=\frac{1}{T} \left( \begin{array}{cc}
\left\langle (\delta \mathcal{Q})^2 \right\rangle & \left\langle (\delta E - \mu \delta \mathcal{Q}) \delta \mathcal{Q} \right\rangle/T  \\
\left\langle (\delta E - \mu \delta \mathcal{Q}) \delta \mathcal{Q} \right\rangle & 
\left\langle (\delta E - \mu \delta \mathcal{Q})^2 \right\rangle/T
\end{array} \right)
\\
 &= \frac{1}{N} \left( \begin{array}{cc}
K &  2 \pi K \mathcal{E} \\
2 \pi K \mathcal{E} T  & (\gamma + 4 \pi^2 \mathcal{E}^2 K ) T
\end{array} \right) 
\end{split}
\end{align}
From Eq.~(\ref{defchis}) we obtain the relationship between the couplings $K$ and $\gamma$ in the effective action
in Eq.~(\ref{pf1}). After application of some thermodynamic identities, we can write these as
\beq
K = \left( \frac{\partial \mathcal{Q}}{\partial \mu} \right)_T \quad , \quad \gamma = - \left( \frac{\partial^2 F}{\partial T^2}
\right)_{\mathcal{Q}}, \label{Kgamma}
\eeq
and also confirm the thermodynamic definitions of $\mathcal{E}$ in Eqs.~(\ref{defE0}), (\ref{dmdt}), and (\ref{edsdq}).

Appendix~\ref{app:fluc} presents another argument for the results in Eq.~(\ref{Kgamma}) without computation of fluctuations
of the effective action.

%********************************************************
\subsection{Higher-dimensional SYK theory}
\label{sec:SYKd}
%********************************************************

%\YG{\bf [YG: I plan to change the structure here, I think it might be better to treat this section as a subsection of the section II, and change title of section II to ``SYK model'', how do you guys think?]} \SA{OK}

Gu {\em et al.\/} have defined a set of higher-dimensional SYK models \cite{GQS16} for Majorana fermions and computed their energy transport properties. Here we extend their results to the case of complex fermions at a general $\mu$, and discuss their thermoelectric transport.

\begin{figure}[t]
\center
\includegraphics[width=6in]{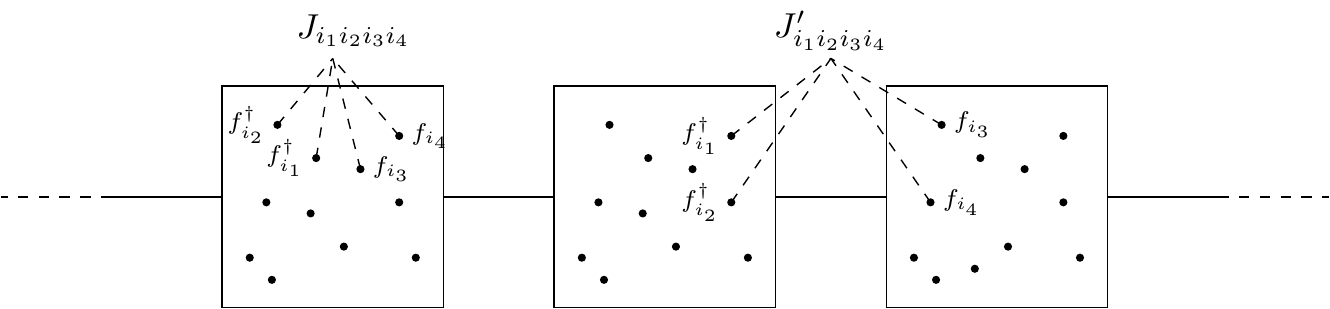}
\caption{%(From \cite{GQS16}) 
A chain of coupled SYK sites with complex fermions (in this figure we draw $q=4$ case): each site contains $N\gg 1$ fermions with on-site
interactions as in (\ref{h}). The coupling between nearest neighbor sites are four fermion interaction with two from each site. In general, one can consider other types of $q$-body interactions ($q=4$ in this caption), e.g. $f^\dagger_{x,i_1} f_{x,i_2} f^\dagger_{x+1,i_3} f_{x+1,i_4}$. Such terms will only change the ratio between $D_1$ and $D_2$, i.e. Eq.~(\ref{Dratio}) by a non-universal coefficient depends on the details of the model. In particular, if we only have $f^\dagger_{x,i_1} f_{x,i_2} f^\dagger_{x+1,i_3} f_{x+1,i_4}$-type terms to couple the nearest neighbour sites, charge diffusion $D_1$ will vanish due to the local charge conservation.}
\label{fig:chainSYK}
\end{figure}

% of the Hamiltonian to the special case $q=4$,\YG{\bf [I will try to write a general q theory to match the style of this paper]} and 
We will limit our presentation
to one spatial dimension (although the results easily generalize to all spatial dimensions). We consider the model
\begin{equation}
H%^{\prime} 
=\sum_{x} \left(H_{x}+\delta H_x \right) \label{tt1}
\end{equation}
The on-site term $H_x$ is equivalent to a copy of Eq.~(\ref{h}) on each site $x$ %\YG{\bf[I modify the model to general q]}
\begin{equation}
H_x = \sum_{\substack{1\leq i_1 < i_2 \ldots i_{q/2} \leq N,\\ 1\leq i_{q/2+1} < i_{q/2+2} \ldots < i_{q} \leq N}} J_{x,i_1, i_2 \ldots i_{q}} \, 
f^\dagger_{x,i_1} % f^\dagger_{i_2} 
\ldots f^\dagger_{x,i_{q/2}}   f_{x,i_{q/2 + 1}}   \ldots 
%f_{x,i_{q-1}} 
f_{x,i_{q}}
\end{equation}
%\begin{equation}
%H_x= \sum_{\substack{1\leq j < k \leq N,\\ %1\leq l < m \leq N}} J_{jklm,x} %c^\dagger_{j,x} c^\dagger_{k,x} c_{l,x} %c_{m,x}. \label{tt2}
%\end{equation}
The nearest neighbor coupling term $\delta H_x$ denotes nearest-neighbor interactions
as shown in Fig~\ref{fig:chainSYK},
\begin{equation}
\delta H_x=  \sum_{\substack{1\leq i_1 < i_2 \ldots i_{q/2} \leq N,\\ 1\leq i_{q/2+1} < i_{q/2+2} \ldots < i_{q} \leq N}} 
J'_{x,i_1, i_2 \ldots i_{q}} \, 
f^\dagger_{x,i_1} % f^\dagger_{i_2} 
\ldots f^\dagger_{x,i_{q/2}}   f_{x+1,i_{q/2 + 1}}   \ldots 
%f_{x,i_{q-1}} 
f_{x+1,i_{q}} + \mbox{H.c.} \,.\label{tt3}
\end{equation}
The couplings
%coupling constant 
$\{J_{x,i_1, i_2 \ldots i_{q}}  \}$ and $\{  J'_{x,i_1, i_2 \ldots i_{q}} \}$ are all independent random variables\footnote{except that $J_{x,i_1, i_2 \ldots i_{q}}$ need to satisfy the hermitian condition as shown in Eq.~(\ref{eqn: hermitian}).} 
%couplings 
with zero mean, and variances given by
\begin{equation}
\overline{|J_{x,i_1, i_2 \ldots i_{q}} |^2}=\frac{J_0^2 (q/2)!^2}{N^{q-1}}\,,\quad \overline{|J'_{x,i_1, i_2 \ldots i_{q}} |^2}=\frac{J_1^2 (q/2)!^2}{N^{q-1}}\,. \label{tt4}
\end{equation}

We remark that the particular interaction we choose here is just one possible example, and this particular choice produces the Wiedemann-Franz ratio discussed in Eq.~(\ref{eq:WFforSYK}). In general, we can choose $p$ fermions from one site (with $p_1$ many $f^\dagger$ and $p_2$ many $f$, $p_1$ and $p_2$ can be chosen arbitrarily) to couple $(q-p)$ fermions in the nearest neighbor site (with $(q/2-p_1)$ many $f^\dagger$ and $(q/2-p_2)$ many $f$). For example, for $q=4$ case, we are allowed to have following couplings between $x$ and $x+1$:
\begin{align*}
p=1: ~~~~&
f^\dagger_{x,i_1} f^\dagger_{x+1,i_2} f_{x+1,i _3} f_{x+1,i_4}~,\quad   
f_{x,i_1}  f^\dagger_{x+1,i_2} f^\dagger_{x+1,i _3} f_{x+1,i_4}~; \\
p=2: ~~~~&
f^\dagger_{x,i_1}  f^\dagger_{x,i_2} f_{x+1,i _3} f_{x+1,i_4}~, \quad
f^\dagger_{x,i_1}  f_{x,i_2} f^\dagger_{x+1,i _3} f_{x+1,i_4}~,\quad
f_{x,i_1}  f_{x,i_2} f^\dagger_{x+1,i _3} f^\dagger_{x+1,i_4}~; \\
p=3: ~~~~&
f^\dagger_{x,i_1} f^\dagger_{x,i_2} f_{x,i _3} f_{x+1,i_4}~, \quad   
f^\dagger_{x,i_1}  f_{x,i_2} f_{x,i _3} f^\dagger_{x+1,i_4}~; 
\end{align*}
If we include these couplings with different coefficient, the Wiedemann-Franz ratio will be non-universal. However, for $q=2$ model, the only term we can add is $f^\dagger_{i_1,x}f_{i_2,x+1}$ and therefore the Wiedemann-Franz ratio goes back to $\pi^2/3$ as discussed previously.

Following the analysis in Ref.~\onlinecite{GQS16}, the effective action for the higher dimensional model can be deduced from
that of the zero dimensional model. Using the results in Appendices~\ref{app:normalmode} and~\ref{app:diffusion}, we find
that the Gaussian action for energy and density fluctuations in higher dimensions generalizes from Eq.~(\ref{pf2}) to 
\begin{align}
\begin{split}
\label{pf3}
\frac{S_{\phi,\epsilon}}{N} &= \frac{K T}{2} \sum_{k,\omega_n \neq 0} |\omega_n| (D_1 k^2 + |\omega_n|) \left| \widetilde{\phi} (k,\omega_n) \right|^2 \,,
\\
&\qquad + \frac{T \gamma}{8 \pi^2} \sum_{k,|\omega_n| \neq 0, 2 \pi T} |\omega_n| (D_2 k^2 + |\omega_n|) (\omega_n^2 - 4 \pi^2 T^2) |\epsilon (k,\omega_n)|^2\,,
\end{split}
\end{align}
where $D_1$ and $D_2$ are the diffusion constants of the conserved charges.
%From the computation 
In Appendix~\ref{app:diffusion} we find that their ratio obeys
\beq
\frac{D_2}{D_1}= \frac{4\pi^2 \Delta^2}{3} \frac{K}{\gamma}. \label{Dratio}
\eeq
Following Ref.~\onlinecite{GQS16}, from Eq.~(\ref{pf3}), and including a contact term as described in Ref.~\onlinecite{Policastro:2002tn}, we 
can obtain the long-wavelength and low frequency dynamic susceptibilities
\begin{align}
\begin{split}
\label{pf4}
\left\langle  \mathcal{Q};  \mathcal{Q} \right\rangle_{k, \omega} &= K \frac{D_1 k^2}{- i \omega + D_1 k^2}\,,
\\
\left\langle  E - \mu  \mathcal{Q};  \mathcal{Q} \right\rangle_{k, \omega} &=  2 \pi K \mathcal{E} T \frac{D_1 k^2}{- i \omega + D_1 k^2}\,,
\\
\left\langle  E - \mu  \mathcal{Q};  E - \mu  \mathcal{Q}  \right\rangle_{k, \omega}/T &= \gamma T \frac{D_2 k^2}{- i \omega + D_2 k^2} + 
4 \pi^2 \mathcal{E}^2 K  T \frac{D_1 k^2}{- i \omega + D_1 k^2}\,.
\end{split}
\end{align}
We note that the form of the thermoelectric correlators in Eq.~(\ref{pf4}) 
is not generic to
incoherent metals \cite{SAH15}, and the simple structure here relies on specific
features of our effective action in Eq.~(\ref{pf3}); in Section~\ref{sec:holotransport}, we will again obtain Eq.~(\ref{pf4}) by holographic methods, where its structure is linked to the presence of AdS$_2$ factor in the near-horizon metric.
Now comparing (\ref{pf4}) with (\ref{ti2}), and using the susceptibility matrix (\ref{defchis}), we can work out the diffusion matrix, $D$, leading to the result presented earlier in Eq.~(\ref{pf5}). Using (\ref{ti3}), we find the conductivity matrix
\beq
 \left( \begin{array}{cc}
 \sigma   & \alpha  \\
 \alpha T &  \overline{\kappa} \end{array} \right) = 
 \left( \begin{array}{cc}
 D_1 K & 2 \pi K \mathcal{E} D_1 \\
 2 \pi K \mathcal{E} D_1 T & (\gamma D_2 + 4 \pi^2 \mathcal{E}^2 K D_1) T 
 \end{array} \right). \label{pf6}
\eeq
From this result we obtain the Seebeck coefficient presented in Eq.~(\ref{seebeck}).
Also, we have the thermal conductivity
\beq
\kappa = \overline{\kappa} - \frac{T \alpha^2}{\sigma} = \gamma D_2  T. \label{tt7}
\eeq
All of these hydrodynamic results are in accord with the linearized equations of motion 
\beq
\left( \begin{array}{c}
\partial \mathcal{Q}/\partial t \\ \partial E/\partial t - \mu \partial \mathcal{Q}/\partial t \end{array}
\right) = D \, \left( \begin{array}{c}
\nabla^2 \mathcal{Q} \\ \nabla^2 E - \mu \nabla^2 \mathcal{Q}  \end{array} \right). \label{d5}
\eeq
with the diffusion matrix as in Eq.~(\ref{pf5}). The dynamic susceptibilities (\ref{pf4}) can be diagonalized by using the operators $\mathcal{Q}$ and $E-\mathcal{Q}(\mu+2\pi\mathcal{E}T)$. The first of these is carried only by the mode with diffusion constant $D_1$, and the second is carried only by the mode with diffusion constant $D_2$. $D_1$ is the charge diffusion constant. We call $D_2$ the thermal diffusion constant as it's very simply related to the thermal conductivity via Eqs.~(\ref{tt7}) and (\ref{Kgamma}).

%Although the dynamic energy susceptibility has poles due to both diffusive modes, we call $D_2$ the energy diffusion constant: from (\ref{d5}) $E$ obeys the diffusion equation with this diffusion constant when there are no fluctuations in the charge density. Also, $D_2$ is simply related to the thermal conductivity via Eqs.~(\ref{tt7}) and (\ref{Kgamma}).

We can also compute the Wiedemann-Franz ratio, $L$, from the above results and the computations in Appendix~\ref{app:diffusion}. Eq.~(\ref{Dratio}) leads directly to 
Eq.~(\ref{eq:WFforSYK}).

The Lyapunov exponent $\lambda_L$ and butterfly velocity $v_B$, which characterize early-time chaotic growth through the growth of connected out-of-time-ordered thermal four-point functions
\begin{equation}
\llangle V^\dagger (t,\vec{x}) W^\dagger (0)V(t,\vec{x}),W(0)\rrangle_{\beta} \sim e^{\lambda_L(t - |\vec{x}|/v_B )}\,,
\end{equation}
can be computed in this model just as in Ref.~\onlinecite{GQS16}.
%The quantum chaos properties of the present model can be computed just as in Ref.~\onlinecite{GQS16}. 
The key observation here is that these properties are associated with the fluctuations of the $\epsilon$ mode, and the $\phi$ mode is mostly a spectator. 
%The Lyapunov time is saturated at the chaos bound
Taking $V$ and $W$ to be the SYK fermions, the present model has a Lyapunov exponent given by
\begin{equation}
\lambda_L = 2\pi T\,,
\end{equation}
saturating the chaos bound of~\cite{maldacena2015abound}. As in the works~\cite{Blake:2016wvh,Blake:2016sud,GQS16,Patel:2016wdy}, we find that the butterfly velocity
%, $v_B$, has a simple relation 
is simply related to the thermal diffusivity as 
\beq
D_2 = \frac{v_B^2}{2 \pi T}\,. \label{tt8}
\eeq
From Eq.~(\ref{Dratio}), we observe that the relationship between $v_B$ and the charge diffusion constant $D_1$ is not universal \cite{Lucas:2016yfl}: it depends upon the specific parameters of the SYK model. 

%********************************************************
\section{Holographic theories}
\label{sec:holotheories}
%********************************************************

The presence, or lack, of translational symmetry has a qualitative impact on the IR transport properties of a system. The disorder present in the higher-dimensional SYK models breaks translational symmetry, resulting in finite thermoelectric conductivities and diffusive transport. 

In this Section we study thermoelectric transport in holographic models which do not have translational symmetry.
%To verify that the SYK transport properties (\ref{seebeck}) can be captured by holography, we must consider gravitational solutions that don't have translational symmetry. 
The simplest holographic theories with this property are ``bottom-up'' models in which translational symmetry is broken isotropically and homogeneously by massless scalar  `axion' fields $\varphi_i$ \cite{Andrade:2013gsa,Donos:2013eha,Gouteraux:2014hca,Donos:2014uba}, with an action~\cite{Bardoux:2012aw,Andrade:2013gsa},
%The simplest theory with solutions of this type has the action \cite{Bardoux:2012aw,Andrade:2013gsa}
\begin{equation}
\label{eq:charged4daction}
S=\int d^4x\sqrt{-g}\left(\mathcal{R}+V-\frac{1}{2}\sum_{i=1}^{2}(\partial\varphi_i)^2-\frac{1}{4}F_{\mu\nu}F^{\mu\nu}\right),
\end{equation}
where $V=6$. Note that the action has a shift symmetry for the axion fields $\varphi_i \rightarrow \varphi_i + \mbox{constant}$. The index $i$ labels the spatial directions of the dual field theory. We will consider more general theories which have AdS$_2$ near-horizon solutions in Appendix~\ref{app:moregeneralads2solutions}. This theory has charged black brane solutions
\begin{equation}
\begin{aligned}
\label{eq:simpleaxionsoln}
&ds^2=-r^2f(r)dt^2+\frac{dr^2}{r^2f(r)}+r^2d\vec{x}^2\,,\;\;\;\;\; \varphi_i=mx^i\,,\;\;\;\;\; A_t(r)=\mu\left(1-\frac{r_0}{r}\right),\\
&f(r)=1-\frac{m^2}{2r^2}-\left(1-\frac{m^2}{2r_0^2}+\frac{\mu^2}{4r_0^2}\right)\frac{r_0^3}{r^3}+\frac{\mu^2r_0^2}{4r^4}\,.
\end{aligned}
\end{equation}
The $x^i$ dependence of the axion fields implies that translational symmetry is broken and 
momentum is no longer conserved. However, because of the shift symmetry of the axion fields, 
the solution is spatially homogeneous, and the metric remains independent of $x^i$.
This simple form of the metric is an
advantage of breaking translational symmetry in this way.

The bottom-up model~\eqref{eq:charged4daction} has not yet been embedded into string theory, and we do not know if it's possible to do so.\footnote{If it were possible, then the dual field theory would be a CFT with two marginal deformations and a flat conformal manifold.}
%and there are reasons to expect that this cannot be done. Among other things, if there was such an embedding, then the fields $\varphi_i$ would be dual to exactly marginal couplings on a trivial conformal manifold, and there are no known CFTs with this property. 
However, as we are interested in transport properties which we expect to be robust for all systems with the same low-energy symmetries, we forge ahead and determine the properties of the putative field theory dual using the usual AdS/CFT dictionary. This would-be field theory is a (2+1)-dimensional CFT, deformed by 
%turning on 
a temperature $T$, a chemical potential $\mu$ for a conserved U(1) charge, and sources $m x^i$ for axion fields. It is these sources, linear in the spatial coordinates, which explicitly break translational symmetry in the field theory. These sources are not spatially disordered, but they can be thought of as capturing the homogeneous $k=0$ mode of a disorder sum, which is the most relevant one at small $m$ \cite{Davison:2013txa}.

The sources $\mu$ and $m$ alter the solution, and the result is a geometry that interpolates between AdS$_4$ near the boundary $r\rightarrow\infty$, and AdS$_2\times\mathbb{R}^2$ near the horizon $r=r_0$. To see the near-horizon AdS$_2$ explicitly, it is convenient to change variables from $m$ to $r_*$, the location of the zero temperature horizon, 
\begin{equation}
12r_*^4=\left(2m^2r_*^2+\mu^2r_0^2\right)\,.
\end{equation}
We then introduce a formal expansion parameter $\epsilon$ and make the change of variables
\begin{equation}
\label{eq:adsnearhorlim}
\zeta=\frac{r-r_*}{\epsilon}\,,\;\;\;\;\;\;\;\;\;\; \zeta_0=\frac{r_0-r_*}{\epsilon}\,,\;\;\;\;\;\;\;\;\;\;\tau=\epsilon t\,,
\end{equation}
and take the $\epsilon\rightarrow0$ limit. This limit gives the geometry near the horizon, at small temperatures. The result is
\begin{equation}
\begin{aligned}
\label{eq:finiteTads2solution}
&ds^2=-\frac{\zeta^2f(\zeta)}{\tilde{L}^2}d\tau^2+\frac{\tilde{L}^2}{\zeta^2f(\zeta)}d\zeta^2+d\vec{x}^2r_*^2+O(\epsilon),\;\;\;\;\;\;\;\;\;\;f(\zeta)=1-\frac{\zeta_0^2}{\zeta^2},\\
&A_\tau=\frac{\mathcal{E}}{\tilde{L^2}}\left(\zeta-\zeta_0\right)+O(\epsilon),\;\;\;\;\;\;\;\;\mathcal{E}=\frac{\mu}{r_*\left(3+\frac{\mu^2}{4r_*^2}\right)},\;\;\;\;\;\;\;\;\varphi_i=mx^i.
\end{aligned}
\end{equation}
At leading order in $\epsilon$, the solution is AdS$_2\times\mathbb{R}^2$ with a non-zero electric field $\mathcal{E}$ and axions $\varphi_i$. The AdS$_2$ radius of curvature is 
\begin{equation}
\label{eq:LtildechargedAdS2}
\tilde{L}^2=\frac{1}{3+\frac{\mu^2}{4r_*^2}}.
\end{equation}
We are working with units in which the radius of the asymptotically AdS$_4$ spacetime is unity. This near-horizon geometry is supported by both the gauge field and the axions, and survives in either of the limits $m\rightarrow0$ or $\mu\rightarrow0$.

As was outlined in the introduction, there is a very close connection between the low-energy physics of the SYK models and gravity in nearly AdS$_2$ spacetimes, as both are governed by the same symmetry-breaking pattern. The presence of the AdS$_2$ factor in this near-horizon geometry 
%of the spacetime 
suggests that the low energy physics of the 
%field theory dual to 
model (\ref{eq:charged4daction}) may coincide with that of the SYK model. This is what we will explore and make more precise in the following Subsections.

%********************************************************
\subsection{Thermodynamics}
%********************************************************
\label{sec:AdSthermodyn}
The thermodynamic properties of the solution (\ref{eq:simpleaxionsoln}) can be determined in the usual way \cite{Andrade:2013gsa}. After an appropriate holographic renormalization, evaluating the on-shell Euclidean action gives the grand potential density
\begin{equation}
\begin{aligned}
\Omega(\mu,T)=-\frac{1}{24}\left(4\pi T+\sqrt{6m^2+16\pi^2T^2+3\mu^2}\right)\left[2m^2+\mu^2+\frac{1}{9}\left(4\pi T+\sqrt{6m^2+16\pi^2T^2+3\mu^2}\right)^2\right]\,,
\end{aligned}
\end{equation}
where the temperature is calculated from regularity of the Euclidean solution at the horizon
\begin{equation}
T=\frac{r_0^2f'(r_0)}{4\pi}=\frac{r_0}{4\pi}\left(3-\frac{m^2}{2r_0^2}-\frac{\mu^2}{4r_0^2}\right)\,,
\end{equation}
and the chemical potential is determined by the value of the gauge field at the AdS$_4$ boundary
\begin{equation}
\label{eq:holomuads4defn}
\mu\equiv\lim_{r\rightarrow\infty}A_t(r).
\end{equation}
The entropy density $\mathcal{S}$ and charge density $\mathcal{Q}$ are given by the usual thermodynamic derivatives
\begin{equation}
\begin{aligned}
\label{eq:ads4entropydensity}
&\mathcal{S}\equiv-\left(\frac{\partial\Omega}{\partial T}\right)_{\mu}=\frac{\pi}{9}\left(3(2m^2+\mu^2)+8\pi T\left(4\pi T+\sqrt{6m^2+16\pi^2T^2+3\mu^2}\right)\right)=4\pi r_0^2,\\
&\mathcal{Q}\equiv-\left(\frac{\partial\Omega}{\partial\mu}\right)_T=\frac{\mu}{6}\left(4\pi T+\sqrt{6m^2+16\pi^2T^2+3\mu^2}\right)=\mu r_0,
\end{aligned}
\end{equation}
These expressions agree with those obtained by using the Bekenstein-Hawking formula to obtain $\mathcal{S}$ in terms of the area of the horizon, and with identifying $\mathcal{Q}$ with the radially conserved electric flux $r^2A_t'(r)$. Finally, the energy density is 
\begin{align}
\label{eq:ads4energydensity}
E&=\Omega+T\mathcal{S}+\mu\mathcal{Q}=2r_0^3\left(1-\frac{m^2}{2r_0^2}+\frac{\mu^2}{4r_0^2}\right)\\
\nonumber
&=\frac{1}{27}\left(4\pi T+\sqrt{6m^2+16\pi^2T^2+3\mu^2}\right)\left(-3m^2+8\pi^2T^2+3\mu^2+2\pi T\sqrt{6m^2+16\pi^2T^2+3\mu^2}\right)\,.
\end{align}
From these expressions, it is straightforward to compute the susceptibility matrix $\chi_s$. In the low-temperature limit, it is
\beq
\label{eq:holoaxionsuscept}
\chi_s = \left( 
\arraycolsep=1.4pt\def\arraystretch{2.0}
\begin{array}{cc} \displaystyle 
\frac{m^2 + \mu^2}{ \sqrt{3(2 m^2 + \mu^2)}} & \displaystyle \frac{2 \pi \mu}{3} \\
\displaystyle \frac{2 \pi T \mu}{3} &  \displaystyle  \frac{8 \pi^2 T \sqrt{2 m^2 + \mu^2}}{3 \sqrt{3}}
\end{array} \right)\,.
\eeq
In the notation of (\ref{Kgamma}), the $T\rightarrow0$ limit of the charge susceptibility $K$ and the specific heat at fixed charge $\gamma$ are
\begin{equation}
\begin{aligned}
\label{eq:ads4suscept}
K = \frac{m^2 + \mu^2}{ \sqrt{3 (2 m^2 + \mu^2)}}\,,\;\;\;\;\;\;\;\;\;\;\;\;\;\;\;\;\;\;\;\;\gamma = \frac{ 4 \pi^2 (2m^2 + \mu^2)^{3/2}}{3 \sqrt{3} (m^2 + \mu^2)}\,.
\end{aligned}
\end{equation}
Finally, it is straightforward to verify that the relation (\ref{defE0}) is true in the $T\rightarrow0$ limit,
\begin{equation}
\label{eq:holozerotlimitthermo}
\lim_{T\to 0}\left(\frac{\partial\mathcal{S}}{\partial\mathcal{Q}}\right)_T=2\pi\mathcal{E}=\frac{2\pi\mu}{\sqrt{3}}\frac{\sqrt{2m^2+\mu^2}}{m^2+\mu^2}\,.
\end{equation}

As we have emphasized, the UV physics of this holographic theory is quite different from that of the SYK models. We expect there to be similarities in their IR properties due to the near-horizon AdS$_2$ geometry. It is then important to determine which of the thermodynamic properties we have just described are universal, \emph{i.e.}~are determined solely by the near-horizon geometry, and which are not and depend upon the UV details of the solution. A naive guess would be that the AdS$_2$ geometry captures the $T\rightarrow0$ limit of the full thermodynamics, but this is not quite right.

As in our analysis of the SYK models, it is much more convenient to work with the canonical free energy density density $F(\mathcal{Q},T)=\Omega+\mu\mathcal{Q}$. In the low-temperature limit, this has the form of (\ref{funiv})
\begin{equation}
\label{eq:fourdF}
F(\mathcal{Q},T)=E_0(\mathcal{Q})-T\mathcal{S}_0(\mathcal{Q})+O(T^2)\,,
\end{equation}
where $E_0$ and $\mathcal{S}_0$ are the $T=0$ limits of the energy (\ref{eq:ads4energydensity}) and entropy (\ref{eq:ads4entropydensity}) densities at fixed $\mathcal{Q}$
\begin{equation}
\begin{aligned}
\label{eq:fourdE0S0}
&E_0(\mathcal{Q})=\mathcal{Q}\sqrt{\sqrt{m^4+12\mathcal{Q}^2}-m^2}-\frac{1}{6\sqrt{3}}\left(\sqrt{m^4+12\mathcal{Q}^2}+m^2\right)^{3/2}\,,\\
&\mathcal{S}_0(\mathcal{Q})=\frac{\pi}{3}\left(\sqrt{m^4+12\mathcal{Q}^2}+m^2\right)\,.
\end{aligned}
\end{equation}

This is more convenient because, from the point of view of the near-horizon geometry, $\mathcal{Q}$ is a much more natural object than $\mu$. Of the four thermodynamic objects $T,\mathcal{S},\mathcal{Q}$ and $\mu$, only the first three can be determined from just the near-horizon solution: $T$ from regularity of the Euclidean near-horizon geometry, $\mathcal{S}$ from the area of the horizon, and -- for a solution with trivial profiles for charged matter fields -- $\mathcal{Q}$ from the radially conserved electric flux, which can be evaluated at the horizon. In contrast, $\mu$ (defined by (\ref{eq:holomuads4defn})) requires knowledge of the UV part of the geometry. For this reason, we will find that thermodynamic quantities involving $\mu$ will in general be non-universal, \emph{i.e.}~they depend upon the UV parts of the geometry. The charge susceptibility $\left(\partial\mathcal{Q}/\partial\mu\right)_T$ is one such non-universal quantity. 

%********************************************************
\subsubsection{Dimensional reduction}
%********************************************************

To make this more precise, we will cut off our geometry at the boundary of the near-horizon spacetime (\ref{eq:finiteTads2solution}), and study the thermodynamics of this solution. So that we may use the usual AdS/CFT dictionary, we will compactify the $\mathbb{R}^2$ part of the geometry on a torus of volume $V_2$, and study the resulting asymptotically AdS$_2$ solution within a 2D theory of gravity. We compactify using the ansatz\footnote{Tildes will denote quantities in the dimensionally reduced theory, and the indices $a,b$ run over the uncompactified directions.} 
\begin{equation}
ds^2=\sqrt{\phi_0/\phi}\,d\tilde{s}^2+\phi d\vec{x}^2\,,\;\;\;\;\;\;\;\;\;\;\;\;\;\;\;\;\;\;\;\;A_\mu dx^\mu=\tilde{A}_a dx^a\,,\;\;\;\;\;\;\;\;\;\;\;\;\;\;\;\;\;\;\;\;\varphi_i=mx^i\,,
\end{equation}
where $\phi_0$ is a constant and the fields do not depend on the torus coordinates. Reducing the theory~\eqref{eq:simpleaxionsoln} on the spatial torus gives the two-dimensional Einstein-Maxwell-dilaton action (up to boundary terms)
\begin{equation}
\label{eq:2dgravityaction}
S_{2D}=V_2\int d^2x\sqrt{-\tilde{g}}\left(\phi\tilde{\mathcal{R}}+V(\phi)-\frac{Z(\phi)}{4}\tilde{F}^2\right),
\end{equation}
where
\begin{equation}
\begin{aligned}
\label{eq:VZaxion}
V(\phi)&=6\sqrt{\phi_0\phi}-\left(6r_*^2-\frac{\mu^2r_0^2}{2r_*^2}\right)\sqrt{\phi_0/\phi}\,,\;\;\;\;\;\;\;\;\;\;Z(\phi)&=\sqrt{\phi^3/\phi_0}\,.
\end{aligned}
\end{equation}
An exact solution of the equations of motion of this action is an AdS$_2$ geometry with constant dilaton and electric field $\mathcal{E}$,
\begin{equation}
\begin{aligned}
&d\tilde{s}^2=-\frac{\zeta^2}{\tilde{L}^2}f(\zeta)dt^2+\frac{\tilde{L}^2}{\zeta^2f(\zeta)}\,,\;\;\;\;\;\;\;f(\zeta)=1-\frac{\zeta_0^2}{\zeta^2}\,,\;\;\;\;\;\;\;\;\;\;\tilde{A}_t=\frac{\mathcal{E}}{\tilde{L}^2}\left(\zeta-\zeta_0\right)\,,\;\;\;\;\;\;\;\;\;\;\phi_0=r_*^2\,, \\
&\mathcal{E}=\frac{\mu r_0\tilde{L}^2}{r_*^2}\,,\;\;\;\;\;\;\;\;\;\;\;\;\;\;\;\;\;\;\;\;\;\;\;\;\;\;\;\;\;\;\;\;\;\;\tilde{L}^2=\frac{1}{3r_*^2+\frac{\mu^2r_0^2}{4r_*^2}}\,.
\end{aligned}
\end{equation}
If we set $r_0=r_*$, this is the compactified version of the near-horizon solution (\ref{eq:finiteTads2solution}).

%********************************************************
\subsubsection{Thermodynamics of AdS$_2$ solutions}
%********************************************************

We will now determine the thermodynamics of these AdS$_2$ spacetimes. The action (\ref{eq:2dgravityaction}), for general $V(\phi)$ and $Z(\phi)$, has charged AdS$_2$ solutions with constant dilaton $\phi=\phi_0$ provided that
\begin{equation}
\label{eq:ads2conditions}
V(\phi_0)=\frac{\mathcal{E}^2}{2\tilde{L}^4}Z(\phi_0),\;\;\;\;\;\;\;\;\;\;\text{and}\;\;\;\;\;\;\;\;\;\;\frac{2}{\tilde{L}^2}=V'(\phi_0)+\frac{\mathcal{E}^2}{2\tilde{L}^4}Z'(\phi_0).
\end{equation}
where the solutions are written as
\begin{equation}
\label{eq:generalads2solution}
d\tilde{s}^2=-\frac{\zeta^2f(\zeta)}{\tilde{L}^2}dt^2+\frac{\tilde{L}^2}{\zeta^2f(\zeta)}d\zeta^2,\;\;\;\;\;\;\;\;\;\;f(\zeta)=1-\frac{\zeta_0^2}{\zeta^2},\;\;\;\;\;\;\;\;\;\; \tilde{A}_t=\frac{\mathcal{E}}{\tilde{L}^2}\left(\zeta-\zeta_0\right).
\end{equation}
Computing the temperature $\tilde{T}$ using regularity of the Euclidean solution and the charge density $\tilde{\mathcal{Q}}$ from the radially conserved electric flux (where densities are now given by dividing by the torus volume $V_2$), we find
\begin{equation}
\tilde{T}=\frac{\zeta_0}{2\pi\tilde{L}^2},\;\;\;\;\;\;\;\;\;\;\;\;\;\;\;\;\;\;\;\;\tilde{\mathcal{Q}}=\frac{\mathcal{E}}{\tilde{L}^2}Z(\phi_0).
\end{equation}

To compute the free energy of these solutions, we must supplement the action (\ref{eq:2dgravityaction}) with boundary terms. For general $V(\phi)$ and $Z(\phi)$ these are given by\footnote{There are typos in the corresponding boundary terms written in Ref.~\onlinecite{KJ16}, which we have corrected here. These boundary terms agree with those in Ref.~\onlinecite{CP16} for the particular theory studied there.}
\begin{equation}
S_{bdy}=V_2\int dt\sqrt{-\tilde{\gamma}}\left(2\phi \tilde{K}-\tilde{L}V(\phi)-Z(\phi)\tilde{A}_a \tilde{F}^{ab}\tilde{n}_b-\frac{\tilde{L}}{4}Z(\phi)\tilde{F}^2\right),
\end{equation}
where $\tilde{n}^a$ is the normal vector to the boundary, $\tilde{\gamma}$ is the induced metric on the boundary with $\tilde{K}$ its extrinsic curvature, and $\tilde{L}$ is the AdS$_2$ radius of curvature. With these counterterms, the Euclidean on-shell action gives the canonical free energy density of the AdS$_2$ solution as
\begin{equation}
\tilde{F}(\tilde{\mathcal{Q}},\tilde{T})=-4\pi\tilde{T}\phi_0.
\end{equation}
The right hand side is a non-trivial function of $\mathcal{Q}$ due to the implicit dependence of $\phi_0$ on $\mathcal{Q}$ given by equations (\ref{eq:ads2conditions}). For small variations, these equations imply that
\begin{equation}
\delta\phi_0=\frac{\mathcal{E}}{2}\delta\tilde{\mathcal{Q}}.
\end{equation}
Taking variations of the canonical free energy density gives the chemical potential and entropy density \cite{KJ16}
\begin{equation}
\begin{aligned}
\tilde{\mu}=\left(\frac{\partial\tilde{F}}{\partial\tilde{\mathcal{Q}}}\right)_{\tilde{T}}=-2\pi\mathcal{E}\tilde{T},\;\;\;\;\;\;\;\;\;\;\text{and}\;\;\;\;\;\;\;\;\;\;\tilde{\mathcal{S}}=-\left(\frac{\partial\tilde{F}}{\partial \tilde{T}}\right)_{\tilde{\mathcal{Q}}}=4\pi \phi_0.
\end{aligned}
\end{equation}
One further application of this formula gives
\begin{equation}
\left(\frac{\partial\tilde{\mathcal{S}}}{\partial\tilde{\mathcal{Q}}}\right)_{\tilde{T}}=-\left(\frac{\partial\tilde{\mu}}{\partial\tilde{T}}\right)_{\tilde{\mathcal{Q}}}=2\pi\mathcal{E}.
\end{equation}
By comparing with (\ref{eq:holozerotlimitthermo}), we see that the AdS$_2$ solution of the 2-dimensional theory (\ref{eq:2dgravityaction}) correctly captures the $T=0$ limit of the thermodynamic function $(\partial\mathcal{S}/\partial\mathcal{Q})_T$ of the full, asymptotically AdS$_4$ solution. It does not capture the small $T$ corrections. 

In fact, for the case (\ref{eq:VZaxion}) which arises from dimensional reduction of the complete solution (\ref{eq:simpleaxionsoln}), we find that the canonical free energy density of the AdS$_2$ geometry can be written 
\begin{equation}
\begin{aligned}
\label{eq:twodF}
\tilde{F}(\tilde{\mathcal{Q}},\tilde{T})&=-\tilde{\mathcal{S}}(\tilde{\mathcal{Q}})\tilde{T},\\
&=-\mathcal{S}_0(\tilde{\mathcal{Q}})\tilde{T},
\end{aligned}
\end{equation}
where $\mathcal{S}_0(\mathcal{Q})$ is (\ref{eq:fourdE0S0}), the zero temperature entropy of the full four dimensional solution. Comparing to (\ref{eq:fourdF}), we see that the linear-in-$T$ part of the free energy density is universal i.e.~it is independent of the UV geometry. This is the holographic analogue of the SYK result (\ref{funiv}).

Comparing (\ref{eq:fourdF}) and (\ref{eq:twodF}), we see that the $T$-independent part of the canonical free energy is not universal: it depends upon how the AdS$_2$ near-horizon geometry is embedded into the full solution. This results in a non-trivial ``renormalization'' of the chemical potential of the AdS$_2$ solution $\tilde{\mu}$ with respect to the chemical potential of the full solution $\mu$,
\begin{equation}
\tilde{\mu}(\mathcal{Q},T)=\mu(\mathcal{Q},T)-\mu_0(\mathcal{Q})+O(T^2)\,,
\end{equation}
where
\begin{equation}
\mu_0(\mathcal{Q})=\frac{\partial E_0}{\partial\mathcal{Q}}=\sqrt{\sqrt{m^4+12\mathcal{Q}^2}-m^2},
\end{equation}
is the $T=0$ chemical potential of the full solution, which depends upon the UV geometry. The linear-in-$T$ components of $\mu$ and $\tilde{\mu}$ agree because they are related to the universal quantity $(\partial\mathcal{S}/\partial\mathcal{Q})_T$ by the Maxwell relation (\ref{defE0}).

One result of this renormalization of $\mu$ is that the low $T$ limit of the charge susceptibility of the full solution, $K$ in equation (\ref{eq:ads4suscept}), is unrelated to the charge susceptibility of the AdS$_2$ solution of the two-dimensional action (\ref{eq:2dgravityaction}). Explicitly, the charge susceptibility of the AdS$_2$ solution is
\begin{equation}
\left(\frac{\partial\tilde{\mathcal{Q}}}{\partial\tilde{\mu}}\right)_{\tilde{T}}=-\frac{1}{2\pi \tilde{T}}\frac{Z(\phi_0)}{\tilde{L}^2}\left(\left(1-\frac{\mathcal{E}^2Z'(\phi_0)}{2\tilde{L}^2}\right)^2-\frac{Z(\phi_0)\mathcal{E}^2}{4}V''(\phi_0)-\frac{Z(\phi_0)\mathcal{E}^4}{8\tilde{L}^4}Z''(\phi_0)\right)^{-1},
\end{equation}
which diverges as $1/T$, unlike $K$. Thus, the $T=0$ charge susceptibility cannot be obtained from an effective two dimensional action for the near-horizon part of the geometry.

Finally, let us address the $O(T^2)$ terms in the free energy (\ref{eq:fourdF}), which are responsible for the $T=0$ limit of the specific heat at constant $\mathcal{Q}$ of the full solution, $\gamma$ in equation (\ref{eq:ads4suscept}). The two-dimensional theory does not have terms like this, and therefore has a vanishing specific heat. For the uncharged case, Ref.~\onlinecite{KJ16,JMDS16b,HV16} showed that the leading contribution to the specific heat can be found by including the correction to the dilaton which grows towards the AdS$_2$ boundary. In principle, the inclusion of corrections like this should also lead to a non-vanishing specific heat in the charged case, but this is beyond the scope of this paper.

%********************************************************
\subsection{Transport}
%********************************************************
\label{sec:holotransport}

The transport properties of the field theory dual to the solution (\ref{eq:simpleaxionsoln}) have been studied in great detail in \cite{Vegh:2013sk,Davison:2013jba,Blake:2013bqa,Davison:2014lua,Kim:2014bza,Donos:2014cya,Davison:2015bea,Blake:2015epa,Blake:2016sud,Withers:2016lft,Amoretti:2014zha,Amoretti:2014ola}. When translational symmetry is broken, which in this case means $m\ne0$, transport of heat and charge over the longest distances and timescales should be be governed by the equations (\ref{ti1}) and (\ref{ti2}) of diffusive hydrodynamics. For the $\mathcal{Q}=0$ case, this was checked numerically in \cite{Davison:2014lua}. Given that the susceptibility matrix is (\ref{eq:holoaxionsuscept}), the remaining quantities characterizing the transport of the system are the three elements of the dc conductivity matrix (\ref{ti3}). With this information, the diffusion matrix (\ref{ti3}) and response functions (\ref{ti2}) are fixed by the theory of diffusive hydrodynamics.

It is not unreasonable to expect a connection between the dc conductivities of the higher-dimensional SYK theory, and those of the holographic theory (\ref{eq:simpleaxionsoln}), because in the holographic case these are determined by the AdS$_2$ horizon. In general, for a given UV gravitational theory with asymptotically AdS solutions and without translational symmetry, the dc conductivities are given by properties of the gravitational solution at the horizon \cite{Blake:2013bqa,Donos:2014cya,Donos:2014yya,Donos:2015gia,Banks:2015wha}. One does not need to know how the near-horizon solutions (which may or may not be AdS$_2$) are embedded into the full solution. 

For the solution (\ref{eq:simpleaxionsoln}), the dc conductivities are
\begin{equation}
\begin{aligned}
\label{eq:holodcconds}
\sigma=1+\frac{\mu^2}{m^2},\;\;\;\;\;\;\;\;\;\;\;\;\;\;\;\;\;\;\alpha=\frac{4\pi\mathcal{Q}}{m^2},\;\;\;\;\;\;\;\;\;\;\;\;\;\;\;\;\;\;\bar{\kappa}=\frac{4\pi \mathcal{S} T}{m^2}.
\end{aligned}
\end{equation}
In the low $T$ limit, the Seebeck coefficient $S \equiv \frac{\alpha}{\sigma}$ is
\begin{equation}
\label{eq:holoseebeckreln}
\lim_{T\to 0}S=\frac{2\pi\mu\sqrt{2m^2+\mu^2}}{\sqrt{3}\left(m^2+\mu^2\right)}=\lim_{T\to 0}\left(\frac{\partial\mathcal{S}}{\partial\mathcal{Q}}\right)_T=2\pi\mathcal{E},
\end{equation}
as advertised in (\ref{seebeck}). This is a non-trivial relation between three quantities associated to the AdS$_2$ near-horizon geometry. Although the final equality can be derived from the simple two-dimensional action (\ref{eq:2dgravityaction}), this action alone is not sufficient for determining the dc conductivities, which depend upon the correlation functions of spatial currents, or equivalently upon the correlation functions of gradients of the charge and energy densities.

Due to the relation (\ref{eq:holoseebeckreln}), the low $T$ response functions of the holographic theory take the same form as those of the SYK model (\ref{pf4}). The low $T$ diffusion constants of the holographic theory are
\begin{equation}
D_1 = \frac{\sqrt{3(2 m^2 + \mu^2)}}{m^2} \quad, \quad D_2 =  \frac{\sqrt{3}}{\sqrt{2 m^2 + \mu^2}}.
\label{eq:holoD12}
\end{equation}
and the charge susceptibility $K$, specific heat $\gamma$, and $\mathcal{E}$ are given in equations (\ref{eq:ads4suscept}) and (\ref{eq:finiteTads2solution}). The functional form of the diffusion constants and thermodynamic functions are different in this holographic model than in the SYK model of Section \ref{sec:SYKd}, but the structure of the response functions is the same. 

The divergence of one of the diffusion constants in the translationally invariant case $m=0$ is a consequence of the fact that diffusive hydrodynamics is not applicable in this limit. One of the diffusive excitations is replaced by a propagating sound wave \cite{Davison:2015bea}. The diffusive mode which survives in this limit corresponds to diffusion of a certain linear combination of the charge and heat currents \cite{Lucas:2015pxa,Davison:2015taa}. When $m=0$, the applicability of the relation (\ref{eq:holoseebeckreln}) is more subtle. In this case, the dc conductivities $\alpha$ and $\sigma$ are infinite due to translational invariance. By studying the optical conductivities, one can cleanly distinguish between an infinite and finite contribution to each dc conductivity. The ratio of the infinite contributions is $\alpha_{inf}/\sigma_{inf}=\mathcal{S}/\mathcal{Q}=2\pi\mathcal{E}$ and so obeys equation (\ref{eq:holoseebeckreln}). The ratio of the finite contributions is $\alpha_f/\sigma_f=-\mu/T$. Note that the right-hand-side is given by the full chemical potential, and thus, in this case, this ratio is not a universal quantity. The $m=0$ theory is special because in this case the conductivities are not simply properties of the AdS$_2$ horizon \cite{Davison:2015taa}. In fact, when $m=0$, the ratio between the finite contributions is fixed by the UV relativistic symmetry \cite{Herzog:2009xv}.

It is simple to obtain the Wiedemann-Franz ratio $L$, defined in Eq.~(\ref{eq:WFforSYK}), which is given by
\begin{equation}
\lim_{T\rightarrow0}L=\frac{4\pi^2m^2\left(2m^2+\mu^2\right)}{3\left(m^2+\mu^2\right)^2}, \label{eq:WFforAdS}
\end{equation}
at zero temperature. Curiously, the prefactor of $4 \pi^2/3$ is the same as that in the SYK model result in Eq.~(\ref{eq:WFforSYK}). Eq.~(\ref{eq:WFforAdS}) vanishes in the translationally invariant limit $m\rightarrow0$, as expected from the general arguments of \cite{Mahajan:2013cja}. We can also define the Wiedemann-Franz-like ratio 
\begin{equation}
    L_\alpha\equiv\frac{\kappa}{\alpha T},
    \label{eq:defnmodifiedWF}
\end{equation}
where we have replaced the electrical conductivity in the usual Wiedemann-Franz ratio with the thermoelectric conductivity. For this holographic theory, the low temperature limit of $L_\alpha$ is given by a simple thermodynamic formula
\begin{equation}
\lim_{T\rightarrow0}L_\alpha=-\mathcal{Q}\frac{\partial}{\partial\mathcal{Q}}\left(\frac{\mathcal{S}_0}{\mathcal{Q}}\right)=\frac{2\pi m^2\sqrt{2m^2+\mu^2}}{\sqrt{3}\mu\left(m^2+\mu^2\right)}.
\label{eq:modifiedWFratioholo}
\end{equation}
The first equality is a consequence of both the `Kelvin formula' (\ref{seebeck}), and the relation $\bar{\kappa}/\alpha T=\mathcal{S}/\mathcal{Q}$. This latter relation is a generic property of holographic theories with homogeneous translational symmetry breaking, and is unrelated to the existence of an AdS$_2$ near-horizon geometry. It is therefore unsurprising that this property, and hence the thermodynamic relation in (\ref{eq:modifiedWFratioholo}), are not shared by the SYK models. But the thermodynamic formula for $L_\alpha$ does extend to more general holographic theories with AdS$_2$ horizons (see appendix \ref{app:moregeneralads2solutions}).

%********************************************************
\section{Conclusions}
%********************************************************

This paper has presented the thermodynamic and transport properties of two classes of solvable models of diffusive metallic states without quasiparticle excitations. Both classes of models conserve total energy and a U(1) charge, $\mathcal{Q}$, but do not conserve total momentum. The first class concerns the higher-dimensional SYK models of fermions with random $q/2$-body interactions. The second class involves a holographic mapping to gravitational theories of black branes with an AdS$_2$ near-horizon geometry. We found that these classes shared a number of common properties:
\begin{itemize}
\item
The low $T$ thermodynamics is described by the free energy in Eq.~(\ref{funiv}), with the entropy $\mathcal{S} (\mathcal{Q})$ universal, and the ground state energy $E_0 (\mathcal{Q})$ non-universal. For the SYK models, universality implies dependence only on the IR scaling dimension of the fermion, and independence from possible higher-order interactions in the Hamiltonian. In the holography, universality implies independence from the geometry far from the AdS$_2$ near-horizon geometry.
\item
The thermoelectric transport is constrained by a simple expression (Eq.~(\ref{seebeck})) equating the Seebeck coefficient to the $\mathcal{Q}$-derivative of the entropy $\mathcal{S}$. This is the `Kelvin formula' proposed in Ref.~\onlinecite{Shastry} by different approximate physical arguments. In our analysis, the Kelvin formula was the consequence of an emergent $\PSL (2, \mathbb{R})$ symmetry shared by both classes of models.
\item As has also been discussed earlier \cite{SS15}, the correlators of non-conserved local operators 
have a form (see Eq.~(\ref{Gsigma2})) constrained by conformal invariance, and characterized by a spectral asymmetry parameter, $\mathcal{E}$, which is defined by Eq.~(\ref{defE0}); see also Appendix~\ref{app:conformal}. In the holographic context, $\mathcal{E}$ also has the interpretation as the strength of an electric field in AdS$_2$.
\item Both classes of models \cite{SSDS14,kitaev2015talk,JMDS16} saturate the bound \cite{maldacena2015abound} on the Lyapunov rate which characterizes the growth of quantum chaos, $\lambda_L = 2 \pi T$.
\item For the SYK models, the butterfly velocity, $v_B$, was found to be universally related to the thermal diffusivity, $D_2$ by Eq.~(\ref{tt8}), as in Ref.~\onlinecite{GQS16}. On the other hand, the SYK models do not display a universal relation between $v_B$ and the charge diffusivity, $D_1$. So the universal connection between and chaos and transport is restricted to energy transport, as was also found in the study of a critical Fermi surface \cite{Patel:2016wdy}. Chaos is naturally connected to energy fluctuations, because the local energy determines the rate of change of the phase of the quantum state, and phase decoherence is responsible for chaos. This physical argument finds a direct realization in the computation on the SYK model. In the holographic axion model with $\mu=0$, the relationship between $D_{1,2}$ and $v_B$ was investigated in Ref.~\onlinecite{Blake:2016sud}, and $D_2$ was found to obey Eq.~(\ref{tt8}). 
\end{itemize}

%********************************************************
\section*{Acknowledgements}
%********************************************************

We thank M.~Blake, B.~Gout\'{e}raux, S.~Hartnoll, C.~P.~Herzog, A.~Kitaev, J.~Maldacena, J.~Mravlje, O.~Parcollet and D.~Stanford for valuable discussions. KJ thanks C.~P.~Herzog for prior collaboration which led to Appendix~\ref{app:conformal}. WF thanks Quan Zhou and Yi-Zhuang You for helpful discussions on the numerics. This research was supported by the NSF under Grant DMR-1360789 and the MURI grant W911NF-14-1-0003 from ARO. Research at Perimeter Institute is supported by the Government of Canada through Industry Canada and by the Province of Ontario through the Ministry of Research and Innovation. The work of RD is supported by the Gordon and Betty Moore Foundation EPiQS Initiative through Grant GBMF\#4306. 
The work of YG is supported by
 a Stanford Graduate Fellowship.
SS also acknowledges support from Cenovus Energy at Perimeter Institute. 

%********************************************************
\appendix
%********************************************************

%********************************************************
\section{Saddle point solution of the SYK model}
\label{app:saddle}
%********************************************************

We follow the condensed matter notation for Green's functions in which 
\beq
G (\tau) = - \langle T_\tau ( f(\tau) f^\dagger (0) ) \rangle.
\eeq
It is useful to make ansatzes for the retarded Green's functions in the complex frequency plane, because then
the constraints from the positivity of the spectral weight are clear.
At the Matsubara frequencies, the Green's function is defined by
\beq
G(i \omega_n) = \int_0^{1/T} d \tau e^{i \omega_n \tau} G(\tau).
\eeq
So the bare Green's function is
\beq
G_{0} (i \omega_n) = \frac{1}{i \omega_n + \mu}.
\eeq 
The Green's functions are continued to all complex frequencies $z$ via
the spectral representation
\beq
G(z) = \int_{-\infty}^{\infty} \frac{d \Omega}{\pi} \frac{\rho (\Omega)}{z - \Omega}. \label{spec}
\eeq
For fermions, the spectral density obeys
\beq
\rho (\Omega ) > 0, \label{fcons}
\eeq
for all real $\Omega$ and $T$.
The retarded Green's function is $G^R (\omega) = G(\omega + i \eta)$ with $\eta$ a positive infinitesimal, while the advanced Green's function is $G^A (\omega) = G(\omega - i \eta)$.
It is also useful to tabulate the inverse Fourier transforms at $T=0$
\bea 
G(\tau) = \left\{
\begin{array}{ccc}
\displaystyle - \int_0^\infty \frac{d \Omega}{\pi} \rho (\Omega) e^{- \Omega \tau} &,& \mbox{for $\tau > 0$ and $T=0$} \\[1em] 
\displaystyle\int_0^\infty \frac{d \Omega}{\pi} \rho (-\Omega) e^{ \Omega \tau} &,& \mbox{for $\tau < 0$ and $T=0$}
\end{array} \right. . \label{ift}
\eea

Using (\ref{ift}) we obtain in $\tau$ space
\bea 
G (\tau) = \left\{
\begin{array}{ccc}
\displaystyle - \frac{C \Gamma (2 \Delta) \sin (\pi \Delta + \theta)}{\pi |\tau|^{2 \Delta}} &,& \mbox{for $\tau > 0$ and $T=0$} \\[1em] 
\displaystyle \frac{C \Gamma (2 \Delta) \sin (\pi \Delta - \theta)}{\pi |\tau|^{2 \Delta}} &,& \mbox{for $\tau < 0$ and $T=0$}
\end{array} \right. . \label{Gftau}
\eea

We also use the spectral representations for the self energies
\beq
\Sigma (z) = \int_{-\infty}^{\infty} \frac{d \Omega}{\pi} \frac{\sigma (\Omega)}{z - \Omega}. \label{sspec}
\eeq

Using (\ref{Gftau}) and (\ref{e2}) we obtain
\bea 
\sigma (\Omega) = \left\{
\begin{array}{ccc}
\displaystyle \frac{\pi J^2 q}{\Gamma (2(q-1) \Delta)} \left[\frac{C \Gamma (2 \Delta)}{\pi} \right]^{q-1} 
\left[\sin(\pi \Delta + \theta) \right]^{q/2} \left[\sin(\pi \Delta - \theta) \right]^{q/2-1}
|\Omega|^{2 \Delta (q-1) - 1} &,& \mbox{for $\Omega > 0$} \\[1em] 
\displaystyle \frac{\pi J^2 q}{\Gamma (2(q-1) \Delta)} \left[\frac{C \Gamma (2 \Delta)}{\pi} \right]^{q-1} 
\left[\sin(\pi \Delta + \theta) \right]^{q/2-1} \left[\sin(\pi \Delta - \theta) \right]^{q/2}
|\Omega|^{2 \Delta (q-1) - 1} &,& \mbox{for $\Omega < 0$}
\end{array} \right. . \label{sbtau}
\eea
Now from (\ref{e3}) we have in the IR limit
\beq
\Sigma (z) - \mu = -\frac{1}{C} e^{i (\pi \Delta + \theta)} z^{(1 - 2 \Delta)} . \label{Sbtau}
\eeq
 So comparing (\ref{sbtau}) and (\ref{Sbtau}), we have the 
 solutions in Eqs.~(\ref{Deltaq}) and (\ref{Cval}), provided $\Sigma (z = 0) = \mu$ at $T=0$ \cite{SY92}.

%********************************************************
\section{Constraints from conformal invariance at nonzero $\mu$}
\label{app:conformal}
%********************************************************

In Eq.~\eqref{ansatz}, we made an ansatz for the form of the low-frequency two-point function of the SYK fermion at nonzero chemical potential. In this Appendix we show that this ansatz follows from the assumption of a low-energy conformal invariance, which unlike in higher-dimensional quantum field theory, can arise in zero or one spatial dimensions.

To see this, it is helpful to imagine coupling a $(0+1)$-dimensional quantum theory with a U(1) global symmetry to an external metric and external U(1) gauge field. Suppose the theory is on the Euclidean line and that the external gauge field corresponds to a chemical potential $\tilde{\mu}$, $A_{\tau} = -i \tilde{\mu}$. When $\tilde{\mu}=0$, this background is invariant under global conformal transformations,
\begin{equation}
\label{E:SL2R}
\tau \to \frac{a \tau+b}{c\tau +d}\,,\qquad a d - b c =1\,.
\end{equation}
The group of global conformal transformations is isomorphic to $\PSL(2,\mathbb{R})$. When $\tilde{\mu}\neq 0$, the coordinate transformation~\eqref{E:SL2R} does not leave the external gauge field invariant, but the combination of~\eqref{E:SL2R} and a gauge transformation %\YG{\bf [YG: How does this acts on the green's function?]}
\begin{equation}
\label{E:SL2Rpt2}
\Lambda = i\tilde{\mu}\left( \frac{a \tau+b}{c\tau+d}- \tau\right)\,,
\end{equation}
does, under the convention that $A_{\tau}$ transforms under gauge trnasformations as $A_{\tau} \to A_{\tau} + \partial_{\tau}\Lambda$. So a $\PSL(2,\mathbb{R})$ global conformal symmetry may be maintained even at nonzero chemical potential.

This global conformal group is generated by a time translation $H$, dilatation $D$, and a special conformal transformation $K$. As we usually do, let a primary operator be one which is annihilated by $K$. Primary operators are labeled by their dimension $\Delta$ and U(1) charge, which we henceforth take to be unity. Using that a conformal transformation is the combination of a coordinate transformation~\eqref{E:SL2R} and gauge transformation~\eqref{E:SL2Rpt2}, the action of an infinitesimal conformal transformation $\delta\tau= f$ and an independent, infinitesimal gauge transformation $\lambda$ on a primary operator $\mathcal{O}$ is given by
\begin{equation}
\delta_f \mathcal{O} = -f (\partial_{\tau} - i A_{\tau}) \mathcal{O} - \Delta (\partial_{\tau}f) \mathcal{O} + i \lambda \mathcal{O}\,. 
\end{equation}
%\KJ{I think this addresses Yingfei's question above (B2).}
Observe that, after Fourier transforming $\tau$ to a Euclidean frequency $\omega_E$, the action of the conformal transformations at $\tilde{\mu}\neq 0$ is the same as at $\tilde{\mu}=0$, but with the substitution $\omega_E \to \omega_E - i \tilde{\mu}$. Thus, up to a change in the normalization, the frequency-space two-point function of $\mathcal{O}$ at nonzero $\tilde{\mu}$ is just given by the two-point function at $\tilde{\mu}=0$ but with this same replacement.

At zero temperature this just gives that the two-point function of $\mathcal{O}$ is proportional to $(\omega_E - i \tilde{\mu})^{2\Delta-1}$, which recovers the ansatz~\eqref{ansatz} with $z^{-1} = \omega_E - i \tilde{\mu}$. At nonzero temperature $T = 1/\beta$, a similar, but lengthier argument shows that the two-point function of $\mathcal{O}$ is given by
\begin{equation}
G(i\omega_n) = - \frac{i C e^{-i\theta}}{\beta^{2\Delta - 1}}\frac{\Gamma\left( \Delta - \frac{\beta}{2\pi}(\omega_n -i\tilde{\mu})\right)}{\Gamma \left(1 - \Delta -\frac{\beta}{2\pi} (\omega_n - i \tilde{\mu})) \right)}\,,
\end{equation}
where $c$ is a constant and $\theta$ is the same phase appearing in~\eqref{ansatz}. This phase is related to $\tilde{\mu}$ and $T$ in the following way. Fourier transforming back to Euclidean time $\tau$, the Euclidean Green's function must be a real function of $\tau$. Using that the Matsubara frequencies for fermions are $\omega_n = 2\pi ( n+\frac{1}{2})$, we find after some algebra that for fermionic $\mathcal{O}$, $\theta$ and $\mu$ are related as
\begin{equation}
e^{-\beta \tilde{\mu}} = \frac{\sin(\pi \Delta + \theta)}{\sin(\pi \Delta - \theta)}\,.
\end{equation}
This coincides with the expression~\eqref{theta} relating $\theta$ and $\mathcal{E}$, provided that we identify
\begin{equation}
\label{E:fromEtotildemu}
2\pi\mathcal{E} = -\frac{\tilde{\mu}}{T}\,.
\end{equation}
For now, take this expression to define the spectral asymmetry $\mathcal{E}$. We conclude this Appendix by arguing that this definition of $\mathcal{E}$ also satisfies~\eqref{dmdt}.

Scale invariance implies that the canonical ensemble free energy has the form 
\begin{equation}
F(\mathcal{Q},T) = - T \mathcal{S}(\mathcal{Q})\,,
\end{equation}
where $\mathcal{S}(\mathcal{Q})$ is the zero-temperature entropy. The chemical potential $\tilde{\mu}$ is then
\begin{equation}
\tilde{\mu}(\mathcal{Q},T) = - T \frac{d\mathcal{S}}{d\mathcal{Q}}\,.
\end{equation}
Eq.~\eqref{E:fromEtotildemu} trivially implies
\begin{equation}
2\pi\mathcal{E} =- \lim_{T\to 0} \frac{\partial^2 F}{\partial T \partial\mathcal{Q}} =- \lim_{T \to 0} \left( \frac{\partial \tilde{\mu}}{\partial T}\right)_{\mathcal{Q}}\,,
\end{equation}
which is what we wanted to show.

%\YG{\bf [YG: I want to understand this part, so this $\tilde{\mu}$ is the linear T part of the $\mu$ defined by Subir? And the argument here shows that the conformal part contributes the linear T part of the full $\mu (T)$?]} \KJ{Yes, I think that's right.}

%********************************************************
\section{Large $q$ expansion of the SYK model}
\label{app:largeq}
%********************************************************

Section~\ref{sec:thermosyk} obtained exact results for the universal parts of the thermodynamic 
observables. However, no explicit results for the non-universal parts dependent upon $J$.
In this appendix we will present the large $q$ expansion of the Hamiltonian in Eq.~(\ref{h}):
the results contain both the universal and non-universal parts.

We begin by recalling the universal results of Section~\ref{sec:thermosyk} in the limit of small $\Delta = 1/q$.
At low $T$, the thermodynamics contains 3 universal quantities which do not undergo any UV renormalization:
they are the density, $\mathcal{Q}$, the entropy $\mathcal{S}$, and the `electric field' $\mathcal{E}$.
All 3 quantities can be expressed in terms of universal expressions of each other.
First, we treat $\mathcal{Q}$ as the independent quantity.
Then, the $T \rightarrow 0$  limit of the entropy is from (\ref{Fres}), (\ref{t6}), (\ref{t7}), 
\beq
\mathcal{S} (\mathcal{Q}) = \mathcal{Q} \ln \left( \frac{1 - 2 \mathcal{Q}}{1 + 2 \mathcal{Q}} \right) + \frac{1}{2} \ln \left( \frac{4}{1 - 4 \mathcal{Q}^2} \right) - \frac{\pi^2}{2} (1- 4 \mathcal{Q}^2) \Delta^2  + \mathcal{O} (\Delta^3). \label{q7}
\eeq
By taking a $\mathcal{Q}$ derivative, we have immediately
\beq
\mathcal{E}(\mathcal{Q}) = \frac{1}{2 \pi} \ln \left(\frac{1 - 2 \mathcal{Q}}{1 +2 \mathcal{Q}}\right) + 2 \pi \mathcal{Q} \Delta^2 + \mathcal{O} (\Delta^3). \label{q4i}
\eeq

Next, we take $\mathcal{E}$ as the independent variable. Then the inverse function (\ref{q4i}) is 
\beq
\mathcal{Q} (\mathcal{E}) = - \frac{1}{2} \tanh ( \pi \mathcal{E}) - \frac{\pi^2 \sinh(\pi \mathcal{E})}{2 \cosh^3 (\pi \mathcal{E})} \Delta^2  + \mathcal{O} (\Delta^3). \label{q4}
\eeq
The entropy is given by (\ref{t7}), $\mathcal{S}(\mathcal{E}) = \mathcal{G}(\mathcal{E}) + 2 \pi \mathcal{E} \mathcal{Q}(\mathcal{E})$, where
\beq
\mathcal{G} (\mathcal{E}) =\ln (2 \cosh(\pi \mathcal{E}))  - \frac{\pi^2}{2 \cosh^2 (\pi \mathcal{E})} \Delta^2 + \mathcal{O} (\Delta^3). \label{q5}
\eeq

Now we turn to the explicit large $q$ expansion to the compute the thermodynamics in terms of microscopic parameters.
The expressions here depend upon the underlying $J$, and the specific form of the Hamiltonian in Eq.~(\ref{h}). We will verify that they are compatible with the universal results presented above.

The large $q$ expansion was presented by Ref.~\onlinecite{JMDS16} at $\mu=0$,
and we follow their analysis here. At $q=\infty$ they showed that the Green's function was that of a
dispersionless free fermion. So, we write
\beq
G_s (\tau) = G_0 (\tau) \left[ 1 + \frac{1}{q} G_1 (\tau) \right] 
\eeq
where the dispersionless free fermion Green's function is
\beq
G_0 (\tau) = \left\{ \begin{array}{c} - e^{\mu \tau} (e^{\mu/T} + 1)^{-1}, \quad, \quad 0 < \tau < 1/T\\
  e^{\mu \tau} (e^{-\mu/T} + 1)^{-1}, \quad, \quad -1/T < \tau < 0 \end{array} \right. .
\eeq
Then from (\ref{e2}) we have the self energy
\beq
\Sigma_s (\tau) = - \frac{q J^2 e^{\mu \tau}}{(e^{\mu/T} + 1)(2 + 2 \cosh(\mu/T))^{q/2-1}}
\left[ 1 + \frac{1}{q} G_1 (\tau) \right]^{q/2}\left[ 1 + \frac{1}{q} G_1 (-\tau) \right]^{q/2-1}
\eeq
Now we define
\beq
\mathcal{J}^2 = \frac{q^2 J^2 }{2(2 + 2 \cosh(\mu/T))^{q/2-1}}.
\eeq
The large $q$ can only be taken if we adjust the bare $J$ so that $\mathcal{J}$ is 
$q$ independent. To the order we shall work, it is legitimate to use the 
$\mathcal{O} (\Delta^0)$ result above, in which case we will find
\beq
\mathcal{J}^2 = \frac{q^2 J^2 }{2(2 + 2 \cosh(2 \pi \mathcal{E}))^{q/2-1}}.
\eeq
As $\mathcal{E}$ is only a function of $\mathcal{Q}$, we find that $\mathcal{J}$ remains finite
as $T \rightarrow 0$.
Then, in the large $q$ limit
\beq
\Sigma_s (\tau) = - \frac{2\mathcal{J}^2}{q}G_0 (\tau) \exp \left( \frac{1}{2}(G_1 (\tau) + G_1 (-\tau)) \right)
\eeq
In this form, the explicit $\mu$ dependence has disappeared. Ref.~\onlinecite{JMDS16} obtained a differential equation
for $G_1$ at $\mu=0$, and so this applies also here; the solution is
\beq
G_1 (\tau) = \ln \left( \frac{\cos^2 (\pi v/2)}{\cos^2 (\pi v ( T |\tau| -1/2))} \right),
\eeq
where $v$ is obtained by the solution of
\beq
\frac{\pi v}{\cos(\pi v/2)} = \frac{\mathcal{J}}{T}. \label{q2}
\eeq
Assuming a fixed $\mathcal{J}$, the low $T$ expansion of $v$ is
\beq
v = 1 - \frac{2 T}{\mathcal{J}} + \frac{4 T^2}{\mathcal{J}^2} + \ldots. \label{vT}
\eeq

To compute the grand potential, $\Omega$, we use the effective action 
\begin{align}
\begin{split}
\label{action}
S[G, \Sigma] &= -\mbox{Tr} \ln \left[ \delta(\tau-\tau') \left(-\frac{\partial}{\partial \tau} + \mu \right) - \Sigma (\tau, \tau') \right] + \frac{\mu}{2T}
\\
&- \int_0^{1/T} d\tau d \tau' \left[ \Sigma (\tau, \tau') G (\tau', \tau) + (-1)^{q/2} J^2 \left[G (\tau,\tau')\right]^{q/2} \left[ G (\tau',\tau) \right]^{q/2} \right]. 
\end{split}
\end{align}

The $G_s (\tau - \tau') $ and $\Sigma_s (\tau - \tau') $ above are the solutions to the saddle-point equations of $S$.
It is simpler to evaluate $d \Omega/d \mathcal{J}$ because
only the last term contributes
\begin{align}
\begin{split}
\mathcal{J} \frac{d \Omega}{d \mathcal{J}} &= - \frac{4 \mathcal{J}^2}{q^2 (2 + 2 \cosh(\mu/T))} \int_0^{1/T} d \tau \exp(G_1 (\tau))
\\
&=  - \frac{4 \mathcal{J}^2 \sin(\pi v)}{q^2 \pi T v (2 + 2 \cosh(\mu/T))},
\end{split}
\end{align}
which implies
\beq
\frac{d \Omega}{dv} = -  \frac{8 \pi T}{q^2 (2 + 2 \cosh(\mu/T))} \tan \left( \frac{\pi v}{2} \right) \left[ 1 + 
\frac{\pi v}{2} \tan \left( \frac{\pi v}{2} \right) \right]
\eeq
Integrating over $v$, we obtain
the grand potential as a function of the bare $\mu$ and $T$
\beq
\Omega (\mu, T) = - T \ln (2 \cosh(\mu/(2T))) - \frac{2 \pi v T}{\cosh^2(\mu/(2T))}  \left[ \tan \left(\frac{\pi v}{2}\right) - \frac{\pi v}{4} \right]\Delta^2
+ \mathcal{O} (\Delta^3). \label{q1}
\eeq
This is the main result of the large $q$ expansion. 

Now we can use thermodynamic relations to determine both universal and non-universal observables.
From the grand potential in (\ref{q1}), we have the density
\bea
\mathcal{Q} &=& \frac{1}{2} \tanh(\mu/(2 T)) - \frac{2 \pi v \sinh(\mu/(2T))}{\cosh^3(\mu/(2T))}  \left[ \tan \left(\frac{\pi v}{2}\right) - \frac{\pi v}{4} \right]\Delta^2   + 
\mathcal{O} (\Delta^3). \label{q3}
\eea
Combining (\ref{q1}) and (\ref{q3}), we can obtain the free energy in the canonical ensemble
\begin{align}
\begin{split}
\label{q3a}
F(\mathcal{Q}, T) &= \Omega (\mu, T) + \mu \mathcal{Q} 
\\
&= -T\left[ \frac{1}{2} \ln \left( \frac{4}{1 - 4 \mathcal{Q}^2} \right) + \mathcal{Q} \ln \left( \frac{1 - 2 \mathcal{Q}}{1 + 2 \mathcal{Q}} \right) \right]
\\
&~~~~~~~~~-2 \pi T (1- 4 \mathcal{Q}^2) \left[ \tan \left(\frac{\pi v}{2}\right) - \frac{\pi v}{4} \right]\Delta^2 + \mathcal{O} (\Delta^3). 
\end{split}
\end{align}
It is more convenient to work with the canonical $F(\mathcal{Q}, T)$, rather than the grand canonical $\Omega (\mu, T)$, because $\mathcal{Q}$
is universal, while $\mu$ is not.
We will use (\ref{q3a}) to verify the universal expressions in Section~\ref{sec:univ}, and also to obtain new non-universal results. 

First, in the fixed $\mathcal{Q}$ ensemble, we can compute the chemical potential $\mu (\mathcal{Q}, T)$ needed to keep $\mathcal{Q}$ fixed.
We find
\begin{align}
\begin{split}
\label{q10}
\mu (\mathcal{Q}, T) &= \left( \frac{\partial F}{ \partial \mathcal{Q}} \right)_T
\\
&= - T \ln \left( \frac{1 - 2 \mathcal{Q}}{1 + 2 \mathcal{Q}} \right) + 16 \pi T \mathcal{Q} v \left[ \tan \left(\frac{\pi v}{2}\right) - \frac{\pi v}{4} \right] \Delta^2 + \mathcal{O} (\Delta^3)
\\
&= \mu_0  -2 \pi \mathcal{E} (\mathcal{Q}) T + \mathcal{O} (T^2)
\end{split}
\end{align}
In the last line, we have taken the low $T$ limit at fixed $\mathcal{J}$ 
using (\ref{vT}), and we find precisely the expression (\ref{mu1}), with 
the universal function $\mathcal{E}(\mathcal{Q})$ given by (\ref{q4i}), and the non-universal bare chemical potential
\beq
\mu_0  = 16 \mathcal{J} \mathcal{Q} \Delta^2 + \mathcal{O} (\Delta^3). \label{q10a}
\eeq
Note that there is no $\mathcal{O} (\Delta^0)$ term in $\mu_0$: this has consequences for the compressibility.
From (\ref{q10}) we can obtain the inverse compressibility, $1/K$, by taking a derivative w.r.t. $\mathcal{Q}$; at low $T$ we have
\beq
\frac{1}{K} = \left(\frac{\partial \mu}{\partial \mathcal{Q}} \right)_T = \frac{4 T}{1 - 4 \mathcal{Q}^2} + (16 \mathcal{J} - 4 \pi^2 T) \Delta^2 + \mathcal{O} (\Delta^3) \label{q11}
\eeq
So we now see that if take the limit $\Delta \rightarrow 0$ first, then the compressibility diverges as $K \sim 1/T $ in the low $T$ limit
at fixed $\mathcal{J}$.
On the other hand, if we take the $T \rightarrow 0$ at non-zero $\Delta$, then $K$ remains finite at $K = q^2/(16 \mathcal{J})$,
as needed for the consistency of the analysis in Section~\ref{sec:SYKf1}. Note that the  large $q$ expansion holds for $1/K$,
and not for $K$, it is the expansion for $1/K$ which establishes the finiteness of $K$ as $T \rightarrow 0$.

We can also obtain the non-universal ground state energy 
\beq
E_0  = F(\mathcal{Q}, T \rightarrow 0) =  -2 \mathcal{J}(1-4 \mathcal{Q}^2) \Delta^2 + \mathcal{O} (\Delta^3). \label{q10b}
\eeq
This is compatible with (\ref{q10a}) and (\ref{t1}). 

Finally, we can compute the entropy, and perform its low $T$ expansion at fixed $\mathcal{J}$; we find
\begin{align}
\begin{split}
\mathcal{S} (\mathcal{Q}, T) &= -  \left( \frac{\partial F}{ \partial T} \right)_\mathcal{Q}
\\
&= \mathcal{S} (\mathcal{Q}) + \gamma \,T + \mathcal{O}(T^2), 
\end{split}
\end{align}
where the universal function $\mathcal{S}(\mathcal{Q})$ agrees with (\ref{q7}), and the non-universal linear-in-$T$ coefficient of the  
specific heat at fixed $\mathcal{Q}$ is
\beq
\gamma = \frac{2 \pi^2 (1 - 4 \mathcal{Q}^2)}{\mathcal{J}} \Delta^2 + \mathcal{O} (\Delta^3) \label{q12}
\eeq
Again, note that there is no $\mathcal{O} (\Delta^0)$ term in $\gamma$. 

%*************************************************
\section{Luttinger-Ward analysis}
\label{app:georges}
%*************************************************
The appendix will generalize the Luttinger-Ward analysis in 
Appendix A of Ref.~\onlinecite{GPS01} (hereafter referred to as GPS)
from $q=4$ to general $q$.
The Luttinger-Ward (LW) functional for general $q$ reads:
\beq
\Phi[G]\,=\,- J^2\,(-1)^{q/2}\,  \int d\tau G(\tau)^{q/2} G(-\tau)^{q/2}
\eeq
such that:
\beq
\Sigma(\tau) = \frac{\delta\Phi}{\delta G(-\tau)} = -(-1)^{q/2} q J^2\,G(\tau)^{q/2} G(-\tau)^{q/2-1}
\label{eq:eqsigma}
\eeq
in accordance with Eq.~(\ref{e2}). 

The low frequency Green's function ansatz in Eq.~(\ref{ansatz}) 
has a prefactor $C$ given in Eq.~(\ref{Cval}). Here, we write the prefactor as
\beq
C(q,\theta)^q\,=\,K(q)\, \left[s_{+}s_{-}\right]^{(2-q)/2}
\label{eq:cq}
\eeq
in which $K(q)$ only depends on $q$, and we will use the notation $s_{\pm}\equiv \sin(\pi/q\pm\theta)$. We also note that 
Eq.~(\ref{ansatz}) implies the following low-frequency behaviour of the spectral function:
\beq
A(\omega)\equiv - \frac{1}{\pi} \mathrm{Im}G(\omega+i0^+) = \frac{C}{\pi} \frac{s_{\pm}}{|\omega|^{1-2/q}}
\label{eq:A}
\eeq
in which the $+$ ($-$) sign applies to positive (negative) frequencies respectively. Hence, the spectral asymmetry is given by: 
$A(\omega)/A(-\omega)=s_{+}/s_{-}$. 

We proceed along the lines of 
Appendix A of GPS. Eq. (A4) is unchanged and reads:
\beq
\mathcal{Q}\,=\,-\frac{\theta}{\pi} - i \int \frac{d\omega}{2\pi}\, G^F(\omega)\partial_\omega \Sigma^F(\omega)e^{i\omega 0^+},
\eeq
where the superscript $F$ indicates Feynman Green's functions at $T=0$ and real frequency.
We will actually not perform a fully explicit calculation of the integral on the r.h.s (`anomalous' term) but instead obtain its value 
from a simple argument. This argument is the one on page 14 (bottom of first column) of GPS, and it turns out that it can be generalized to arbitrary $q$.

Imagine one makes an explicit calculation of the anomalous term, along the lines of Appendix A of GPS. Then, one would have 
a sum of terms which all involve a product of $q$ spectral functions because the LW functional is a polynomial of degree $q$ in $G$. 
The spectral functions can either be for negative or positive 
frequency (see Eq. (A.11) in GPS) and hence at the end of the computation, using the low-frequency form (\ref{eq:A}) we get a sum of terms: 
\beq
C^q\,\sum_{n=0}^{q/2} c_n\, \left[s_{+}^{q/2+n} s_{-}^{q/2-n}-s_{+}^{q/2-n} s_{-}^{q/2+n} \right]
\eeq
We have used the fact that this must be an odd function of $\theta$ (hence the antisymmetry) and have assumed that the anomalous 
term only depends on the IR properties (this is the part which needs a detailed proof by regularisation as in GPS). The coefficients 
$c_n$ depend a priori on $q$  but not on $\theta$ because all $\theta$-dependence is contained in $C(q,\theta)$ and $s_{\pm}$. 

Let us examine these terms. The $n=0$ one vanishes by symmetry. The $n=1$ term yields:
\beq
c_1\, C^q\, \left[s_{+}^{q/2+1} s_{-}^{q/2-1}-s_{+}^{q/2-1} s_{-}^{q/2+1} \right] = c_1 C^q (s_{+}s_{-})^{q/2-1} (s_{+}^2-s_{-}^2)
\eeq
Using Eq.~(\ref{eq:cq}), this simplifies to:
\beq
c_1 K(q) (s_{+}^2-s_{-}^2) = c_1 K(q) \sin(\frac{2\pi}{q})\,\sin 2\theta 
\label{eq:term1}
\eeq
The important point here is that the only $\theta$-dependence is in the $\sin 2\theta$ term. 

Let us now consider the terms with $n>1$. It is easily seen that {\it all} these terms involve a combination of $s_{\pm}$ which 
has a {\it divergence} at either $\theta=\pi/q$ or $-\pi/q$, the reason being that the factor 
$\left[s_{+}s_{-}\right]^{(2-q)/2}$ in the prefactor $C^q$ no longer cancels (note that $2-q < 0$). These terms are not admissible because at $\theta=\pm \pi/q$, 
the fermion occupation number should either vanish or go to unity, and cannot diverge. Hence, these terms should not appear and all $c_n$'s 
with $n>1$ should be zero. 
For $q=4$, the only such term is $C^4(s_{+}^4-s_{-}^4)$, which we eliminated 
for the same reason in the heuristic argument of Appendix A of GPS. It extends here to all $n=2,\cdots, q/2$.

As a result, this argument shows that a full calculation of the anomaly will yield (with a simple redefinition of $c_1$):
\beq
\mathcal{Q}\,=\,-\frac{\theta}{\pi} - \tilde{c}_1(q) \sin 2\theta
\eeq
 Fixing the constant is straightforward: we note that for $\theta=+\pi/q$ the negative-frequency spectral function vanishes and thus we should 
 get the smallest fermion number ($\mathcal{Q}=-1/2$). Hence
 \beq
 -\frac{1}{2} = -\frac{1}{q} - \tilde{c}_1 \sin \frac{2\pi}{q},
 \eeq
 and we finally obtain Eq.~(\ref{Qtheta}).

%*******************************************************
\section{Numerical solution of the SYK model}
\label{app:numerics}
%*******************************************************

This appendix describes our numerical solution of Eqs.~(\ref{e2}) and (\ref{e3}) for the case $q=4$.

We worked in the frequency domain by writing  Eq.~(\ref{e2}) as a convolution
\beq
\Sigma(i\omega_n)=-\frac{J^2}{\beta^2}\sum_{\omega_n=\omega_1+\omega_2-\omega_3}G(i\omega_1)G(i\omega_2)G(i\omega_3)
\eeq
We used the function package {\tt conv\_fft2} in Matlab to perform the convolution. We restricted the frequency argument in $G(i\omega_n)$ to be ${2\pi}{T}(n+\frac{1}{2})$ where $-N\leqslant  n\leqslant N-1$. After the convolution, we cut off the frequency argument in $\Sigma(i\omega_n)$ to be within the same regime. Finally, we updated the Green's function in a weighted way:
\beq
G_j(i\omega_n)=(1-\alpha)G_{j}(i\omega_n)+\alpha\frac{1}{i\omega_n+\mu-\Sigma_{j-1}(i\omega_n)}
\eeq
where we choose the weight $\alpha=0.2$, and $j$ denotes the iteration step. 

We also used a second numerical approach in which we directly evaluate Eqs.~(\ref{e2}) and (\ref{e3}) in frequency space and time space separately, and then use fast Fourier transforms (FFT) between them. But there is a subtlety: when considering the transformation from $\tau$ space to $\omega_n$ space, we are doing a discrete sum to represent the numerical integral. For a sensible discrete sum, we do not want the exponential phase to vary too much between the two adjacent discrete points. So we want $\omega_n(\tau_j-\tau_{j-1})\ll 1$. With $N_\tau$ and $N_\omega$ the number of points of $\tau$ and $\omega$, we need ${N_{\omega}}/{N_{\tau}}\ll1$. We found $N_{\omega}=2^{18}, N_{\tau}=2^{20}$ gave accurate results.

\begin{figure}
\center
\includegraphics[width=4in]{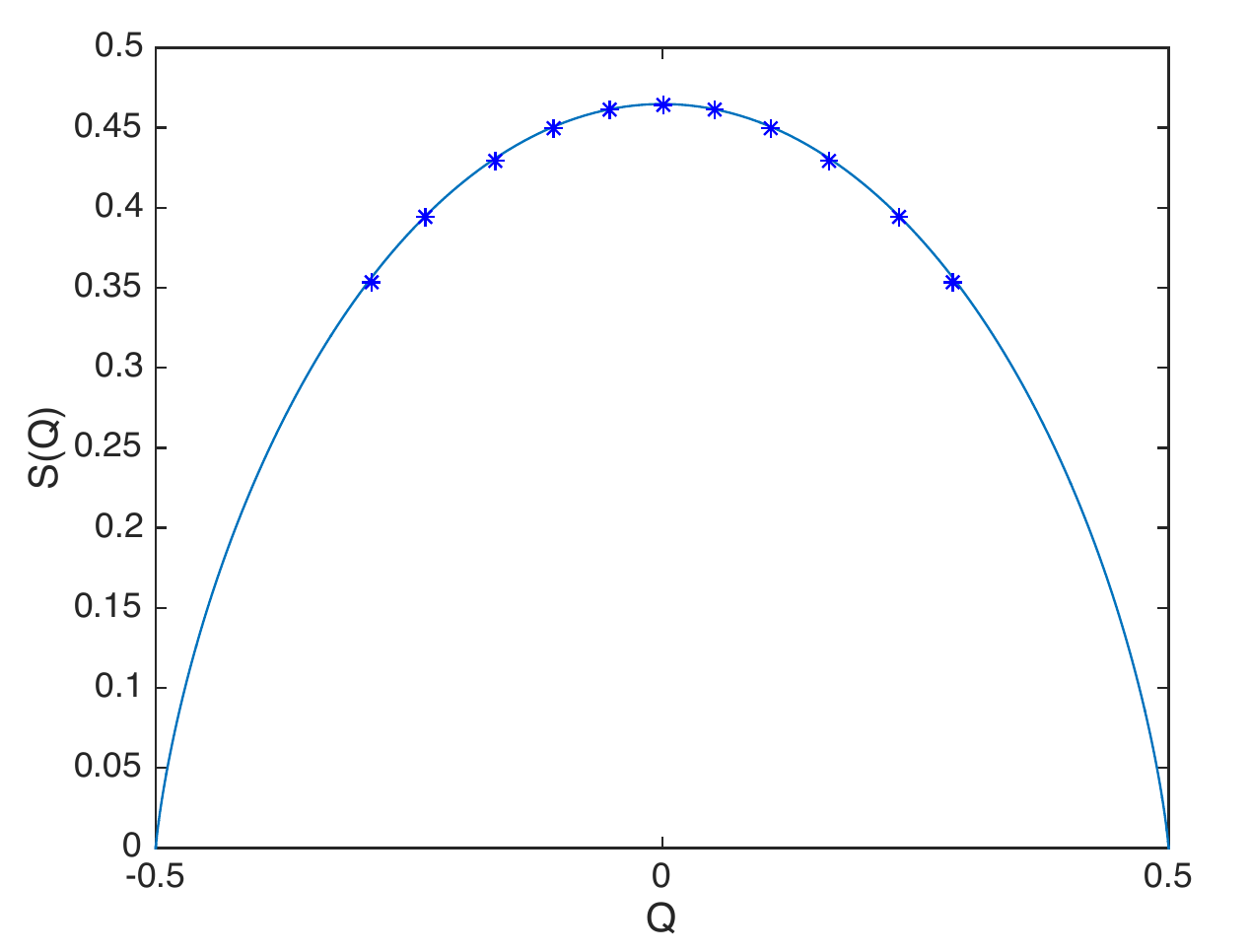}
\caption{The entropy $\mathcal{S}(\mathcal{Q})$ obtained from the exact results \cite{GPS01} in Section~\ref{sec:entropy} (full line), and by the numerical
solutions (stars).}
\label{numS(Q)}
\end{figure}
From the numerical solution for Green's function and self energy, we obtain the grand potential by evaluating Eq.~(\ref{action}). In practice, we want to subtract the grand potential of a free theory and then add it back to obtain a convergent sum over frequencies. So we wrote the first term in Eq.~(\ref{action}) as
\beq
T\sum_n\log{\left[G(i\omega_n)/G_0(i\omega_n)\right]}+T\log{\left[1+e^{\mu/T}\right]}.
\eeq
By the equations of motion, the second integral can be written as
\beq
-\frac{3}{4\beta}\sum_n \Sigma(i\omega_n)G(i\omega_n)
\eeq
Then we put the solution into these two terms and obtained the grand potential $\Omega(\mu,T)$. The density, $\mathcal{Q}$, the compressibility, $K$, and the entropy, $\mathcal{S}$ were then obtained from  suitable thermodynamic derivatives\footnote{$\mathcal{Q}$ can also be obtained from $G(\tau=0^-)$, we have checked that it is consistent with the derivative method}. 
Our numerical results for $\mathcal{S}(\mathcal{Q})$, obtained by both methods are shown in Fig.~\ref{numS(Q)}, they are in excellent agreement with the exact analytic results \cite{GPS01}. In the frequency domain computation, we used the cutoff $N=2 \times 10^6$. The points in Fig.~\ref{numS(Q)} are at moderate values of $\mathcal{Q}$, 
and our numerics did not converge for $|\mathcal{Q}|$ near $1/2$.\footnote{At large $\mu$, we always find the free Green's function $G_0=\frac{1}{i\omega_n+\mu}$ to be solution. The reason can be understood by the self-energy obtained from the free solution
\begin{displaymath}
\Sigma_0(i\omega_n)=-\frac{J^2}{\beta^2}\sum_{\omega_n=\omega_1+\omega_2-\omega_3}G_0(i\omega_1)G_0(i\omega_2)G_0(i\omega_3)=-\frac{J^2}{i\omega_n+\mu}\frac{1}{(2\cosh{\frac{\beta\mu}{2}})^2}
\end{displaymath}
Notice the exponential suppression at low temperature. This means at any finite $\mu$, at zero temperature, the free one is always a solution. Numerically we are always at small finite temperature to represent the zero temperature result, but when $\mu$ becomes large, the exponential suppression will make the free Green's function converge well within the fixed tolerance. }

For the compressibility, numerically near $\mu=0$ and at $T=0$, we find that $K={1.04}/{J}={1.04}/({\sqrt{2}\mathcal{J}})$; With $q=4$, this is of the same order of the large $q$ result: $K={q^2}/({16\mathcal{J}})={1}/{\mathcal{J}}$. 

%********************************************************
\section{Normal mode analysis of the SYK model}
\label{app:normalmode}
%********************************************************

This appendix will generalize the analysis of Maldacena and Stanford \cite{JMDS16}, and describe the structure of the effective action for fluctuations directly from the action in Eq.~(\ref{action}). We will work here in an angular variable
\beq
\varphi = 2 \pi T \tau \label{n1}
\eeq
which takes values on a temporal circle of unit radius. We also use the notation $\varphi_{12} \equiv \varphi_1 - \varphi_2$.

We begin with the saddle-point solution of Eq.~(\ref{action}), the Green's function $G_s (\varphi)$. In the scaling limit, this is given by Eq.~(\ref{Gsigma2}). We write this here as
\begin{equation}
G_s(\varphi)=b \frac{e^{-\mathcal{E} \varphi}}{\left(\sin \frac{\varphi}{2} \right)^{2\Delta}}, \quad \Delta=\frac{1}{q},\quad \varphi \in [0,2\pi ) \label{n2}
\end{equation}
where the prefactor $b$ is specified in Eq.~(\ref{Gsigma2}). We now 
expand the effective action Eq.~(\ref{action}) to quadratic order of the fluctuations $\delta G(\varphi_1, \varphi_2) = G(\varphi_1, \varphi_2) - G_s (\varphi_{12})$, $\delta \Sigma(\varphi_1, \varphi_2) = \Sigma(\varphi_1, \varphi_2) - \Sigma_s (\varphi_{12})$ and further integrate over $\delta \Sigma$. For convenience we use renormalized form of the fluctuation: 
%Inserting this into  leads to a quadratic form for fluctuations which is not symmetric. As argued
%in Ref.~\onlinecite{JMDS16}, the quadratic form can be made symmetric by writing instead
\beq
g(\varphi_1, \varphi_2)= \left[-G_s(\varphi_{12})G_s(\varphi_{21})\right]^{q/4} G_s(\varphi_{12})^{-1} \delta G(\varphi_1, \varphi_2) \label{n4}
\eeq
and obtaining the action (to quadratic order) in $g$:
\begin{align}
\frac{S_{\eff}}{N} = \frac{1}{2} \int d^{4} \varphi \, g(\varphi_1,\varphi_2) Q(\varphi_1,\varphi_2; \varphi_3,\varphi_4) g(\varphi_3,\varphi_4), \label{n5}
\end{align}
where $Q$ is a quadratic form on the space of functions with two time variables.

We now focus on just the zero mode fluctuations specified by the transformations in Eq.~(\ref{GGs}).
First, examine the infinitesmal reparameterization mode, with an accompanying U(1) transformation satisfying Eq.~(\ref{phif})
\beq
f(\varphi) = \varphi + \epsilon (\varphi) \quad , \quad \phi (\varphi) = -i \mathcal{E} \epsilon (\varphi). \label{n6}
\eeq
Notice that under this mode, $\widetilde{\phi}=0$ in Eq.~(\ref{deftildephi}). Inserting Eq.~(\ref{n6}) into Eq.~(\ref{GGs}), 
using 
Eq.~(\ref{n4}) to get the normalized fluctuations for each Fourier mode
$
\epsilon(\varphi) = \frac{1}{2\pi} \sum_n e^{-in\varphi} \epsilon_n, \label{n7}
$. 
we find that the linear order change in $g$ is 
\beq
g^\epsilon_n(\varphi_1,\varphi_2) =  \frac{i \Delta b^{q/2} e^{-2 \pi q \mathcal{E}/4} }{\pi} \left(\frac{  f_n(\varphi_{12})}{\left\vert \sin \frac{\varphi_{12}}{2} \right\vert} \right)\epsilon_n e^{-in \frac{\varphi_1+\varphi_2}{2}}. \label{n8}
\eeq
The functions $f_n (\varphi_{12})$ is a symmetric function of two variables $\varphi_1$, $\varphi_2$:
\beq
f_n(\varphi) = \frac{\sin n \frac{\varphi}{2}}{\tan \frac{\varphi}{2}} - n \cos n\frac{\varphi}{2} \quad , \quad \int_0^{2 \pi} d\varphi \left(  \frac{f_n(\varphi	)}{\sin \frac{\varphi }{2}} \right)^2 = \frac{2\pi}{3} |n|(n^2-1). \label{n9}
\eeq
Similarly, we can examine the
$\UU(1)$ fluctuation mode, under which $\epsilon$ is unchanged but $\phi$ changes:
\begin{equation}
g^\phi_n(\varphi_1,\varphi_2) = \frac{ b^{q/2} e^{-2\pi q\mathcal{E}/4} }{\pi}  \left( \frac{ \sin n \frac{\varphi_{12}}{2} }{\left\vert\sin \frac{\varphi}{2}\right\vert}  \right) \phi_n e^{-in \frac{\varphi_1+\varphi_2}{2}}, \label{n10}
\end{equation}
which implies that the phase fluctuation is anti-symmetric in two time variable. It is also useful to notice the following equation:
\beq
\int_0^{2 \pi} d\varphi \left(  \frac{ \sin n \frac{\varphi}{2} }{\sin \frac{\varphi}{2}} \right)^2 = 2\pi |n|. \label{n11}
\eeq

Turning to the structure of the quadratic form, $Q$, we now make the key observation that evaluating $Q$ from Eq.~(\ref{action}) and the conformal Green's function in Eq.~(\ref{n2}) leads to a vanishing action of $Q$ on the normal
modes described above. This is a direct consequence of the invariance of Eq.~(\ref{GGs}) under reparameterization and
$\UU(1)$ transformations. Ref.~\onlinecite{JMDS16} argued that going beyond the conformal limit will lead to 
a shift in the eigenvalue of $Q$ of order $|n| T/J$ in the first order perturbation theory. Assuming this applies here to both modes discussed above,\footnote{One can justify this statement by a renormalization theory argument in Ref.~\onlinecite{kitaevRussia}} we have
\begin{equation}
Q \cdot g^\phi_n = \alpha_\phi \frac{|n| T}{J} g^\phi_n \quad , \quad Q \cdot g^\epsilon_n = \alpha_\epsilon \frac{|n| T}{ J} g^\epsilon_n, \label{n12}
\end{equation}
where the numerical coefficients $\alpha_\phi$ and $\alpha_\epsilon$ cannot be obtained analytically,
but can be computed
in the large $q$ expansion. Here, we can fix them by comparing with the large $q$ results already obtained in Appendix~\ref{app:largeq}. 

Inserting Eq.~(\ref{n12}) into (\ref{n5}), and using the explicit form of the fluctuations in Eqs.~(\ref{n8}) and (\ref{n10}), we obtain the effective
action to quadratic order:
\begin{equation}
\frac{S_{\eff}}{N}= \frac{1}{2} \sum_n  \left\lbrace   c_\phi  n^2 |\widetilde{\phi}_n |^2  + c_\epsilon n^2 \left( n^2 - 1\right) |\epsilon_n|^2  \right\rbrace.
\label{quadact}
\end{equation}
where $c_\phi$ and $c_\epsilon$ are coefficients of order $\frac{T}{J}$ and proportional to $\alpha_\phi$ and $\alpha_\epsilon$.  
We confirm that this is of the form in Eq.~(\ref{pf2}), and 
we can further express the ratio of $K$ and $\gamma$ in terms of the numerical coefficients here
\begin{align}
\frac{K}{\gamma}= \frac{c_\phi}{4\pi^2 c_\epsilon} = \frac{3\alpha_\phi}{4\pi^2 \Delta^2 \alpha_\epsilon}
 \label{cvals}
\end{align}

%so the coefficients are given by
%\begin{align}
%c_\phi % \frac{ b^{q} e^{-2\pi q\mathcal{E}/2} }{\pi^2} \times 2\pi \times 2\pi  \times \frac{\alpha_\phi}{\beta J}
%= \frac{4 b^{q} e^{-2\pi q\mathcal{E}/2} T \alpha_\phi }{ J} %\Leftrightarrow \frac{KT}{2},\quad
%c_\epsilon=
%\frac{b^{q} e^{-2\pi q \mathcal{E}/2 } \Delta^2 }{\pi^2} \times 2\pi \times  \frac{2\pi}{3}  \times \frac{\alpha_\epsilon}{\beta J} =
%\frac{4b^{q} e^{-2\pi q \mathcal{E}/2 } \Delta^2 T %\alpha_\epsilon}{3  J} \Leftrightarrow \frac{T\gamma}{8\pi^2}. %\label{cvals}
%\end{align}

Using the effective action Eq.~(\ref{quadact}) we can also extract an order-one piece of the free energy which arises from the 1-loop calculation. In addition to the Schwarzian part that has been discussed in Ref.~\onlinecite{JMDS16}, we have a new piece from phase fluctuations $\widetilde{\phi}$:
\begin{equation}
Z_{\widetilde{\phi}} (\beta) = \sqrt{\det B}^{-1}, \quad B_{n,m}= \delta_{n+m} \frac{Nc_{\phi}}{2} n^2
\end{equation}
We can evaluate the determinant using the zeta function regularization :
\begin{equation}
\log Z_{\widetilde{\phi}} = - \left( \sum_{n=1}^\infty \log \frac{N c_{\phi}}{2} n^2 \right) = \frac{1}{2} \log \frac{N c_{\phi}}{8\pi^2} \sim -\frac{1}{2}\log{\beta J}
\end{equation}
Together with the contribution from Schwarzian ($ \sim - \frac{3}{2} \log \beta J $), we conclude that the partition function $Z(\beta)$ is proportional to $\beta^{-2}$ at large $\beta$. From this, one can further extract the low energy density of state $\rho(E)$ from inverse Laplace transformation of $Z(\beta)$, and show that $\rho(E)$ is proportional to $E$ at small $E\ll\frac{J}{N}$.

\begin{figure}
\center
\includegraphics[width=4in]{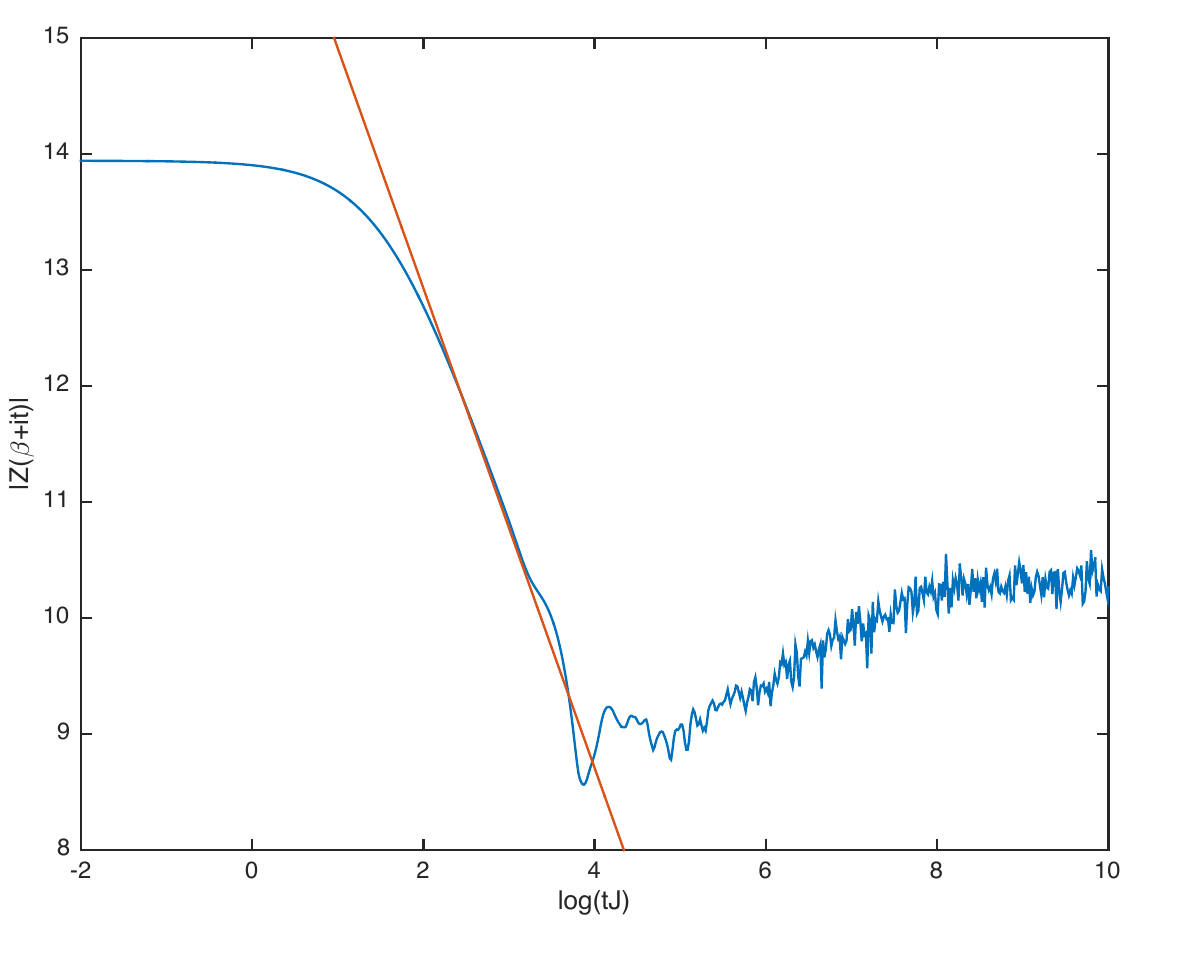}
\caption{Blue line is $|Z(\beta+it)|$ computed from exact diagonalization data for $N=15$ and $1000$ realizations at $\beta J=5$. The orange line is the linear fit for the "slope" regime and the slope is around $-2.07$.}
\label{1loop}
\end{figure}

We have also numerically computed a variation of partition function $|Z(\beta+it)|$ as in Ref.~\onlinecite{Cotler:2016fpe} using exact diagonalization, the result is shown in Fig.~\ref{1loop}. The slope is around $-2.07$ in the "slope" regime which is naively outsite the validity of the one-loop computation ($1\ll |\beta+i t|\ll N$), this is an indication of the one-loop exactness\cite{SWtoappear} of the complex SYK model.

%The ``$2\pi$'' in the middle come from the integral over the average time. 

%********************************************************
\section{Couplings in effective action of the SYK model}
\label{app:fluc}
%********************************************************

This appendix will present another derivation for the values of the couplings in the Schwarzian and phase fluctuation effective action in Eq.~(\ref{pf1}). Here, we will only obtain the leading quadratic terms in the gradient expansion, which have two temporal derivatives, although Eq.~(\ref{pf1}) contains many higher order terms. Just by matching these low order terms we will fix the couplings as in Eq.~(\ref{Kgamma}).

First we examine phase fluctuations, under which by Eq.~(\ref{GGs})
\begin{align}
\begin{split}
\label{Gphase}
G(\tau, \tau') &= e^{i \phi (\tau)} G_s(\tau- \tau') e^{- i \phi (\tau')} \\
\Sigma(\tau, \tau') &= e^{i \phi (\tau)} \Sigma_s (\tau- \tau') e^{- i \phi (\tau')} 
\end{split}
\end{align}
We insert the ansatz (\ref{Gphase}) into the action (\ref{action}), and perform a gradient expansion in derivatives
of $\phi (\tau)$. It is evident that the entire contribution comes from the $\mbox{Tr} \ln$ term, as the other terms are
independent of $\phi$. Furthermore, we can use the identity
\begin{align}
\begin{split}
\label{detid}
& \mbox{Tr} \ln \left[ \delta(\tau-\tau') \left(-\frac{\partial}{\partial \tau} + \mu \right) - e^{i \phi (\tau)} \Sigma_s (\tau- \tau') e^{- i \phi (\tau')} \right]
\\
&~~~ = \mbox{Tr} \ln \left[ \delta(\tau-\tau') \left(-\frac{\partial}{\partial \tau} + \mu + i \partial_\tau \phi (\tau) \right) -  \Sigma_s (\tau- \tau')  \right],
\end{split}
\end{align}
which is easily derived by a gauge transformation of the fermion fields that were integrated to obtain the determinant. 
In a gradient expansion about a saddle point at a fixed $\mu$, after all other modes (other than the reparameterization mode mentioned below) have been integrated 
out, we expect an effective action of the form
\beq
\frac{S_\phi}{N} = \frac{K}{2} \int_0^{1/T} d \tau (\partial_\tau \phi)^2 . \label{Sphi1}
\eeq
We can determine $K$ by evaluating the effective action for the special case where $\partial_\tau \phi$ a constant; under these conditions, we note from (\ref{detid}) that all we
have to do in the effective action is to make a small change in $\mu$ by  $ i \partial_\tau \phi$. Therefore, we have established that
\beq
K = - \left( \frac{\partial^2 \Omega}{\partial \mu^2} \right)_T
\eeq
is indeed the compressibility, as in Eq.~(\ref{Kgamma}).

A similar argument can made for energy fluctuations. Now we consider the temporal reparameterization 
\beq
\tau \rightarrow \tau + \epsilon (\tau) \label{taumap}
\eeq
After integrating out all other high energy modes at a fixed chemical potential (other than the phase mode above), we postulate an effective action for $\epsilon (\tau)$, and
assume that the lowest order gradient expansion leads to
\beq
\frac{S_\epsilon}{N} = \frac{\widetilde{K}}{2} \int_0^{1/T} d \tau (\partial_\tau \epsilon)^2 . \label{Sepsilon1}
\eeq
We can now relate the coefficient to a thermodynamic derivative. As for (\ref{Sphi1}), consider the case where
$\partial_\tau \epsilon$ is a constant.  Then (\ref{taumap}) implies a change in temperature
\beq
- \frac{\delta T}{T^2} = \frac{\partial_\tau \epsilon}{T} \label{deltaT}
\eeq
Inserting (\ref{deltaT}) into (\ref{Sepsilon1}), we conclude that 
\beq
\frac{\widetilde{K}}{T^2}  = \left( \frac{\partial^2 \Omega}{\partial T^2} \right)_\mu
\eeq

Finally, we can also fix the cross term by a similar argument, and so obtain the complete Gaussian effective action for $\phi$
and $\epsilon$ fluctuations, after all other modes have been integrated out
\beq
\frac{S_{\phi\epsilon}}{N} = \int_0^{1/T}  d\tau \left[ - \frac{1}{2} \left(\frac{\partial^2 \Omega}{\partial \mu^2} \right)_T (\partial_\tau \phi)^2 -
i T \frac{\partial^2 \Omega}{\partial T \partial \mu}   (\partial_\tau \epsilon)( \partial_\tau \phi) 
+ \frac{T^2}{2} \left(\frac{\partial^2 \Omega}{\partial T^2}\right)_\mu (\partial_\tau \epsilon)^2
\right].
\eeq
After application of thermodynamic identities, this is found to agree with the second order temporal derivatives in Eq.~(\ref{pf1}),
and the identifications in Eqs.~(\ref{defE0}) and (\ref{Kgamma}).

%********************************************************
\section{Diffusion constants of the higher-dimensional SYK model}
\label{app:diffusion}
%********************************************************

The generalization of the zero-dimensional SYK results in Appendix~\ref{app:normalmode}
to the higher dimensional models closely follows the lines of Ref.~\onlinecite{GQS16}.
In high dimensional models, the quadratic form $Q$ acquires a spatial dependence, formally we have $Q \rightarrow Q_{xy}$
where $Q_{xy}$ contains a hopping matrix for the fluctuations, 
which can be easily diagonalized by going to $k$-space. For long wavelength limit, we can expand its eigenvalue around $k=0$:
$
Q (k)= Q (0) + ck^2 +\ldots
$
where $c$ is a constant depends on $J_0$ and $J_1$ that captures the band structure at long wavelength, and $Q(0)$ is the quadratic form at $k=0$ which reproduces the quadratic form in $(0+1)$-dimension. % for the model in Eq.~(\ref{tt1}). 
In general, the hopping matrix acts differently on anti-symmetric fluctuation $g^{\phi}$ and symmetric fluctuation $g^\epsilon$, which will induce two different band structures $
Q(k)^\phi= Q(0)^\phi+ c_1 k^2+ \ldots
$ and $
Q(k)^\epsilon= Q(0)^\epsilon+ c_2 k^2+ \ldots
$ for charge and energy fluctuation respectively.\footnote{This is different from the SYK with Majorana fermions, where we have symmetries in Green's function when exchanging two time variables. More details about the properties of the fluctuations in complex SYK model will be discussed in Ref.~\onlinecite{FGS}. }

Inserting this back into the effective action derivation in Appendix~\ref{app:normalmode}, 
we notice that for the 
the $\phi$ modes, we need to replace the UV correction for $Q$ from
 $
 Q^\phi(0) \sim 
\alpha_\phi \frac{|n| T}{ J}$ to $ Q^\phi(k)= \alpha_\phi \frac{|n| T}{ J} +c_1 k^2
 $
Similarly, for 
$\epsilon$ modes, we need to replace 
$ \alpha_\epsilon \frac{|n|T}{ J} $ to $ \alpha_\epsilon \frac{|n|T}{ J} + c_2 k^2, 
$ where $J=\sqrt{J_0^2+J_1^2}$.
This replacement leads to the effective action in Eq.~(\ref{pf3}) with
\beq
D_1= \frac{2\pi c_1 J}{\alpha_\phi} \quad , \quad D_2= \frac{2\pi c_2 J}{\alpha_\epsilon} . 
\eeq
For the specific model we discussed in main text, the special form of the hopping term Eq.~(\ref{tt3}) leads to $c_1=c_2$.
Using Eq.~(\ref{cvals}), we then obtain the ratio of the diffusion constants
\begin{equation}
\frac{D_2}{D_1}= 
\frac{\alpha_\phi}{\alpha_\epsilon}
=
%\frac{\Delta^2 c_\phi }{3c_\epsilon}
%=
\frac{4\pi^2 \Delta^2}{3} \frac{K}{\gamma}
\end{equation}
which was
presented in Eq.~(\ref{Dratio}).

%********************************************************
\section{More general AdS$_2$ solutions}
\label{app:moregeneralads2solutions}
%********************************************************

The field theory dual to the solution (\ref{eq:simpleaxionsoln}) shares the property (\ref{eq:holoseebeckreln}) with the SYK model because of the AdS$_2$ factor in its near-horizon geometry. To further validate this, this Appendix will look at more complicated gravitational theories which also have solutions that break translational symmetry and have AdS$_2$ factors in their near-horizon geometry. The UV details of these differ from those of the solution (\ref{eq:simpleaxionsoln}), but we will find that the relation (\ref{seebeck}) is nevertheless obeyed. We will consider only homogeneous solutions for which we can write down analytic solutions. It would be interesting to see how far this result generalizes, particularly to cases where translational symmetry is broken inhomogeneously.

%********************************************************
\subsection{Asymptotically AdS$_4$}
%********************************************************

We will study a more general class of gravitational actions than (\ref{eq:charged4daction}), by including a new scalar field $\Phi$ in the four dimensional action. By choosing the potential and the gauge field coupling appropriately, one can find a whole class of solutions which are asymptotically AdS$_4$ and have a near-horizon AdS$_2$ geometry \cite{Gouteraux:2014hca}. The action is
\begin{equation}
S=\int d^4x\sqrt{-g}\left(\mathcal{R}-\frac{1}{2}(\partial\Phi)^2+V(\Phi)-\frac{1}{2}\sum_{i=1}^2\left(\partial\varphi_i\right)^2-\frac{Z(\Phi)}{4}F_{\mu\nu}F^{\mu\nu}\right),
\end{equation}
where $V(\Phi)$ and $Z(\Phi)$ are a family of functions depending on a single parameter $\delta$
\begin{equation}
\begin{aligned}
V(\Phi)=\frac{16\delta^2}{(1+\delta^2)^2}e^{\frac{(\delta^2-1)}{2\delta}\Phi}+\frac{2(3-\delta^2)}{(1+\delta^2)^2}e^{\delta\Phi}-\frac{2(1-3\delta^2)}{(1+\delta^2)^2}e^{-\Phi/\delta},\;\;\;\;\;\;\;\;\;\;Z(\Phi)=e^{-\delta\Phi},\\
\end{aligned}
\end{equation}
The asymptotically AdS$_4$ solutions which break translational symmetry homogeneously are
\begin{equation}
\begin{aligned}
&ds^2=-r^2f(r)h(r)^{-\frac{2}{1+\delta^2}}dt^2+\frac{dr^2}{r^2f(r)h(r)^{-\frac{2}{1+\delta^2}}}+r^2h^{\frac{2}{1+\delta^2}}d\vec{x}^2,\;\;\;\;\; \varphi_i=mx^i,\\
&f(r)=h(r)^{\frac{4}{1+\delta^2}}-\frac{r_0^3}{r^3}h(r_0)^{\frac{4}{1+\delta^2}}-\frac{m^2}{2r^2}\left(1-\frac{r_0}{r}\right),\;\;\;\;\;\;\;\;\;\;\;\;\;\;\;\;\;\;\;\;\;\;\; h(r)=1+\frac{Q}{r},\\
&A_t(r)=2\sqrt{\frac{Qr_0}{1+\delta^2}\left(h(r_0)^{2\frac{1-\delta^2}{1+\delta^2}}-\frac{m^2}{2r_0^2h(r_0)}\right)}\frac{\left(1-\frac{r_0}{r}\right)}{h(r)},\;\;\;\;\;\;\;\;\;\;\Phi=-\frac{2\delta}{1+\delta^2}\log h(r).
\end{aligned}
\end{equation}
When $\delta=0$, the scalar field vanishes, and the solution reduces to (\ref{eq:simpleaxionsoln}). When $\delta=1/\sqrt{3}$, the solution reduces to that studied in \cite{Davison:2013txa}, which is conformal to AdS$_2\times\mathbb{R}^2$ near the horizon at zero temperature, and has a linear-in-temperature entropy at small $T$. For any value $0\leq\delta<1/\sqrt{3}$, the solution has a near-horizon AdS$_2\times\mathbb{R}^2$ geometry at zero temperature, and we will restrict to this parameter range from now on, assuming that both $Q$ and $r_0$ are positive. The thermodynamic properties of this solution are
\begin{equation}
\begin{aligned}
\label{eq:moregeneralads2thermo}
&T=\frac{r_0}{4\pi}h(r_0)^{-\frac{2}{1+\delta^2}}\left(\frac{4-(1-3\delta^2)h(r_0)}{1+\delta^2}h(r_0)^{\frac{3-\delta^2}{1+\delta^2}}-\frac{m^2}{2r_0^2}\right),\;\;\;\;\;\;\;\;\;\;\mathcal{S}=4\pi r_0^2h(r_0)^\frac{2}{1+\delta^2},\\
&\mu=2\sqrt{\frac{Qr_0}{1+\delta^2}\left(h(r_0)^{2\frac{1-\delta^2}{1+\delta^2}}-\frac{m^2}{2r_0^2h(r_0)}\right)},\;\;\;\;\;\;\;\;\;\;\;\;\;\;\;\;\;\;\;\;\;\;\;\;\;\;\;\;\;\;\;\;\;\;\;\mathcal{Q}=\mu(Q+r_0),
\end{aligned}
\end{equation}
so that $T=0$ corresponds to the condition
\begin{equation}
\label{eq:zeroTads4ex}
\frac{4-(1-3\delta^2)h(r_0)}{1+\delta^2}h(r_0)^{\frac{3-\delta^2}{1+\delta^2}}=\frac{m^2}{2r_0^2}.
\end{equation}
To obtain the $T=0$ near-horizon geometry, one should perform the coordinate transformation
\begin{equation}
\label{eq:zerotcoordtransform}
\zeta=\frac{r-r_*}{\epsilon},\;\;\;\;\;\;\;\;\;\; \tau=\epsilon t,
\end{equation}
followed by the small $\epsilon$ limit to give
\begin{equation}
\begin{aligned}
&ds^2=\left(-\frac{\zeta^2}{\tilde{L}^2}d\tau^2+\frac{\tilde{L}^2}{\zeta^2}d\zeta^2\right)+r_0^2h(r_0)^\frac{2}{1+\delta^2}d\vec{x}^2+O(\epsilon),\\
&A_\tau=\frac{\mathcal{E}}{\tilde{L}^2}\zeta+O(\epsilon),\;\;\;\;\;\;\;\;\;\; \varphi_i=mx^i,\;\;\;\;\;\;\;\;\;\;\Phi=-\frac{2\delta}{1+\delta^2}\log(h(r_0)).
\end{aligned}
\end{equation}
The AdS$_2$ radius of curvature is
\begin{equation}
\tilde{L}^2=\frac{\left(1+\delta^2\right)^2 h(r_0)^\frac{2\delta^2}{1+\delta^2}}{2(3-\delta^2)-h(r_0)(1-3\delta^2)\left(4-h(r_0)(1-\delta^2)\right)},
\end{equation}
and the appropriately normalized AdS$_2$ electric field in these units is
\begin{equation}
\mathcal{E}=\tilde{L}^2A_\tau'({\zeta}) =2(1+\delta^2)\frac{\sqrt{(h(r_0)-1)(-(3-\delta^2)+h(r_0)(1-3\delta^2))}}{2(3-\delta^2)-h(r_0)(1-3\delta^2)(4-h(r_0)(1-\delta^2))}.
\end{equation}

The dc conductivities are given by properties of the solution at the horizon \cite{Blake:2013bqa,Donos:2014cya}. Explicitly, they are
\begin{equation}
\begin{aligned}
\sigma=\frac{4\pi\mathcal{Q}^2}{m^2\mathcal{S}}+Z(\Phi(r_0)),\;\;\;\;\;\;\;\;\;\;\alpha=\frac{4\pi\mathcal{Q}}{m^2},\;\;\;\;\;\;\;\;\;\;\bar{\kappa}=\frac{4\pi\mathcal{S}T}{m^2}.
\end{aligned}
\end{equation}
In the limit $T\rightarrow0$, the Seebeck coefficient is
\begin{equation}
S\equiv\frac{\alpha}{\sigma}=4\pi\left(1+\delta^2\right)\frac{\sqrt{(h(r_0)-1)\left(-(3-\delta^2)+h(r_0)(1-3\delta^2)\right)}}{2(3-\delta^2)-h(r_0)(1-3\delta^2)\left(4-h(r_0)(1-\delta^2)\right)}=2\pi\mathcal{E}.
\end{equation}

To verify the relation (\ref{seebeck}), we require the thermodynamic susceptibilities of these solutions. In general, it's not possible to invert (\ref{eq:moregeneralads2thermo}) to obtain closed form expressions for $\mathcal{S}(\mathcal{Q},T,m)$ etc. It is convenient to use $r_0,h(r_0)$ and $m$ as our independent parameters, such that (for fixed $m$),
\begin{equation}
\begin{aligned}
\delta \mathcal{S}=\frac{\partial \mathcal{S}}{\partial r_0}\Biggr|_{h(r_0)}\delta r_0+\frac{\partial \mathcal{S}}{\partial h(r_0)}\Biggr|_{r_0}\delta h(r_0),
\end{aligned}
\end{equation}
and similarly for other thermodynamic objects. Variations at fixed $T$ therefore correspond to the condition
\begin{equation}
\delta h(r_0)=-\frac{\partial T}{\partial r_0}\Biggr|_{h(r_0)}\left(\frac{\partial T}{\partial h(r_0)}\Biggr|_{r_0}\right)^{-1}\delta r_0,
\end{equation}
and so the relevant thermodynamic susceptibility can be written
\begin{equation}
\left(\frac{\partial \mathcal{S}}{\partial \mathcal{Q}}\right)_{T}=\frac{\frac{\partial T}{\partial h(r_0)}|_{r_0}\frac{\partial\mathcal{S}}{\partial r_0}|_{h(r_0)}-\frac{\partial\mathcal{S}}{\partial h(r_0)}|_{r_0}\frac{\partial T}{\partial r_0}|_{h(r_0)}}{\frac{\partial T}{\partial h(r_0)}|_{r_0}\frac{\partial\mathcal{Q}}{\partial r_0}|_{h(r_0)}-\frac{\partial\mathcal{Q}}{\partial h(r_0)}|_{r_0}\frac{\partial T}{\partial r_0}|_{h(r_0)}}.
\end{equation}
Evaluating this in the limit $T\rightarrow0$ gives
\begin{equation}
S=\left(\frac{\partial \mathcal{S}}{\partial \mathcal{Q}}\right)_{T}=2\pi\mathcal{E},
\end{equation}
in agreement with (\ref{seebeck}). For any non-zero $m$, the low energy correlators of the dual field theory should be those of diffusive hydrodynamics, and because of the relation (\ref{seebeck}), they will have the same form (\ref{pf4}) as those of the higher dimensional SYK model in the small $T$ limit, with the parameters
\begin{equation}
\begin{aligned}
r_0D_1&=h(r_0)^{-\frac{2}{1+\delta^2}}\frac{2(3-\delta^2)-h(r_0)(1-3\delta^2)(3-\delta^2(1-2h(r_0))}{\left(2(1-\delta^2)-h(r_0)(1-3\delta^2)\right)\left(4-h(r_0)(1-3\delta^2)\right)},\\
r_0D_2&=h(r_0)^{\frac{\delta^2-1}{\delta^2+1}}\frac{1+\delta^2}{2(1+h(r_0)\delta^2)},\\
r_0^{-1}K&=\frac{h(r_0)}{1+\delta^2}\frac{\left(2(1-\delta^2)-h(r_0)(1-3\delta^2)\right)\left(2(3-\delta^2)-h(r_0)(1-3\delta^2)(4-h(r_0)(1-\delta^2))\right)}{2(3-\delta^2)-h(r_0)(1-3\delta^2)(3-\delta^2+2\delta^2h(r_0))},\\
r_0^{-1}\gamma&=\frac{16\pi^2h(r_0)(1+\delta^2)(1+\delta^2h(r_0))}{2(3-\delta^2)-h(r_0)(1-3\delta^2)(4-h(r_0)(1-\delta^2))}.
\end{aligned}
\end{equation}
The zero temperature limit of the Wiedemann-Franz ratio, and of the modified Wiedemann-Franz ratio (\ref{eq:defnmodifiedWF}), for these solutions is
\begin{equation}
\begin{aligned}
\lim_{T\rightarrow0}L&=\frac{8\pi^2h(r_0)\left(1+\delta^2\right)^3\left(4-h(r_0)(1-3\delta^2)\right)}{\left(2(3-\delta^2)-h(r_0)(1-3\delta^2)\left(4-h(r_0)(1-\delta^2)\right)\right)^2},\\
\lim_{T\rightarrow0}L_\alpha&=-\mathcal{Q}\frac{\partial}{\partial\mathcal{Q}}\left(\frac{\mathcal{S}_0}{\mathcal{Q}}\right)\\
&=\frac{2\pi\left(1+\delta^2\right)^2h(r_0)(4-h(r_0)(1-3\delta^2))}{\sqrt{(h(r_0)-1)(-3+\delta^2+h(r_0)(1-3\delta^2))\left(2(3-\delta^2)-h(r_0)(1-3\delta^2)(4-h(r_0)(1-\delta^2))\right)}},
\end{aligned}
\end{equation}
which both vanish in the translationally invariant limit $m\rightarrow0$. The zero temperature `equation of state' $\mathcal{S}_0(\mathcal{Q})$ is given by the solution to
\begin{equation}
\begin{aligned}
&2\pi m^2\left(\frac{2(1-\delta^2)\mathcal{S}_0+(1+\delta^2)\sqrt{\mathcal{S}_0^2+4\pi^2(1-3\delta^2)\mathcal{Q}^2}}{\mathcal{S}_0\left(1-3\delta^2\right)}\right)^{\frac{\delta^2-1}{\delta^2+1}}\\
&+\sqrt{\mathcal{S}_0^2+4\pi^2\left(1-3\delta^2\right)\mathcal{Q}^2}-2\mathcal{S}_0=0.
\end{aligned}
\end{equation}

It naively appears that the relation (\ref{seebeck}) is true independently on the value of $\delta$. In particular it seems to apply outside the range $0\le\delta<1/\sqrt{3}$, where the solutions no longer have AdS$_2$ horizons. This is not the case -- the condition (\ref{eq:zeroTads4ex}) only corresponds to the $T=0$ limit of the system when $\delta$ is in this range. For example, for $\delta=1/\sqrt{3}$ (when the near-horizon geometry is conformal to AdS$_2\times \mathbb{R}^2$) the condition (\ref{eq:zeroTads4ex}) leads to an imaginary value of the chemical potential. The $T=0$ limit of the $\delta=1/\sqrt{3}$ solution is when $r_0=0$, and so the relation (\ref{seebeck}) is not true in this case. 

%********************************************************
\subsection{Asymptotically AdS$_5$}
%********************************************************

There are an analogous class of solutions which are asymptotically AdS$_5$ \cite{Gouteraux:2014hca}. In this case, the action is
\begin{equation}
S=\int d^5x\sqrt{-g}\left(\mathcal{R}-\frac{1}{2}(\partial\Phi)^2+V(\Phi)-\frac{1}{2}\sum_{i=1}^3\left(\partial\varphi_i\right)^2-\frac{Z(\Phi)}{4}F_{\mu\nu}F^{\mu\nu}\right),
\end{equation}
and the functions $V(\Phi)$ and $Z(\Phi)$ depend on a single parameter $\delta$
\begin{equation}
\begin{aligned}
V(\Phi)=\frac{18\delta^2(6\delta^2-1)}{\left(1+3\delta^2\right)^2}e^{-\frac{2\Phi}{3\delta}}+\frac{108\delta^2}{\left(1+3\delta^2\right)^2}e^{\frac{\Phi}{3\delta}(3\delta^2-1)}-\frac{6(3\delta^2-2)}{\left(1+3\delta^2\right)^2}e^{2\delta\Phi},\;\;\;\;\;\;\;\;\;\; Z(\Phi)=e^{-2\delta\Phi}.
\end{aligned}
\end{equation}
The asymptotically AdS$_5$ solutions that break translational symmetry homogeneously are
\begin{equation}
\begin{aligned}
&ds^2=-f(r)h(r)^{-\frac{2}{1+3\delta^2}}dt^2+\frac{dr^2}{f(r)h(r)^{-\frac{1}{1+3\delta^2}}}+r^2h^{\frac{1}{1+3\delta^2}}d\vec{x}^2,\;\;\;\;\;\;\;\;\;\; \varphi_i=mx^i,\\
&f(r)=r^2\left(h(r)^{\frac{3}{1+3\delta^2}}-\frac{r_0^4}{r^4}h(r_0)^{\frac{3}{1+3\delta^2}}\right)-\frac{m^2}{4}\left(1-\frac{r_0^2}{r^2}\right),\;\;\;\;\;\;\;\;\;\;\;\;\;\;\; h(r)=1+\frac{Q}{r^2},\\
&A_t(r)=\sqrt{\frac{3Q}{1+3\delta^2}\left(h(r_0)^{\frac{1-6\delta^2}{1+3\delta^2}}-\frac{m^2}{4r_0^2h(r_0)}\right)}\frac{\left(1-\frac{r_0^2}{r^2}\right)}{h(r)},\;\;\;\;\;\;\;\;\;\;\;\;\;\;\;\;\;\Phi=-\frac{3\delta}{1+3\delta^2}\log h(r).
\end{aligned}
\end{equation}
We will assume that both $Q$ and $r_0$ are positive. The thermodynamic properties of these solutions are
\begin{equation}
\begin{aligned}
&T=\frac{r_0}{2\pi}h(r_0)^{-\frac{3}{2(1+3\delta^2)}}\left(\frac{3-(1-6\delta^2)h(r_0)}{(1+3\delta^2)}h(r_0)^{\frac{2-3\delta^2}{1+3\delta^2}}-\frac{m^2}{4r_0^2}\right),\;\;\;\;\;\;\;\;\;\mathcal{S}=4\pi r_0^3h(r_0)^{\frac{3}{2(1+3\delta^2)}},\\
&\mu=\sqrt{\frac{3Q}{1+3\delta^2}\left(h(r_0)^{\frac{1-6\delta^2}{1+3\delta^2}}-\frac{m^2}{4r_0^2h(r_0)}\right)},\;\;\;\;\;\;\;\;\;\;\;\;\;\;\;\;\;\;\;\;\;\;\;\;\;\;\;\;\;\;\;\;\;\;\;\;\;\;\;\mathcal{Q}=2\mu(Q+r_0^2).
\end{aligned}
\end{equation}
For $0\le\delta<1/\sqrt{6}$, the $T=0$ geometries are found by imposing the condition
\begin{equation}
\frac{3-(1-6\delta^2)h(r_0)}{1+3\delta^2}h(r_0)^{\frac{2-3\delta^2}{1+3\delta^2}}=\frac{m^2}{4r_0^2},
\end{equation}
and we will restrict to these values of $\delta$ from now on. After changing coordinates to
\begin{equation}
r=r_0+\epsilon h(r_0)^{\frac{1}{2(1+3\delta^2)}}\zeta,\;\;\;\;\;\;\;\;\;\;\;\;\;\;\;\;\;\;\;\; t=\frac{\tau}{\epsilon},
\end{equation}
and taking the near-horizon limit $\epsilon\rightarrow0$, we find a $T=0$ charged AdS$_2\times\mathbb{R}^3$ geometry
\begin{equation}
\begin{aligned}
&ds^2=\left(-\frac{\zeta^2}{\tilde{L}^2}d\tau^2+\frac{\tilde{L}^2}{\zeta^2}d\zeta^2\right)+r_0^2h(r_0)^{\frac{1}{1+3\delta^2}}d\vec{x}^2+O(\epsilon),\\
&A_\tau=\frac{\mathcal{E}}{\tilde{L}^2}\zeta+O(\epsilon),\;\;\;\;\;\;\;\;\;\;\varphi_i=mx^i,\;\;\;\;\;\;\;\;\;\;\Phi=-\frac{3\delta}{1+3\delta^2}\log(h(r_0)),
\end{aligned}
\end{equation}
with the AdS$_2$ radius of curvature
\begin{equation}
\tilde{L}^2=\frac{(1+3\delta^2)^2h(r_0)^{\frac{6\delta^2}{1+3\delta^2}}}{2\left(3(2-3\delta^2)-h(r_0)(1-6\delta^2)\left(6-h(r_0)(2-3\delta^2)\right)\right)},
\end{equation}
and the electric field
\begin{equation}
\mathcal{E}=(1+3\delta^2)\frac{\sqrt{3(h(r_0)-1)\left(-2+3\delta^2+h(r_0)(1-6\delta^2)\right)}}{3(2-3\delta^2)-h(r_0)(1-6\delta^2)\left(6-h(r_0)(2-3\delta^2)\right)}.
\end{equation}
The dc conductivities can be computed using the usual techniques \cite{Blake:2013bqa,Donos:2014cya}, and are given by 
\begin{equation}
\sigma=\frac{\mathcal{S}}{4\pi g_{xx}(r_0)}Z(\Phi(r_0))+\frac{4\pi\mathcal{Q}^2}{m^2\mathcal{S}},\;\;\;\;\;\;\;\;\;\;\alpha=\frac{4\pi\mathcal{Q}}{m^2},\;\;\;\;\;\;\;\;\;\;\bar{\kappa}=\frac{4\pi\mathcal{S}T}{m^2}.
\end{equation}
The $T\rightarrow0$ limit of the Seebeck coefficient is then
\begin{equation}
S\equiv\frac{\alpha}{\sigma}=2\pi(1+3\delta^2)\frac{\sqrt{3(h(r_0)-1)\left(-2+3\delta^2+h(r_0)(1-6\delta^2)\right)}}{3(2-3\delta^2)-h(r_0)(1-6\delta^2)\left(6-h(r_0)(2-3\delta^2)\right)}.
\end{equation}
Computing $(\partial\mathcal{S}/\partial\mathcal{Q})_T$ in a similar manner to the previous subsection, we find that equation (\ref{seebeck}) is true. This is further evidence that (\ref{seebeck}) is a consequence of the AdS$_2$ part of the near-horizon geometry. The low energy correlators of the field theory states dual to these solutions will have the form (\ref{pf4}) in the small $T$ limit, with
\begin{equation}
\begin{aligned}
r_0D_1&=h(r_0)^{-\frac{3}{2(1+3\delta^2)}}\frac{3(2-3\delta^2)-h(r_0)(1-6\delta^2)\left(4-3\delta^2(2-3h(r_0))\right)}{2\left(3(1-3\delta^2)-2h(r_0)(1-6\delta^2)\right)\left(3-h(r_0)(1-6\delta^2)\right)},\\
r_0D_2&=h(r_0)^{\frac{-1+6\delta^2}{2(1+3\delta^2)}}\frac{1+3\delta^2}{3(1+3h(r_0)\delta^2)},\\
r_0^{-2}K&=\frac{2h(r_0)}{1+3\delta^2}\frac{\left(3(1-3\delta^2)-2h(r_0)(1-6\delta^2)\right)\left(3(2-3\delta^2)-h(r_0)(1-6\delta^2)(6-h(r_0)(2-3\delta^2))\right)}{3(2-3\delta^2)-h(r_0)(1-6\delta^2)\left(2(2-3\delta^2)+9h(r_0)\delta^2\right)},\\
r_0^{-2}\gamma&=\frac{12\pi^2h(r_0)(1+3\delta^2)(1+3h(r_0)\delta^2)}{3(2-3\delta^2)-h(r_0)(1-6\delta^2)(6-h(r_0)(2-3\delta^2))}.\\
\end{aligned}
\end{equation}
It is straightforward to calculate the Wiedemann-Franz ratio $L$, and the modified ratio $L_\alpha$ ,which have the zero temperature values
\begin{equation}
\begin{aligned}
\lim_{T\rightarrow0}L&=\frac{4\pi^2h(r_0)(1+3\delta^2)^3\left(3-h(r_0)(1-6\delta^2)\right)}{\left(3(2-3\delta^2)-h(r_0)(1-6\delta^2)\left(6-h(r_0)(2-3\delta^2)\right)\right)^2},\\
\lim_{T\rightarrow0}L_\alpha&=-\mathcal{Q}\frac{\partial}{\partial\mathcal{Q}}\left(\frac{\mathcal{S}_0}{\mathcal{Q}}\right)\\
&=\frac{2\pi\left(1+3\delta^2\right)^2h(r_0)(3-h(r_0)(1-6\delta^2))}{\sqrt{3(h(r_0)-1)(-2+3\delta^2+h(r_0)(1-6\delta^2)}{\left(3(2-3\delta^2)-h(r_0)(1-6\delta^2)(6-h(r_0)(2-3\delta^2))\right)}},
\end{aligned}
\end{equation}
for these solutions. These vanish when translational invariance is restored ($m\rightarrow0$). The zero temperature `equation of state' $\mathcal{S}_0(\mathcal{Q})$ is given by the solution to the equation
\begin{equation}
\begin{aligned}
&6^{\frac{3}{1+3\delta^2}}\pi m^3\mathcal{S}_0^{1/2}\left(\frac{9\mathcal{S}_0(1-3\delta^2)+(1+3\delta^2)\sqrt{9\mathcal{S}_0^2+48\pi^2(1-6\delta^2)\mathcal{Q}^2}}{\mathcal{S}_0(1-6\delta^2)}\right)^{\frac{3(3\delta^2-1)}{2(3\delta^2+1)}}\\
&-2\left(9\mathcal{S}_0-\sqrt{9\mathcal{S}_0^2+48\pi^2(1-6\delta^2)\mathcal{Q}^2}\right)^{3/2}=0.
\end{aligned}
\end{equation}

\bibliography{syk}

\end{document}